\documentclass[iop]{emulateapj}

\shorttitle{Physical properties of 15 quasars at $z\gtrsim 6.5$}
\shortauthors{Mazzucchelli et al.}

\begin{document}


\title{Physical properties of 15 quasars at $z\gtrsim 6.5$}


\author{Mazzucchelli C.\altaffilmark{1},
Ba\~{n}ados E.\altaffilmark{1,2,$\dagger$},
Venemans B. ~P.\altaffilmark{1},
Decarli R.\altaffilmark{1,3},
Farina E. ~P.\altaffilmark{1},
Walter F.\altaffilmark{1},
Eilers A.-C.\altaffilmark{1},
Rix H.-W.\altaffilmark{1},
Simcoe R.\altaffilmark{4},
Stern D.\altaffilmark{5},
Fan X.\altaffilmark{6},
Schlafly E.\altaffilmark{7,$\ddagger$},
De Rosa G.\altaffilmark{8},
Hennawi J.\altaffilmark{1,9},
Chambers K. C.\altaffilmark{10},
Greiner J.\altaffilmark{11}, 
Burgett W.\altaffilmark{10},
Draper P.W.\altaffilmark{12},
Kaiser N.\altaffilmark{10},
Kudritzki R.-P.\altaffilmark{10},
Magnier E.\altaffilmark{10},
Metcalfe N.\altaffilmark{12},
Waters C.\altaffilmark{10},
Wainscoat R. ~J.\altaffilmark{10}
}
\affil{\altaffilmark{1}Max-Planck-Institut f$\rm \ddot{u}$r Astronomie, K{\"o}nigstuhl 17, D-69117 Heidelberg, Germany}
\affil{\altaffilmark{2}The Observatories of the Carnegie Institute of Washington, 813 Santa Barbara Street, Pasadena, CA 91101, USA}
\affil{\altaffilmark{3} INAF $-$ Osservatorio Astronomico di Bologna, via Gobetti 93/3, I-40129, Bologna, Italy}
\affil{\altaffilmark{4}MIT-Kavli Center for Astrophysics and Space Research, 77 Massachusetts Avenue, Cambridge, MA 02139, USA}
\affil{\altaffilmark{5}Jet Propulsion Laboratory, California Institute of Technology, 4800 Oak Grove Drive, Pasadena, CA 91109, USA}
\affil{\altaffilmark{6}Steward Observatory, The University of Arizona, 933 North Cherry Avenue, Tucson, AZ 85721–0065, USA}
\affil{\altaffilmark{7}Lawrence Berkeley National Laboratory, Berkeley, CA, 94720, USA}
\affil{\altaffilmark{8}Space Telescope Science Institute, 3700 San Martin Drive, Baltimore, MD 21218, USA}
\affil{\altaffilmark{9}Physics Department, University of California, Santa Barbara, CA 93106-9530, USA}
\affil{\altaffilmark{10}Institute for Astronomy, University of Hawaii, 2680 Woodlawn Drive, Honolulu,HI 96822, USA}
\affil{\altaffilmark{11}Max-Planck-Institut f$\rm \ddot{u}$r extraterrestrische Physik, Giessenbachstrasse 1, D-85748 Garching, Germany}
\affil{\altaffilmark{12}Department of Physics, Durham University, South Road, Durham DH1 3LE, UK}
\affil{\altaffilmark{$\dagger$}Carnegie-Princeton Fellow}
\affil{\altaffilmark{$\ddagger$}Hubble Fellow}





\begin{abstract}
Quasars are galaxies hosting accreting supermassive black holes; due to their brightness, they are unique probes of the early universe. To date, only few quasars have been reported at $z > 6.5$ ($<$800 Myr after the Big Bang).
In this work, we present six additional $z \gtrsim 6.5$ quasars discovered using the Pan-STARRS1 survey.
We use a sample of 15 $z \gtrsim 6.5$ quasars to perform a homogeneous and comprehensive analysis of this highest-redshift quasar population.
We report four main results:
(1) the majority of $z\gtrsim$6.5 quasars show large blueshifts of the broad C$\,${\scriptsize IV} 1549\AA$\,$emission line compared to the systemic redshift of the quasars, with a median value $\sim$3$\times$ higher than a quasar sample at $z\sim$1;
(2) we estimate the quasars' black hole masses (M$\rm _{BH}\sim$0.3$-$5 $\times$ 10$^{9}$ M$_{\odot}$) via modeling of the Mg$\,${\scriptsize II} 2798\AA$\,$emission line and rest-frame UV continuum; we find that quasars at high redshift accrete their material
(with $\langle (L_{\mathrm{bol}}/L_{\mathrm{Edd}}) \rangle = 0.39$) at a rate comparable to a luminosity-matched sample at lower$-$redshift, albeit with significant scatter ($0.4$ dex);
(3) we recover no evolution of the Fe$\,${\scriptsize II}/Mg$\,${\scriptsize II} abundance ratio with cosmic time;
(4) we derive near zone sizes; together with measurements for $z\sim6$ quasars from recent work, we confirm a shallow evolution of the decreasing quasar near zone sizes with redshift. Finally, we present new millimeter observations of the [CII] 158 $\mu$m emission line and underlying dust continuum from NOEMA for four quasars, and provide new accurate redshifts and [CII]/infrared luminosities estimates.
The analysis presented here shows the large range of properties of the most distant quasars.
\end{abstract}

\section{Introduction} \label{secIntro}
Quasars are massive galaxies hosting fast accreting supermassive black holes ($\gtrsim 10^8$ M$_{\odot}$) in their centers. They are the most luminous, non-transient sources in the sky, and hence they can be observed at extremely large cosmological look-back times ($z>$6, $<$1 Gyr after the Big Bang), where normal star-forming galaxies are often too faint to be comprehensively studied. Quasars are therefore unique lighthouses, illuminating a number of open issues regarding the very early stages of the universe.

First, their very presence at such primeval cosmic times challenges models of the formation and growth of supermassive black holes (e.g. \citealt{Volonteri10}, \citealt{Latif16}).
The current preferred models include the formation of black hole seeds from the direct collapse of massive gaseous reservoirs (e.g., \citealt{Haehnelt93}, \citealt{Latif&Schleicher15}), the collapse of Population III stars (e.g., \citealt{Bond84}, \citealt{Alvarez09}, \citealt{Valiante16}), the co-action of dynamical processes, gas collapse and star formation (e.g., \citealt{Devecchi&Volonteri09}), or the rapid growth of stellar-mass seeds via episodes of super-Eddington, radiatively inefficient accretion (e.g., \citealt{Madau14}, \citealt{Alexander14}, \citealt{Pacucci15}, \citealt{Volonteri16}, \citealt{Lupi16}, \citealt{Pezzulli16}, \citealt{Begelman17}).
From black hole growth theory, we know that black holes can evolve very rapidly from their initial seed masses $M_{\mathrm{BH,seed}}$ to the final mass $M_{\mathrm{BH,f}}$ ($M_{\mathrm{BH,f}}\sim M_{\mathrm{BH,seed}}\, e^{9 \times t\,[\mathrm{Gyr}]/0.45}$; assuming 
accretion at the Eddington limit and an efficiency of 10\%; \citealt{VolonteriRees05}).
For instance, in the seemingly short redshift range $z \sim 6.0 - 6.5$, corresponding to $\sim90$ Myr, a black hole can grow by a factor of six.
From the observational perspective, the discovery of quasars at $z\gtrsim$6.5 can give stronger constraints on the nature of black hole seeds than the quasar population at $z\sim 6$.  

Moreover, several studies show that quasars at $z\sim$6 are hosted in massive, already chemically evolved galaxies (e.g., \citealt{Barth03}, \citealt{Stern03}, \citealt{Walter03}, \citealt{DeRosa11}). These galaxies contain a conspicuous amount of cool gas and dust, as observed through the detection of the bright [CII] 158 $\mu$m emission line and its underlying continuum, falling in the millimeter regime at $z \gtrsim 5.5$ (e.g., \citealt{Maiolino09}, \citealt{Walter09}, \citealt{Wang13}, \citealt{Willott15}, \citealt{Venemans16}, \citealt{DecarliSub}; for a review see \citealt{Carilli13}).
Sampling the cool gas content of high redshift quasar host galaxies with millimeter observations is therefore of great importance in pinpointing the gas content of massive galaxies in the universe at early ages.

Finally, the bright quasar emission has been used as background light to study the conditions of the intergalactic medium (IGM) at the epoch of reionization (EoR), when the universe transitioned from being neutral to the current, mostly ionized state.
The current best constraints on the EoR are derived from the Cosmic Microwave Background (CMB) and quasar spectra.
In the former case, recent CMB measurements by the \cite{Planck16} set a redshift of $z\sim8.8$ for the EoR (under the hypothesis that the reionization is instantaneous).
In the latter case, several studies investigate the evolution of the IGM ionized fraction during the EoR through high$-z$ quasar emission, e.g. by measuring transmission spikes in the Ly$\alpha$ forest (e.g., \citealt{Fan06}, \citealt{Becker15}, \citealt{Barnett17}), and by computing the Ly$\alpha$ power spectrum (e.g., \citealt{PalanqueDelabrouille13}).
Another method is based on measurements of near zone sizes, e.g. regions around quasars which are ionized by emission from the central objects. Their evolution with redshift has been studied to investigate the evolution of the IGM neutral fraction with cosmic time (e.g., \citealt{Fan06}, \citealt{Carilli10}, \citealt{Venemans15}).
However, the modest-sized and non-homogeneous quasar samples at hand, the large errors due to uncertain redshifts, and the limited theoretical models available have inhibited our understanding of these measurements to date, i.e. do they trace the evolution of the ionized gas fraction or are they dominated by degeneracies (e.g. quasars lifetimes)?
Recently, \cite{Eilers17} addressed some of these caveats, deriving near zone sizes of 34 quasars at $5.77 \lesssim z \lesssim 6.54$.
They find a less pronounced evolution of near zone radii with redshift than what has been reported by previous studies (e.g. \citealt{Carilli10} and \citealt{Venemans15}).
Measurements from quasars at higher redshift are required to test if this trend holds far deeper into the EoR.
To further progress in all the issues reported above, it is of paramount importance to identify new quasars, especially at the highest redshifts, and study their properties comprehensively.

Color selection techniques, which rely on multi-wavelength broad band observations, are among the most commonly used methods to find high redshift quasars. The quasar flux at wavelengths shorter than the Ly$\alpha$ emission line (at rest frame $\lambda_{rf}=1215.67$ \AA) is absorbed by the intervening neutral medium, causing an extremely red ($i-z$) or ($z-y$) color if the source is at $z\gtrsim 6$ ($i-$dropouts) or $z\gtrsim 6.4$ ($z-$dropouts), respectively. 
In the last two decades $\sim$200 quasars have been discovered at $5.4 < z < 6.4$, mainly thanks to the advent of large-area surveys: e.g., the Sloan Digital Sky Survey (SDSS; \citealt{Fan00}, \citeyear{Fan03}, \citeyear{Fan06}, \citealt{Jiang16}, \citealt{WangF16}, \citeyear{FWang17}); the Canada-France High-redshift Quasar Survey (CFHQS; \citealt{Willott07}, \citeyear{Willott09}, \citeyear{Willott10a},b); the UK Infrared Deep Sky Server (UKIDSS;\citealt{Venemans07}, \citealt{Mortlock09}); the Dark Energy Survey (DES; \citealt{Reed15}, \citeyear{Reed17}); the Very Large Telescope Survey Telescope (VST) ATLAS Survey (\citealt{Carnall15}); the ESO public Kilo Degree Survey (KiDS; \citealt{Venemans15b}) and the Panoramic Survey Telescope and Rapid Response System (Pan-STARRS1 or PS1; \citealt{Morganson12}, \citealt{Banados14}, \citeyear{Banados15b}, \citeyear{Banados16}).
However, the search for sources at even higher redshift ($z\gtrsim$6.4; age of the universe $<$0.80 Gyr) has been extremely challenging, and only a few quasars have been discovered at such distance prior to the results presented here: three from the VISTA Kilo-Degree Infrared Galaxy Survey (VIKING; \citealt{Venemans13}), four from PS1 (\citealt{Venemans15}, \citealt{Tang17}), one from the Subaru Hyper Suprime-Cam-SPP Survey (HSC-SPP; \citealt{Matsuoka16}, \citeyear{Matsuoka17})
; so far, only one quasar has been found at $z>7$ in the UKIDSS survey \citep{Mortlock11}.

In this work, we describe our search for $z\gtrsim 6.5$ quasars in the Pan-STARRS1 survey (\citealt{Kaiser02}, \citeyear{Kaiser10}, \citealt{Chambers16}, \citealt{Magnier16a},b,c, \citealt{Waters16}, \citealt{Flewelling16}), which imaged the entire sky at Decl.$>-30^{\circ}$ in five filters ($g_{\mathrm{P1}}$, $r_{\mathrm{P1}}$, $i_{\mathrm{P1}}$, $z_{\mathrm{P1}}$, $y_{\mathrm{P1}}$). We use here the third internal release of the 3$\pi$ stacked catalog (PS1 PV3, in the internal naming convention). The 5$\sigma$ AB magnitude limits are ($g_{\mathrm{P1}}$, $r_{\mathrm{P1}}$, $i_{\mathrm{P1}}$, $z_{\mathrm{P1}}$, $y_{\mathrm{P1}}$)=(23.3, 23.2, 23.1, 22.4, 21.4)\footnote{See also \cite{Chambers16}, Table 11.}.
We present six newly discovered $z-$dropouts from this search, at $z\sim$6.5 ($6.42 < z < 6.59$).
We then provide a comprehensive  analysis of the sample of the known $z \gtrsim 6.5$ quasars (15 objects)\footnote{We do not consider the quasars VDESJ0224$-$4711 ($z=$6.50; \citealt{Reed17}), DELS J104819.09$-$010940.2 ($z=$6.63; \citealt{FWang17}) and J1429$-$0104 ($z=6.80$; \citealt{Matsuoka17}), which were reported during the final stages of the preparation of this manuscript.}.
Our goal is to implement a coherent investigation of several key quasar properties (i.e., black hole mass, accretion rate, near zone size and infrared luminosity), and compare them to lower redshift samples.
The paper is organized as follows:
in Section \ref{secCandSel} we present our method for selecting quasar candidates from the PS1 PV3 database together with other publicly available surveys;
in Section \ref{secFolUpObs} we report the imaging and spectroscopic follow up observations obtained to confirm the quasar nature of our candidates; we also present new NIR/optical spectroscopy of quasars from the literature, and new observations of the [CII] 158 $\mu$m emission line and underlying continuum for 4 quasars.
In Section \ref{secNotesPS1} we discuss the properties of each of the new PS1 quasars presented here.
In Section \ref{secAnalysis} we present our quasar sample at $z\gtrsim 6.5$: redshifts ($\S$\ref{subsecRedTempl}), absolute magnitudes at rest-frame wavelength 1450 \AA$\,$ ($\S$\ref{subsecM1450}), C$\,${\scriptsize IV} $\lambda$1549.06 \AA$\,$ broad emission line characteristics ($\S$\ref{subsecCIV}), black hole masses, bolometric luminosities and accretion rates ($\S$\ref{secQSOCont}$-\S$\ref{subsecBHM}), iron-to-magnesium flux ratios ($\S$\ref{subsecFeMgII}), infrared and [CII] luminosities ($\S$\ref{subsecCII}) and near zone sizes ($\S$\ref{subsecNZ}).
Finally, in Section \ref{secDiscConc} we discuss and summarize our findings.
The International Astronomical Union imposes in its naming convention that all non-transient sources discovered in the PS1 survey are named ``PSO JRRR.rrrr$\pm$DD.dddd'', with RRR.rrrr and DD.dddd right ascension and declination in decimal degrees (J2000), respectively. For simplicity, in this paper we will refer to the PS1 quasars as ``PSORRR+DD'', and to sources from other surveys, e.g. VIKING, UKIDSS and HSC, as ``VIKhhmm'', ``ULAShhmm'' and ``HSChhmm''.

We consider throughout the paper the PS1 PSF magnitudes ($g_{\mathrm{P1}}$, $r_{\mathrm{P1}}$, $i_{\mathrm{P1}}$, $z_{\mathrm{P1}}$, $y_{\mathrm{P1}}$).
The magnitudes reported in this work are all in the AB system. We use a $\rm \Lambda$CDM cosmology with H$\rm _{0}=$70 km s$^{-1}$ Mpc$^{-1}$, $\rm \Omega_{m}=$0.3, and $\rm \Omega_{\Lambda}=$ 0.7.

\section{Candidate Selection} \label{secCandSel}
We perform a search for $z-$dropouts in the Pan-STARRS1 survey using the PS1 PV3 catalog (see Section \ref{secIntro}). We follow and expand the selection illustrated both in \cite{Banados16}, which was focused on lower redshift objects ($z\sim$ 6), and in \cite{Venemans15}.\\
Samples of high redshift quasar candidates selected through broad-band imaging and optical color criteria are highly contaminated by the numerous cool dwarf stars in our Galaxy (mainly M/L/T-dwarfs), which present similar colors and morphology. We therefore compile our sample and clean it from contaminants through the following steps:
\begin{itemize}
\item initial search based on the PS1 PV3 catalog and cross-match with known cool dwarf and quasar lists;
\item cross-match with other infrared public surveys;
\item forced photometry on the stacked and single epoch PS1 images;
\item fit of the spectral energy distribution (SED); and
\item visual inspection.
\end{itemize}
Afterwards, we follow up the selected candidates with dedicated photometric campaigns, followed by spectroscopy of the remaining targets to confirm (or discard) their quasar nature (see Section \ref{secFolUpObs}).
\subsection{PS1 Catalog} \label{secCandSelPS1}
The flux of high$-$redshift quasars at wavelengths shorter than the Ly$\alpha$ emission line is strongly absorbed by the intervening intergalactic medium. Therefore, we expect to recover little or no flux in the bluer bands, and to observe a strong break of the continuum emission.
We base our selection of z-dropouts on the $y_{\mathrm{P1}}$ magnitude, and require the objects to have S/N$(y_{\mathrm{P1}})>7$, where S/N is the signal-to-noise ratio. Then, we require S/N$(g_{\mathrm{P1}}$, $r_{\mathrm{P1}})<$3 and S/N$(i_{\mathrm{P1}})<$5, or, in case the latter criterion is not satisfied, a color $(i_{\mathrm{P1}}-y_{\mathrm{P1}}) > 2.2$.
Furthermore, we require a ($z_{\mathrm{P1}}-y_{\mathrm{P1}}$) color criterion as:
\begin{eqnarray}
   &\mathrm{S/N}(z_{\mathrm{P1}})>3  \quad \mathrm{and} \quad z_{\mathrm{P1}}-y_{\mathrm{P1}}>1.4 \quad \mathrm{or}\\
   &\mathrm{S/N}(z_{\mathrm{P1}})<3 \quad \mathrm{and} \quad z_{\mathrm{P1,lim}}-y_{\mathrm{P1}}>1.4
\end{eqnarray}
In order to reject objects with an extended morphology, we require:
\begin{equation}
| y_{\mathrm{P1}}-y_{\mathrm{P1,aper}} | < 0.3
\end{equation}
where $y_{\mathrm{P1,aper}}$ is the aperture magnitude in the PS1 catalog. This cut was implemented based on a test performed on a sample of spectroscopically confirmed stars and galaxies (from SDSS-DR12, \citealt{Alam15}), and quasars at $z>2$ (from SDSS-DR10, \citealt{Paris14}). Using this criterion, we are  able to select a large fraction of point-like sources (83\% of quasars and 78\% of stars) and reject the majority of galaxies (94\%; see \citealt{Banados16} for more details on this approach). Additionally, we discard objects based on the quality of the $y_{\mathrm{P1}}$ band image using the flags reported in the PS1 catalog (e.g., we require that the peak of the object is not saturated, and that it does not land off the edge of the chip or on a diffraction spike; for a full summary, see \citealt{Banados14}, Appendix A). We require also that 85\% of the expected PSF-weighted flux in the $z_{\mathrm{P1}}$ and $y_{\mathrm{P1}}$ bands falls in a region of valid pixels (the catalog entry \texttt{PSF\_QF} $>$0.85).
We exclude objects in regions of high Galactic extinction ($E(B-V)>0.3$), following the extinction map of \cite{Schlegel98}; we also exclude the area close to M31 (00:28:04$<$R.A.$<$00:56:08 and $37^{\circ}<$Decl.$< 43^{\circ}$).
We clean the resulting sample by removing known quasars at $z\geq$5.5 (see references in \citealt{Banados16}, Table 7). 
The total number of candidates at this stage is $\sim$781$^{\cdot}$000.
\subsection{Cross-match with public surveys} \label{secCandSelSurv}
We take advantage of the information provided by other public surveys, when their sky coverage overlaps with Pan-STARRS1.
We here consider solely the sources with a detection in the WISE catalog.
\subsubsection{ALLWISE Survey}
We consider the ALLWISE catalog\footnote{http://wise2.ipac.caltech.edu/docs/release/allwise/}, resulting from the combination of the all-sky \textit{Wide-field Infrared Survey Explorer} mission (\textit{WISE} mission; \citealt{Wright10}) and the NEOWISE survey (\citealt{Mainzer11}). The 5$\sigma$ limiting magnitudes are $W$1=19.3, $W$2= 18.9 and $W$3=16.5. We use a match radius of 3\arcsec, requiring S/N$>$3 in $W$1 and $W$2. We further impose:
\begin{eqnarray}
  &-0.2 < W\mathrm{1}-W\mathrm{2}< 0.86\\ 
  &W\mathrm{1}-W\mathrm{2}>(-1.45\times(({y_{\mathrm{P1}}}-W\mathrm{1})-0.1)-0.6)
\end{eqnarray}
For candidates with S/N($W\mathrm{3}$)$>3$ we prioritize sources with $W\mathrm{2}-W\mathrm{3}>0$. These selection criteria help exclude the bulk of the L-dwarf population (\citealt{Banados16}).
The aforementioned color criteria were solely used to prioritize sources for follow-up observations, but not to reject them.
\subsubsection{UKIDSS and VHS Surveys}
We cross-match our sample using a 2\arcsec$\,$matching radius with the UKIDSS Large Area Survey (UKIDSS LAS, \citealt{Lawrence07}) data release 10 (http://surveys.roe.ac.uk/wsa/dr10plus\_release.html), and the VISTA Hemisphere Survey (VHS, \citealt{McMahon13}). UKIDSS and VHS provide $Y$, $J$, $H$ and $K$ images over areas of $\sim$4000 deg$^{2}$ and $\sim$8000 deg$^{2}$, respectively. The UKIDSS survey mapped regions of the sky within coordinates 00:32:04$<$R.A.$<$01:04:07 and -1.0$^{\circ}<$Decl.$<$16$^{\circ}$, and 00:32:04$<$R.A.$<$01:04:07 and 20$^{\circ}<$Decl.$<$40$^{\circ}$, to 5$\sigma$ limiting magnitudes of $Y$=20.8, $J$=20.5, $H$=20.2, $K$=20.1. The VHS survey aims to cover the southern hemisphere, avoiding the Milky Way footprint, and to reach a depth $\sim$30 times fainter than 2MASS. In this work, we reject objects from our initial selection in case they were detected in these catalogs and had $Y-J>0.6$ and/or $y_{\mathrm{P1}}-J>1$ (e.g., typical colors of brown dwarfs; see \citealt{Best13}).
\subsubsection{DECaLS}
The Dark Energy Camera Legacy Survey (DECaLS\footnote{http://legacysurvey.org/decamls/}) is an on-ongoing survey which will image $\sim$6700 deg$^{2}$ of the sky in the northern hemisphere, up to Decl.$<$30$^{\circ}$, in $g_{\mathrm{decam}}$, $r_{\mathrm{decam}}$ and $z_{\mathrm{decam}}$, using the Dark Energy Camera on the Blanco Telescope. We consider the Data Release 2 (DR2\footnote{http://legacysurvey.org/dr2/description/}), which covers only a fraction of the proposed final area (2078 deg$^{2}$ in $g_{\mathrm{decam}}$, 2141 deg$^{2}$ in $r_{\mathrm{decam}}$ and 5322 deg$^{2}$ in $z_{\mathrm{decam}}$), but is deeper than PS1 ($g_{\mathrm{decam}, 5\sigma}=24.7$, $r_{\mathrm{decam}, 5\sigma}=23.6$, $z_{\mathrm{decam},5\sigma}=22.8$). We use a match radius of 2\arcsec. We reject all objects detected in $g_{\mathrm{decam}}$ and/or $r_{\mathrm{decam}}$, or that present an extended morphology (e.g. with catalog entry \texttt{type} different than `PSF').
\\[2mm]

\begin{figure}
\centering
\includegraphics[width=\columnwidth]{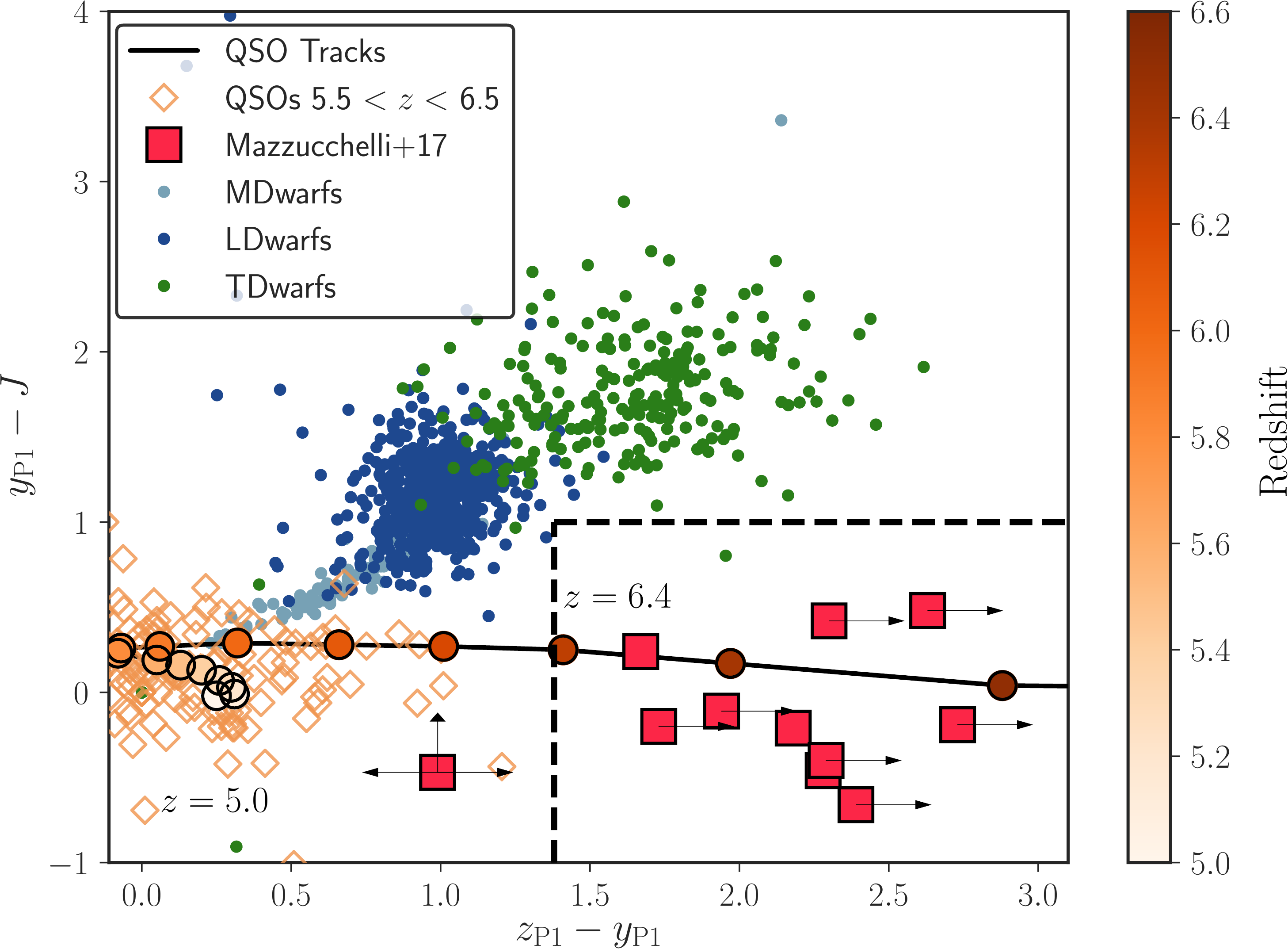}
\caption{Color-color diagram ($y_{\mathrm{P1}}-J$ vs $z_{\mathrm{P1}}-y_{\mathrm{P1}}$) used in our search for high$-$redshift quasars.
We show the predicted quasar track (black solid line and points color-coded with respect to redshift, in steps of $\mathrm{\Delta} z=$0.1), obtained by convolving the high$-$redshift quasar composite template reported by \citeauthor{Banados16} (\citeyear{Banados16}; see also Section \ref{secSEDFit}) with the filters considered here.
Observed colors of L/T dwarfs, taken from the literature (see text for references), are reported with blue and green points, while we consider for M dwarfs the colors calculated convolving a collection of spectra with the filters used here (see Section \ref{secSEDFit}). We show also the location of known quasars at $5.5<z<6.5$ (orange empty diamonds; see Section \ref{subsecM1450} for references), and the objects studied in this work (red squares with black right-pointing arrow in case they only have lower limits in the $z_{\mathrm{P1}}$ band from the PS1 PV3 catalog, see Table \ref{tabPhotCatQSOs}). We do not show quasars from the VIKING survey, which are not present in the PS1 catalog, and PSO006+39, for which we do not possess J$-$band photometry. For HSC1205 we use the 3$\sigma$ limits in $z_{\mathrm{P1}}$ and $y_{\mathrm{P1}}$ obtained from the forced photometry on the PS1 PV3 stacked images.
Our selection box is highlighted with dashed black lines.
}
\label{figColorColor}
\end{figure}

In Figure \ref{figColorColor} we show one of the color-color plots ($y_{\mathrm{P1}}-J$ vs $z_{\mathrm{P1}}-y_{\mathrm{P1}}$) used at this stage of the candidate selection.
\subsection{Forced photometry on PS1 images}
Next, we perform forced photometry on both the stacked and single epoch images from PS1 of our remaining candidates. This is to confirm the photometry from the PS1 PV3 stacked catalog and to reject objects showing a large variation in the flux of the single epoch images which would most probably indicate spurious detections (for further details on the cuts used at this stage, see \citealt{Banados14}). 
\subsection{SED Fit} \label{secSEDFit}
We implement a SED fitting routine to fully exploit all the multi-wavelength information provided by the surveys described in Sections \ref{secCandSelPS1} and \ref{secCandSelSurv}.
We compare the observations of our candidates with synthetic fluxes, obtained by interpolating quasar and brown dwarf spectral templates through different filter curves, in the 0.7$-$4.6 $\mu$m observed wavelength range.\\
We consider 25 observed brown dwarf spectra taken from the SpeX Prism Library\footnote{http://pono.ucsd.edu/~adam/browndwarfs/spexprism/} (\citealt{Burgasser14}), and representative of typical M4-M9, L0-L9 and T0-T8 stellar types. These spectra cover the wavelength interval 0.65$-$2.55 $\mu$m (up to $K$ band). The corresponding \textit{W}1 (3.4 $\mu$m) and \textit{W}2 (4.6 $\mu$m) magnitudes are obtained following \cite{Skrzypek15}, who exploit a reference sample of brown dwarfs with known spectral and photometric information to derive various color relations. For each brown dwarf template, we derive the \textit{WISE} magnitudes using the synthetic $K$ magnitude and scaling factors ($K\_W\mathrm{1}$ and $W\mathrm{1}\_W\mathrm{2}$) which depend on the stellar spectral type\footnote{The scaling factors $K\_W\mathrm{1}$ and $W\mathrm{1}\_W\mathrm{2}$ for the different M/T/L stellar types can be found in Table 1 of \cite{Skrzypek15}.}. We apply the following relations:
\begin{eqnarray}
  &W\mathrm{1} = K - K\_W\mathrm{1} - 0.783\\
  &W\mathrm{2} = W\mathrm{1} - W\mathrm{1}\_W\mathrm{2} - 0.636
\end{eqnarray}

For the quasar models, we use four different observed composite spectra: the SDSS template, obtained from a sample of $1 \lesssim z \lesssim 2$ quasars \citep{Selsing16}, and three composite of $z \gtrsim 5.6$ quasars by \cite{Banados16}, the first one based on 117 sources (from PS1 and other surveys), the second obtained considering only the 10\% of objects with the largest rest-frame Ly$\alpha$+N$\,${\scriptsize V} equivalent width (EW), and the last using the 10\% of sources with smallest EW (Ly$\alpha$+N$\,${\scriptsize V}). These different templates allow us to take into account color changes due to the Ly$\alpha$ emission line strength. However, the three models from \cite{Banados16} cover only up to rest-frame wavelength $\lambda_{rf}\sim 1500$ \AA, so we use the template from \cite{Selsing16} to extend coverage into the NIR region.
We shift all the quasar templates over the redshift interval $5.5\leq z \leq 9.0$, with $\mathrm{\Delta}z=$0.1.
We consider the effect of the IGM absorption on the SDSS composite spectrum using the redshift-dependent recipe provided by \cite{Meiksin06}.
For the quasar templates from \cite{Banados16}, we implement the following steps: we correct each model for the IGM absorption as calculated at redshift $z$=$z_{\mathrm{median}}$ of the quasars used to create the composite, obtaining the reconstructed emitted quasar spectra. Then, we re-apply the IGM absorption to the corrected models at each redshift step, again using the method by \cite{Meiksin06}.
The total number of quasar models is 140.

\begin{figure}
\centering
\includegraphics[width=\columnwidth]{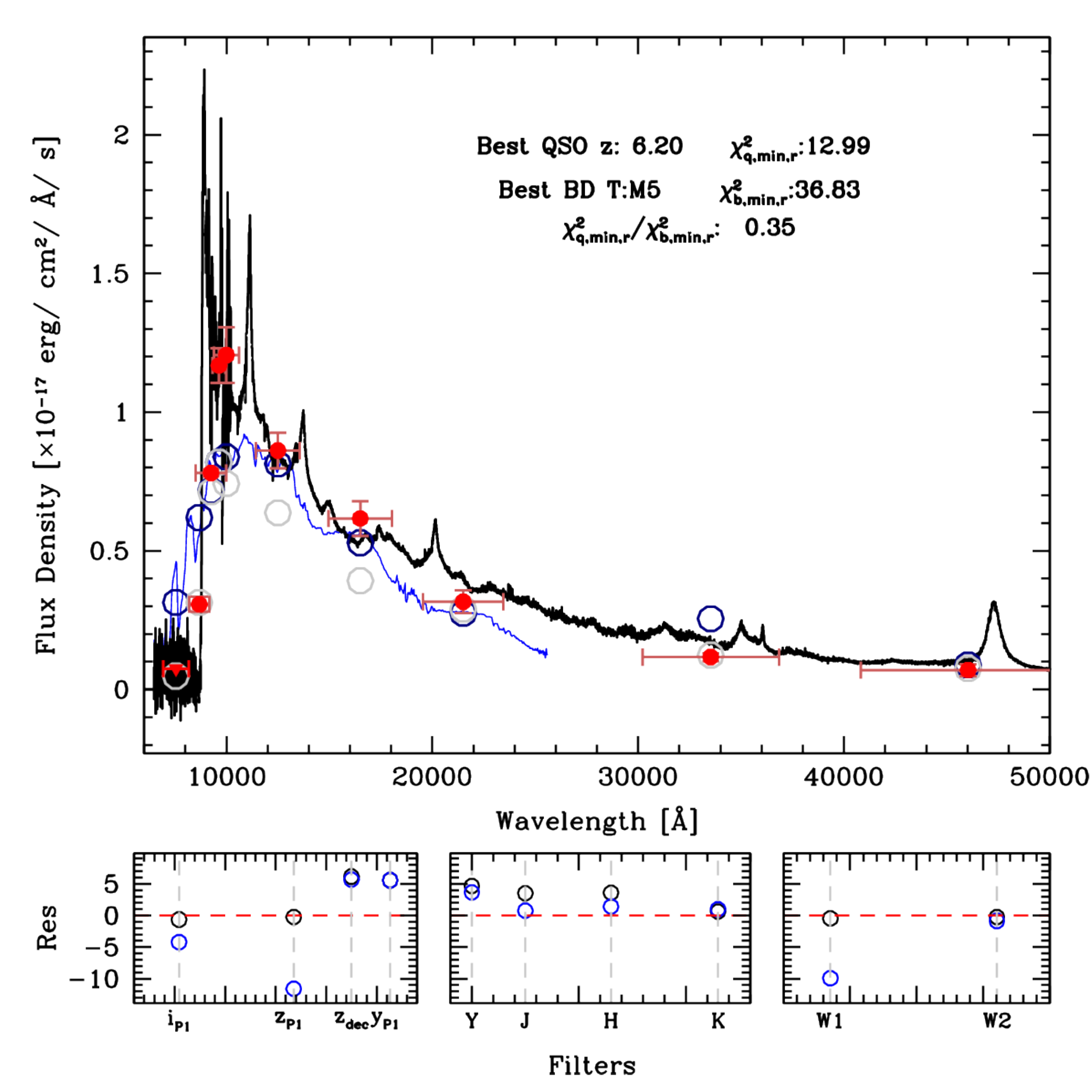}
\caption{Example of SED fit for one of our candidates, confirmed to be a quasar at $z$=6.4377 (PSO183+05, see Table \ref{tabSampleQSOs}). In the upper panel, we show the photometric information taken from public surveys (red points and down-pointing arrows in case of non-detections at 3$\sigma$ significance, see Section \ref{secCandSel}), the best quasar template (the weak-lined PS1 quasar template at $z=6.2$; black solid line) and the best brown dwarf template (M5; blue solid line). The synthetic fluxes of the best quasar and brown dwarf templates, obtained by convolving the models to the filters considered here, are shown with light grey and blue points, respectively. In the bottom panels, the residuals, e.g. $(\mathrm{flux}_{data,f}-\mathrm{flux}_{best model,f,q/b})/\sigma_{f}$, are displayed, for each $f$ band used here. Blue and black empty circles indicate the best brown dwarf and quasar template, respectively.
}
\label{figSEDex}
\end{figure}

For each quasar candidate from our selection, after having normalized the brown dwarf and quasar templates to the candidate observed flux at $y_{\mathrm{P1}}$, we find the best models that provide the minimum reduced $\chi^{2}$, $\chi^{2}_{b,min,r}$ and $\chi^{2}_{q,min,r}$, for brown dwarf and quasar templates, respectively.
We assume that the candidate is best fitted by a quasar template if $R=\chi^{2}_{q,min,r}/\chi^{2}_{b,min,r} < 1$. In our search, we prioritize for further follow-up observations sources with the lowest $R$ values. Though we do not reject any object based on this method, candidates with $R>1$ were given the lowest priority.
An example of the best quasar and brown dwarf models for one of our newly discovered quasars is shown in Figure \ref{figSEDex}.
\\[2mm]

Finally, we visually inspect all the stacked and single epoch PS1 frames, together with the images from the other public surveys, when available ($\sim$4000 objects). This is to reject non-astronomical or spurious sources (e.g. CCD defects, hot pixels, moving objects).
We then proceed with follow-up of the remaining targets ($\sim$1000). 
\section{Observations} \label{secFolUpObs}
We first obtain imaging follow-up observations of our quasar candidates, and then we take spectra of the most promising objects.
\subsection{Imaging and spectroscopic confirmation}
We perform follow-up imaging observations in order to both confirm the catalog magnitudes, and to obtain missing NIR and deep optical photometry, crucial in identifying contaminant foreground objects.

We take advantage of different telescopes and instruments: MPG 2.2m/GROND \citep{Greiner08}, NTT/EFOSC2 \citep{Buzzoni84}, NTT/SofI \citep{Moorwood98}, du Pont/Retrocam\footnote{http://www.lco.cl/telescopes-information/irenee-du-pont/instruments/website/retrocam}, Calar Alto 3.5m/Omega2000 \citep{Bailer-Jones00}, Calar Alto 2.2m/CAFOS\footnote{http://www.caha.es/CAHA/Instruments/CAFOS/index.html}.
In Table \ref{tabFollowUpPhotSpec} we report the details of our campaigns, together with the filters used.

The data were reduced using standard data reduction procedures \citep{Banados14}. We refer to \cite{Banados16} for the color conversions used to obtain the flux calibration. In case we collect new $J$ band photometry for objects undetected or with low S/N in NIR public surveys, we consider as good quasar candidates the ones with $-1 < y_{\mathrm{P1}}-J < 1$, while the sources with very red or very blue colors were considered to be stellar contaminants or spurious/moving objects, respectively.
For a sources with good NIR colors (from either public surveys or our own follow-up photometry), we collected deep optical imaging.\\
We the took spectra of all the remaining promising candidates using VLT/FORS2 \citep{Appenzeller98}, P200/DBSP \citep{OkeGunn82}, MMT/Red Channel \citep{Schmidt89}, Magellan/FIRE (\citealt{Simcoe08}, \citeyear{Simcoe10}) and LBT/MODS \citep{Pogge10} spectrographs. Standard techniques were used to reduce the data (see \citealt{Venemans13}, \citealt{Banados14}, \citeyear{Banados16}, \citealt{Chen16}). Six objects, out of nine observed candidates, were confirmed as high$-$redshift quasars: we present them and provide further details in Section \ref{secNotesPS1}. 
We list the spectroscopically rejected objects (Galactic sources) in Appendix \ref{secApSpecRej}.
Further information on these observations,
together with the additional spectroscopic observations for other objects in our quasar sample (see Section \ref{subsecSpcFU}),
are reported in Table \ref{tabFollowUpSpec}.
In Table \ref{tabPhotCatQSOs}, we provide photometric data from catalogs for all the objects in our high$-$redshift quasars sample. Also, photometry from our own follow-up campaigns for the new six quasars is listed in Table \ref{tabPhotFollowUpQSOs}. Table \ref{tabApFilt} in Appendix \ref{secAppFilt} lists the information (central wavelength and width, $\rm \lambda_{c}$ and $\rm \Delta \lambda$) of the various filters used in this work, both from public surveys and follow-up photometry. 
\subsection{Spectroscopic follow-up of $z\gtrsim$6.44 quasars} \label{subsecSpcFU}
Once the high$-$redshift quasar nature of candidates is confirmed, we include them in our extensive campaign of follow-up observations aimed at characterizing quasars at the highest redshifts.

\begin{figure*}
\centering
\includegraphics[width=\textwidth]{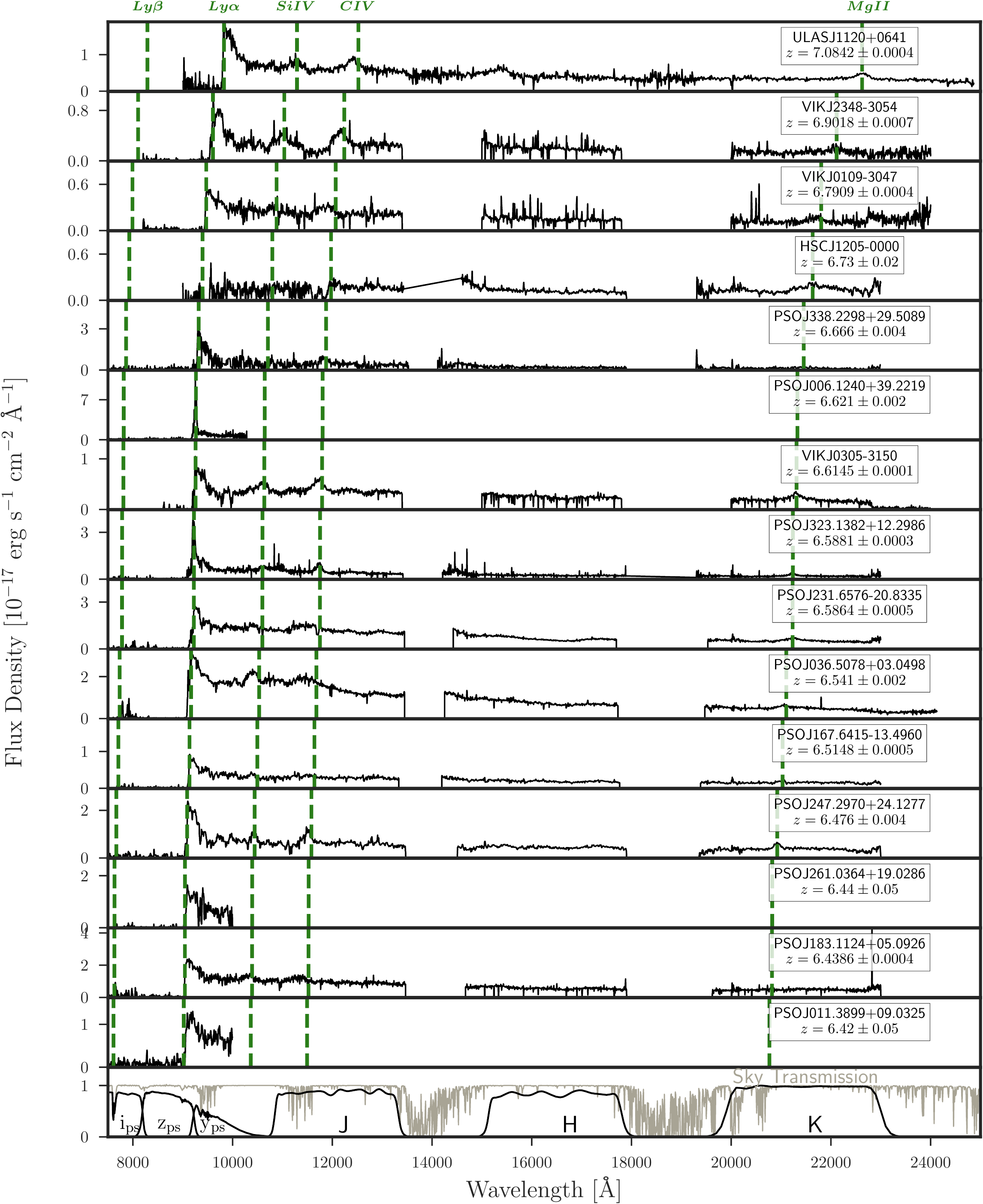}
\caption{Binned spectra of the 15 $z\gtrsim6.5$ quasars in the sample considered here. The quasars PSO323+12, PSO231-20, PSO247+24, PSO011+09, PSO261+19 and PSO183+05 are newly discovered from the PS1 PV3 survey; the other objects are taken from the literature (see Table \ref{tabSampleQSOs}). The locations of key emission lines (Ly$\beta$, Ly$\alpha$, Si$\,${\scriptsize IV}, C$\,${\scriptsize IV} and Mg$\,${\scriptsize II}) are highlighted with dashed, green lines.}
\label{figSpecAll}
\end{figure*}

Here we present new optical/NIR spectroscopic data for nine quasars, the six objects newly discovered from PS1 and three sources from the literature (PSO006+39, PSO338+29 and HSC1205). These observations have been obtained with a variety of telescopes and spectrographs: VLT/FORS2, P200/DBSP, MMT/Red Channel, Magellan/FIRE, VLT/X-Shooter \citep{Vernet11}, Keck/LRIS (\citealt{Oke95} and \citealt{Rockosi10}) and GNT/GNIRS \citep{Dubbeldam00}.
We take the remaining spectroscopic data from the literature.
The details (i.e.$\,$observing dates, instruments, telescopes, exposure times and references) for all the spectra presented here are reported in Table \ref{tabFollowUpSpec}.
In case of multiple observations an one object, we use the weighted mean of the spectra. We scale the spectra to the observed $J$ band magnitudes (see Table \ref{tabPhotCatQSOs}), with the exceptions of PSO006+39, for which we do not have this information, and PSO011+09 and PSO261+19, that only have optical coverage; in these cases, we normalize the spectra to the $y_{\mathrm{P1}}$ magnitudes. We also correct the data for Galactic extinction, using the extinction law provided by \cite{Calzetti00}. 
The reduced spectra are shown in Figure \ref{figSpecAll}. 
\\
\begin{deluxetable*}{lccc}
\tablecaption{Imaging follow$-$up observation campaigns for PS1 high$-$redshift quasar candidates.\label{tabFollowUpPhotSpec}}
\tablewidth{0pt}
\tablehead{
\colhead{Date} & \colhead{Telescope/Instrument} & \colhead{Filters} & \colhead{Exposure Time} 
}
\startdata
2014 May 9              & CAHA 3.5m/Omega2000 & $z_{\mathrm{O2K}}$, $Y_{\mathrm{O2K}}$, $J_{\mathrm{O2K}}$ & 300s\\
2014 Jul 23$-$27        & NTT/EFOSC2          & $I_{\mathrm{E}}$,$Z_{\mathrm{E}}$ & 600s\\
2014 Jul 25             & NTT/SofI            & $J_{\mathrm{S}}$ & 600s \\
2014 Aug 7 and 11$-$13  & CAHA 3.5m/Omega2000 & $Y_{\mathrm{O2K}}$, $J_{\mathrm{O2K}}$ & 600s \\
2014 Aug 22$-$24        & CAHA 2.5m/CAFOS     & $i_{\mathrm{w}}$ & 1800s \\
2015 Feb 22             & NTT/SofI            & $J_{\mathrm{S}}$ & 300s \\
2016 Jun 5$-$13         & MPG 2.2m/GROND      & $g_{\mathrm{G}}$, $r_{\mathrm{G}}$, $i_{\mathrm{G}}$, $z_{\mathrm{G}}$, $J_{\mathrm{G}}$, $H_{\mathrm{G}}$, $K_{\mathrm{G}}$ & 1440s \\
2016 Sep 11$-$13        & NTT/EFOSC2          & $I_{\mathrm{E}}$ & 900s \\
2016 Sep 16$-$25        & MPG 2.2m/GROND      & $g_{\mathrm{G}}$, $r_{\mathrm{G}}$, $i_{\mathrm{G}}$, $z_{\mathrm{G}}$, $J_{\mathrm{G}}$, $H_{\mathrm{G}}$, $K_{\mathrm{G}}$ & 1440s \\
2016 Sep 18$-$21        & du Pont/Retrocam     & $Y_{\mathrm{retro}}$ & 1200s 
\enddata
\end{deluxetable*}
\begin{deluxetable*}{lcccccc}
\tablecaption{Spectroscopic observations of the $z\gtrsim6.5$ quasars presented in this study. We present optical/NIR spectra for all the newly discovered objects and for some known sources. We also gather data from the literature. The references are: (1) \cite{Venemans13}; (2) \cite{DeRosa14}, (3) \cite{Venemans15}, (4) \cite{Chen16}, (5) \cite{Mortlock11}, and (6) this work. \label{tabFollowUpSpec}}
\tablewidth{0pt}
\tablehead{
\colhead{Object} & \colhead{Date} & \colhead{Telescope/Instrument} & \colhead{$\lambda$ range}
& \colhead{Exposure Time} & \colhead{Slit Width}
& \colhead{Reference}
\\
& & & [$\mu$m] & [s] & & 
}
\startdata
PSO J006.1240+39.2219 & 2016 Jul 5 &  Keck/LRIS & 0.55$-$1.1 & 1800 & 1\farcs0 & (6)\\
PSO J011.3899+09.0325 & 2016 Nov 20 & Magellan/FIRE & 0.82$-$2.49 & 600 & 1\farcs0 & (6) \\
 & 2016 Nov 26 & Keck/LRIS  & 0.55$-$1.1 & 900 & 1.\arcsec0 & (5)\\
VIK J0109--3047 & 2011 Aug-Nov & VLT/X-Shooter & 0.56$-$2.48  & 21600 & 0\farcs9$-$1\farcs5 & (1,2)\\
PSO J036.5078+03.0498 & 2015 Dec 22$-$29 & VLT/FORS2 & 0.74$-$1.07  & 4000. & 1\farcs0 & (5) \\
 & 2014 Sep 4$-$6 & Magellan/FIRE & 0.82$-$2.49 & 8433 & 0.\arcsec6 & (3) \\ 
VIK J0305--3150 & 2011 Nov $-$2012 Jan & Magellan/FIRE & 0.82$-$2.49 & 26400 & 0\farcs6 & (1,2) \\  
PSO J167.6415--13.4960 & 2014 Apr 26 & VLT/FORS2 & 0.74$-$1.07 & 2630 & 1\farcs3 & (3)\\
 & 2014 May 30-Jun 2 & Magellan/FIRE & 0.82$-$2.49 & 12004 & 0\farcs6 & (3) \\  
ULAS J1120+0641 & 2011 & GNT/GNIRS & 0.90$-$2.48 &  & 1\farcs0 & (5)\\
HSC J1205--0000 & 2016 Mar 14  & Magellan/FIRE & 0.82$-$2.49 & 14456 & 0\farcs6 & (6) \\
PSO J183.1124+05.0926 & 2015 May 8 & VLT/FORS2 & 0.74$-$1.07  & 2550 & 1\farcs3 & (6)\\
 &  2015 Apr 6 & Magellan/FIRE & 0.82$-$2.49  &  11730 & 0\farcs6  & (4,5)\\
PSO J231.6576--20.8335 & 2015 May 15 & VLT/FORS2 & 0.74$-$1.07  & 2600 & 1\farcs3 & (6)\\ 
 &  2015 Mar 13 & Magellan/FIRE & 0.82$-$2.49  &  9638 & 0\farcs6  & (4,5)\\
PSO J247.2970+24.1277 & 2016 Mar 10 & VLT/FORS2 & 0.74$-$1.07  & 1500 & 1\farcs0  & (6)\\
 &  2016 Mar 31 & Magellan/FIRE & 0.82$-$2.49 &  6626 & 0\farcs6  & (4,6)\\
PSO J261.0364+19.0286 & 2016 Sep 12  & P200/DBSP & 0.55$-$1.0 & 3600 &  1\farcs5 & (6)\\
PSO J323.1382+12.2986 & 2015 Nov 5 & VLT/FORS2 & 0.74$-$1.07  & 1500 & 1\farcs0  & (6)\\
 & 2016 Aug 15 & Magellan/FIRE & 0.82$-$2.49 &  3614 & 0\farcs6  & (6)\\
PSO J338.2298+29.5089 & 2014 Oct 19 & MMT/Red Channel & 0.67$-$1.03 & 1800 & 1\farcs0 &(3)\\
 & 2014 Oct 30 & Magellan/FIRE & 0.82$-$2.49 & 7200 & 0\farcs6  &(3)\\
 & 2014 Nov 27 & LBT/MODS & 0.51$-$1.06 & 2700 & 1\farcs2  & (3)\\
VIK J2348--3054 & 2011 Aug 19$-$21 & VLT/X-Shooter & 0.56$-$2.48 & 8783 & 0\farcs9$-$1\farcs5 & (1,2)
\enddata
\end{deluxetable*}
\begin{deluxetable*}{lcccccccc}
\tabletypesize{\small}
\tablecaption{PS1 PV3, $z_{\mathrm{decam}}$, $J$ and \textit{WISE} photometry and Galactic $E(B-V)$ values (from \citealt{Schlegel98}) of the quasars analysed here. The limits are at 3$\sigma$ significance. The $J-$band information is from (1) UKIDSS, (2) VHS, (3) \cite{Venemans13}, (4) \cite{Venemans15}, (5) \cite{Matsuoka16}, (6) this work (in case we have follow up photometry on the quasar, we report the magnitude with the best S/N; see also Table \ref{tabPhotFollowUpQSOs}). The $z_{\mathrm{decam}}$ information is taken from the last DECaLS DR3 release. The \textit{WISE} data are from ALLWISE or, in case the object was present in DECaLS DR3, from the UNWISE catalog (\citealt{Lang14} and \citealt{Meisner16}).\label{tabPhotCatQSOs}}
\tablewidth{0pt}
\tablehead{ \colhead{name} & \colhead{$z_{\mathrm{P1}}$} & \colhead{$y_{\mathrm{P1}}$}
& \colhead{$z_{\mathrm{decam}}$} & \colhead{$J$} & \colhead{$J_{\mathrm{ref}}$}
& \colhead{$W$1} & \colhead{$W$2} & \colhead{$E(B-V)$}
}
\startdata
PSO J006.1240+39.2219 & $>$23.02 & 20.06 $\pm$ 0.07 & $-$ & $-$ & $-$ & $-$ & $-$ & 0.075\\
PSO J011.3899+09.0325 & $>$22.33 & 20.60 $\pm$ 0.09 & $-$ & 20.80 $\pm$ 0.13 & (6) & 20.19 $\pm$ 0.19 & $-$ & 0.059\\
VIK J0109--3047 & $-$ & $-$ & $-$ & 21.27 $\pm$ 0.16 & (3) & 20.96 $\pm$ 0.32 & $-$ & 0.022\\
PSO J036.5078+03.0498 & 21.48 $\pm$ 0.12 & 19.30 $\pm$ 0.03 & 20.01 $\pm$ 0.01 & 19.51 $\pm$ 0.03 & (4) & 19.52 $\pm$ 0.06 & 19.69 $\pm$ 0.14 & 0.035 \\
VIK J0305--3150 &  $-$ & $-$ & $-$ & 20.68 $\pm$ 0.07 & (3) & 20.38 $\pm$ 0.14 & 20.09 $\pm$ 0.24 & 0.012\\
PSO 167.6415--13.4960  & $>$22.94 & 20.55 $\pm$ 0.11 & $-$ & 21.21 $\pm$ 0.09 & (4) & $-$ & $-$ & 0.057\\
ULAS J1120+0641\footnote{The PS1 magnitudes are taken from the PV2 catalog, since the object is not detected in PV3, having S/N$<$5 in all bands. However, forced photometry on the $y_{\mathrm{P1}}$ PV3 stack image at the quasar position reveals a faint source with S/N=4.3 in PV3.} & $>$23.06 & 20.76 $\pm$ 0.19 & 22.38 $\pm$ 0.1 & 20.34 $\pm$ 0.15 & (1) & 19.81 $\pm$ 0.09 & 19.96 $\pm$ 0.23 & 0.052\\
HSC J1205--0000\footnote{This object does not appear in the PS1 PV3 catalog. The PS1 magnitudes are obtained by performing forced photometry on the $z_{\mathrm{P1}}$ and $y_{\mathrm{P1}}$ PV3 images.}  & $>$22.47 & $>$21.48 & $-$ & 21.95 $\pm$ 0.21 & (5) & 19.98 $\pm$ 0.15 & 19.65 $\pm$ 0.23 & 0.0243 \\
PSO J183.1124+05.0926 & 21.68 $\pm$ 0.10 & 20.01 $\pm$ 0.06 & 20.53 $\pm$ 0.02 & 19.77 $\pm$ 0.08 & (6) & 19.74 $\pm$ 0.08 & 20.03 $\pm$ 0.24 & 0.0173\\
PSO J231.6576--20.8335 & $>$22.77 & 20.14 $\pm$ 0.08 & $-$ & 19.66 $\pm$ 0.05 & (6) & 19.91 $\pm$ 0.15 & 19.97 $\pm$ 0.35 & 0.133 \\
PSO J247.2970+24.1277 & $>$22.77 & 20.04 $\pm$ 0.07 & 20.82 $\pm$ 0.03 & 20.23 $\pm$ 0.09 & (6) & 19.46 $\pm$ 0.04 & 19.28 $\pm$ 0.08 & 0.053 \\
PSO J261.0364+19.0286 & $>$22.92 & 20.98 $\pm$ 0.13 & $-$ & 21.09 $\pm$ 0.18 & (6) & 20.61 $\pm$ 0.21 & $-$ & 0.045\\
PSO J323.1382+12.2986 & 21.56 $\pm$ 0.10 & 19.28 $\pm$ 0.03 & $-$ & 19.74 $\pm$ 0.03 & (6) & 19.06 $\pm$ 0.07 & 18.97 $\pm$ 0.12 & 0.108 \\
PSO J338.2298+29.5089 & $>$22.63 & 20.34 $\pm$ 0.1 & 21.15 $\pm$ 0.05 & 20.74 $\pm$ 0.09 & (4) & 20.51 $\pm$ 0.14 &  $-$ & 0.096\\
VIK J2348--3054 & $-$ & $-$ & $-$ & 21.14 $\pm$ 0.08 & (3) & 20.36 $\pm$ 0.17 & $-$ & 0.013
\enddata
\end{deluxetable*}
\begin{deluxetable*}{lc}
\tabletypesize{\small}
\tablecaption{Photometry from our follow-up campaigns for the newly discovered PS1 quasars; the limits are at 3$\sigma$. (see also Table \ref{tabFollowUpPhotSpec}).\label{tabPhotFollowUpQSOs}}
\tablewidth{0pt}
\tablehead{ \colhead{name} & }
\startdata
PSO J011.3899+09.0325  & $i_{\mathrm{G}}>$23.36; $z_{\mathrm{G}}$=22.38 $\pm$ 0.16; $Y_{\mathrm{retro}}$=20.81 $\pm$ 0.07; $J_{\mathrm{G}}$=20.80 $\pm$ 0.13\\
PSO J183.1124+05.0926  & $I_{\mathrm{E}}$=23.51 $\pm$ 0.21; $Z_{\mathrm{E}}$=20.93 $\pm$ 0.09; $J_{\mathrm{S}}$=19.77 $\pm$ 0.08 \\
PSO J231.6576--20.8335 &  $I_{\mathrm{E}}>$23.81; $J_{\mathrm{S}}$=19.66 $\pm$ 0.05 \\
PSO J247.2970+24.1277  & $i_{\mathrm{w}}>$22.36; $i_{\mathrm{MMT}}>$22.69; $z_{\mathrm{O2k}}$=20.89 $\pm$ 0.07; $Y_{\mathrm{O2k}}$=20.04 $\pm$ 0.24; $J_{\mathrm{O2k}}$=20.23 $\pm$ 0.09 \\
PSO J261.0364+19.0286  & $i_{\mathrm{G}}>$23.40; $I_{\mathrm{E}}>$24.01; $z_{\mathrm{G}}$=22.18 $\pm$ 0.12; $J_{\mathrm{G}}$=21.09 $\pm$ 0.18; $H_{\mathrm{G}}$=20.92 $\pm$ 0.30\\
PSO J323.1382+12.2986  & $z_{\mathrm{O2k}}$=20.14 $\pm$ 0.05; $Y_{\mathrm{O2k}}$=19.45 $\pm$ 0.07; $J_{\mathrm{S}}$=19.74 $\pm$ 0.03
\enddata
\end{deluxetable*}
\subsection{NOEMA observations} \label{subsecNOEMAobs}
Four quasars in our sample (PSO323+12, PSO338+29, PSO006+39 and HSC1205)
have been observed with the NOrthern Extended Millimeter Array (NOEMA):
this data, together with the one retrieved from the literature
(\citealt{Venemans12}, \citeyear{Venemans16},
\citealt{Banados15b}, \citealt{DecarliSub};
see also Table \ref{tabSampleQSOs} and Section \ref{subsecRedTempl})
completes coverage of the [CII]~158 $\mu$m emission line
for all currently published $z\geq$6.5 quasars.
The NOEMA observations were carried out in the compact array configuration, 
for which the primary beam at 250 GHz is $\sim 20"$ (full
width at half power). Data were processed with the latest release of the
software \textsf{clic} in the \textsf{GILDAS} suite, and analyzed using
the software \textsf{mapping} together with a number of custom routines
written by our group.

PSO323+12 was observed in a Director's Discretionary Time program (project
ID: E15AD) on 2015 December 28, with seven 15m antennae arranged in the 7D
configuration. The source MWC349 was used for flux calibration,
while the quasar 2145+067 was used for phase and amplitude calibration.
The system temperature was in the range 110--160\,K. Observations were
performed with average precipitable water vapour conditions ($\sim
2.2$\,mm). The final cube includes 5159 visibilities, corresponding to
3.07\,hr on source (7 antennas equivalent). After collapsing the entire
3.6\,GHz bandwidth, the continuum rms is 0.146 mJy\,beam$^{-1}$.

PSO338+29 was observed on 2015 December 3 (project ID: W15FD) 
in the 7C array configuration. MWC349 was observed for flux calibration, while
the quasar 2234+282 was targeted for phase and amplitude calibration. The
typical system temperature was 85-115\,K. Observations were carried out
in good water vapour conditions (1.7-2.0 mm).  The final data cube
consists of 4110 visibilities, corresponding to 2.45 hr on source (7
antennas equivalent). The synthesized beam is 1\farcs35$\times$0\farcs69.
The rms of the collapsed data cube is 0.215 mJy\,beam$^{-1}$.

PSO006+39 was observed in two visits, on 2016 May 20 and 2016 July 7, as part
of the project S16CO, with 5-7 antennae. The May visit was hampered by
high precipitable water vapour ($\sim$3 mm) yielding high system temperature
(200--300\,K). The July track was observed in much better
conditions, with precipitable water vapor (pwv) $\sim$ 1.3\,mm and $T_{\rm sys}$=105-130\,K.
The final cube consists of 2700 visibilities, corresponding to 2.25 hr on source
(six antennas equivalent), with a continuum sensitivity of 0.178
mJy\,beam$^{-1}$. The synthesized beam is 1.19\farcs$\times$0\farcs61.

HSC1205 was also observed as part of project S16CO, on 2016 October 29,
using the full 8-antennae array. MWC349 was observed for flux calibration,
while the quasar 1055+018 served as phase and amplitude calibrator. The
precipitable water vapout was low ($\sim 1.3$\,mm), and the system
temperature was 120-180 K. The final cube consists of 2489 visibilities,
or 1.11 hr on source (8 antennae equivalent). The synthesized beam is
1.19\farcs$\times$0\farcs61 and the continuum rms is 0.176 mJy\,beam$^{-1}$.
\section{Individual notes on six new quasars from PS1} \label{secNotesPS1} 
We present six new quasars at $z\sim 6.5$ discovered from the PS1 survey: we here present brief observational summaries of each source.
\subsection{PSO J011.3899+09.0325 @ z=6.42}
Follow-up imaging data for PSO011+09 were acquired with MPG 2.2m/GROND and du Pont/Retrocam in September 2016; its quasar nature was confirmed with a short 600s low-resolution prism mode spectrum using Magellan/FIRE on 2016 November 20. We then obtained a higher S/N, higher resolution optical spectrum with Keck/LRIS. We consider in this work only the latter spectroscopic observation (see Figure \ref{figSpecAll}) because the FIRE spectrum has a very limited S/N and over-exposed H and K bands.
It is a relatively faint object, with $J_{\mathrm{G}}$=20.8, and presents a very flat $Y_{\mathrm{retro}}-J_{\mathrm{G}}$ color of 0.01 (see Table \ref{tabPhotFollowUpQSOs}). This quasar does not show strong emission lines. Through a comparison with SDSS quasar templates (see Section \ref{subsecRedTempl}), we calculate a redshift of $z=6.42$, with an uncertainty of $\mathrm{\Delta} z$=0.05.
\subsection{PSO J183.1124+05.0926 @ z=6.4386}
PSO183+05 was first followed up with the SofI and EFOSC2 instruments at the NTT, in February 2015. The discovery spectrum was taken with the Red Channel spectrograph at the MMT; higher quality spectra were later acquired with Magellan/FIRE and VLT/FORS2, in April and May 2015, respectively.
Evidence was found for the presence of a very proximate Damped Lyman Absorber (DLA; $z\sim$6.404) along the same line-of-sight (see also \citealt{Chen16}).
An in-depth study of this source will be presented in \cite{BanadosInPrep}.
\subsection{PSO J231.6576--20.8335 @ z=6.5864}
The imaging follow-up for PSO231$-$20 was also undertaken with EFOSC2 and SofI at the NTT in February 2015. It was spectroscopically confirmed with Magellan/FIRE on the 2015 March 13, and we acquired a VLT/FORS2 spectrum on the 2015 May 15. With a $J-$band magnitude of 19.66, this quasar is the brightest newly discovered object, and one of the brightest known at $z>6.5$, alongside PSO036+03 and VDESJ0224-4711.
\subsection{PSO J247.2970+24.1277 @ z=6.476}
We acquired follow-up photometric observations of PSO247+24 with CAFOS and Omega2000 at the 2.2m and 3.5m telescope at CAHA, respectively. We confirmed its quasar nature with VLT/FORS2 in March 2016 and we obtained NIR spectroscopy with Magellan/FIRE in the same month. This quasar presents prominent broad emission lines (see Figure \ref{figSpecAll}).
\subsection{PSO J261.0364+19.0286 @ z=6.44}
We used the 2.2m MPG/GROND and SofI at the NTT in June$-$September 2016 to acquire follow-up photometry for PSO261+19. Spectroscopic observations with the DBSP at the Palomar Observatory in September 2016 confirmed that the object is a quasar at $z=6.44 \pm 0.05$ (redshift from SDSS quasar template fitting; see Section \ref{subsecRedTempl}). Similar to PSO11+09, this is a relatively faint quasar, with $J_{\mathrm{G}}=21.09$.
\subsection{PSO J323.1382+12.2986 @ z=6.5881}
Imaging follow-up of PSO323+12 was acquired with CAHA 3.5m/Omega2000 and NTT/SofI in August 2014 and February 2015, respectively. Spectroscopic observations with FORS2 at the VLT in December 2015 confirmed that the source is a high redshift quasar. The NIR spectrum was later obtained with Magellan/FIRE, in August 2016.
This quasar is the one at the highest redshift among the newly discovered objects ($z=$6.5881; see Section \ref{subsecRedTempl}).
\section{Analysis} \label{secAnalysis}
We next present a comprehensive study of the quasar population at the highest redshifts currently known ($z\gtrsim6.42$). We consider a total sample of 15 quasars, six newly presented here and discovered in our search in the PS1 catalog (see Section \ref{secCandSel} and \ref{secFolUpObs}) and 9 sources from the literature (one from UKIDSS, three from VIKING, four from PS1 and one from HSC). We report their coordinates, redshifts and discovery references in Table \ref{tabSampleQSOs}.
Due to the variety of the data collected (e.g. we do not have NIR spectra or [CII] observations for all the objects in this work), we consider different sub-samples of quasars in the following sections, depending on the physical parameters that we could measure.
\begin{deluxetable*}{lccccccc}[h]
\tabletypesize{\small}
\tablecaption{Sample of quasars at $z\gtrsim6.42$ considered in this study. The objects were discovered by several studies: (1) \cite{Mortlock11}, (2) \cite{Venemans13}, (3) \cite{Venemans15}, (4) \cite{Matsuoka16}, (5) \cite{Tang17} and (6) this work. In addition to this work (6), the redshifts measurements are taken from: (7) \cite{Venemans12}, (8) \cite{Venemans16}, (9) \cite{Banados15b}, (10) \cite{DecarliSub}. \label{tabSampleQSOs}}
\tablewidth{0pt}
\tablehead{ \colhead{name}& \colhead{R.A.(J2000)} & \colhead{Decl. (J2000)} & \colhead{$z$} & \colhead{$z_{err}$} 
 & \colhead{$z$ method} & \colhead{Ref Discovery} & \colhead{Ref $z$} 
}
\startdata
PSO J006.1240+39.2219 & 00:24:29.772 &	+39:13:18.98 &  6.621 & 0.002 & [CII] &  (5) & (6) \\
PSO J011.3899+09.0325 & 00:45:33.568 & +09:01:56.96  & 6.42 & 0.05 & template &(6) & (6)\\
VIK J0109--3047 & 01:09:53.131 & --30:47:26.32 & 6.7909 & 0.0004 & [CII] &  (2) & (8)\\
PSO J036.5078+03.0498 & 02:26:01.876 &	+03:02:59.39 & 6.541 & 0.002 & [CII] &  (3) & (9) \\
VIK J0305--3150 & 03:05:16.916 & --31:50:55.90 & 6.6145 & 0.0001 & [CII] &   (2) & (8)\\
PSO J167.6415--13.4960 & 11:10:33.976 & --13:29:45.60 & 6.5148 & 0.0005 & [CII] &  (3) & (10) \\
ULAS J1120+0641  & 11:20:01.479 & +06:41:24.30 & 7.0842 & 0.0004 & [CII] &  (1) & (7)\\
HSC J1205--0000 & 12:05:05.098	& --00:00:27.97 & 6.73 & 0.02 & Mg$\,${\scriptsize II} &  (4) & (6)  \\
PSO J183.1124+05.0926 & 12:12:26.981 &	+05:05:33.49 & 6.4386 & 0.0004 & [CII] &  (6) & (10) \\
PSO J231.6576--20.8335 & 15:26:37.841 & --20:50:00.66 & 6.5864 & 0.0005 & [CII] &  (6) & (10) \\
PSO J247.2970+24.1277 & 16:29:11.296 &	+24:07:39.74 & 6.476 & 0.004 & Mg$\,${\scriptsize II} &  (6) & (6) \\
PSO J261.0364+19.0286 & 17:24:08.743 & +19:01:43.12 & 6.44 & 0.05 & template &  (6) & (6) \\
PSO J323.1382+12.2986 & 21:32:33.191 &	+12:17:55.26 & 6.5881 & 0.0003 & [CII] &  (6) & (6) \\
PSO J338.2298+29.5089 & 22:32:55.150 &	+29:30:32.23 & 6.666 & 0.004 & [CII] &  (3) & (6)  \\
VIK J2348--3054 & 23:48:33.334	& --30:54:10.24  & 6.9018 &0.0007 &  [CII] & (2) & (8)
\enddata
\end{deluxetable*}
\subsection{Redshifts} \label{subsecRedTempl}
\begin{figure}
\centering
\includegraphics[width=\columnwidth]{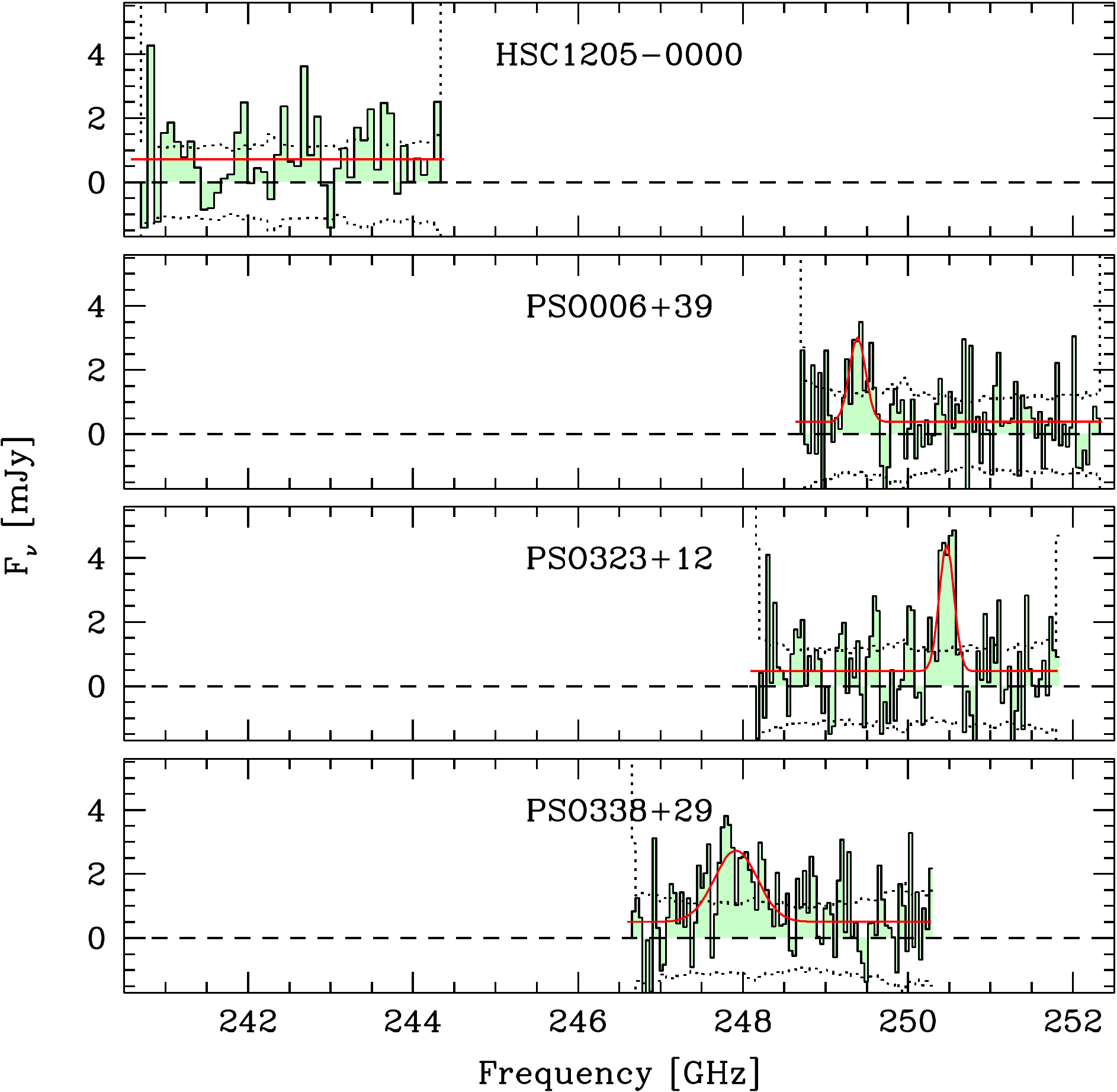}
\caption{NOEMA 1.2 mm observations of the [CII] 158 $\mu$m emission line and underlying dust continuum for four objects in our sample. The extracted spectra are fitted with a flat continuum and Gaussian function. We detect [CII] 158 $\mu$m emission for all the objects except HSC1205, whose observations were tuned based on the initial redshift range reported by \cite{Matsuoka16}: our Mg$\,${\scriptsize II} emission line detection, consistent with the new redshift in \cite{Matsuoka17}, positions its [CII] emission line out of the covered band (see text for details). We still detect the dust continuum from this quasar.}
\label{figCIIline}
\end{figure}

An accurate measurement of high$-$redshift quasar systemic redshifts is challenging. Several techniques have been implemented, and previous studies have shown that redshift values obtained with different indicators often present large scatters or substantial shifts (e.g. \citealt{DeRosa14}, \citealt{Venemans16}).
In general, the most precise redshift indicators (with measurement uncertainties of $\Delta z < 0.004$) are the atomic or molecular narrow emission lines, originating from the interstellar medium of the quasar host galaxy. This emission, in particular the [CII] lines, and the underlying dust continuum emission is observable in the millimeter wavelength range at $z\sim6$.
When available, we adopt $z_{\mathrm{[CII]}}$ measurements for the objects in our sample (11 out of 15).
We take advantage of our new NOEMA observations of four quasars (see Section \ref{subsecNOEMAobs}) to estimate their systemic redshifts from the [CII] 158 $\mu$m emission line. A flat continuum and a Gaussian profile are fitted to the spectra, as shown in Figure \ref{figCIIline}, allowing us to derive $z_{\mathrm{[CII]}}$ for PSO006+39, PSO323+12 and PSO338+29. The frequency of the observations of HSC1205 was tuned for a redshift of $z=$6.85, in the range of redshifts originally reported in the discovery paper \citep{Matsuoka16}. 
No [CII] emission line is detected from the quasar, possibly due to our frequency tuning not being centered on the true redshift of the source. This scenario is supported by our own new NIR observations of the Mg$\,${\scriptsize II} line, which place HSC1205 at $z=6.73 \pm 0.02$ (see below, Table \ref{tabSampleQSOs} and Section \ref{secFitMgII}); this is also consistent with the new redshift reported in \citeauthor{Matsuoka17} (\citeyear{Matsuoka17}; $z$=6.75). At this redshift, the [CII] emission line falls at an observed frequency of 245.87 GHz, outside the range probed in the NOEMA data (see top panel of Figure \ref{figCIIline}). 
The redshifts of PSO231-20, PSO167-13 and PSO183+05 are measured from the [CII] line, observed in our ALMA survey of cool gas and dust in $z\gtrsim6$ quasars (\citealt{DecarliSub}). We take the values of $z_{\mathrm{[CII]}}$ for ULAS1120, VIK2348, VIK0109, VIK0305 and PSO036+03 from the literature (\citealt{Venemans12}, \citeyear{Venemans16}, \citealt{Banados15b}).

\begin{figure}
\centering
\includegraphics[width=\columnwidth]{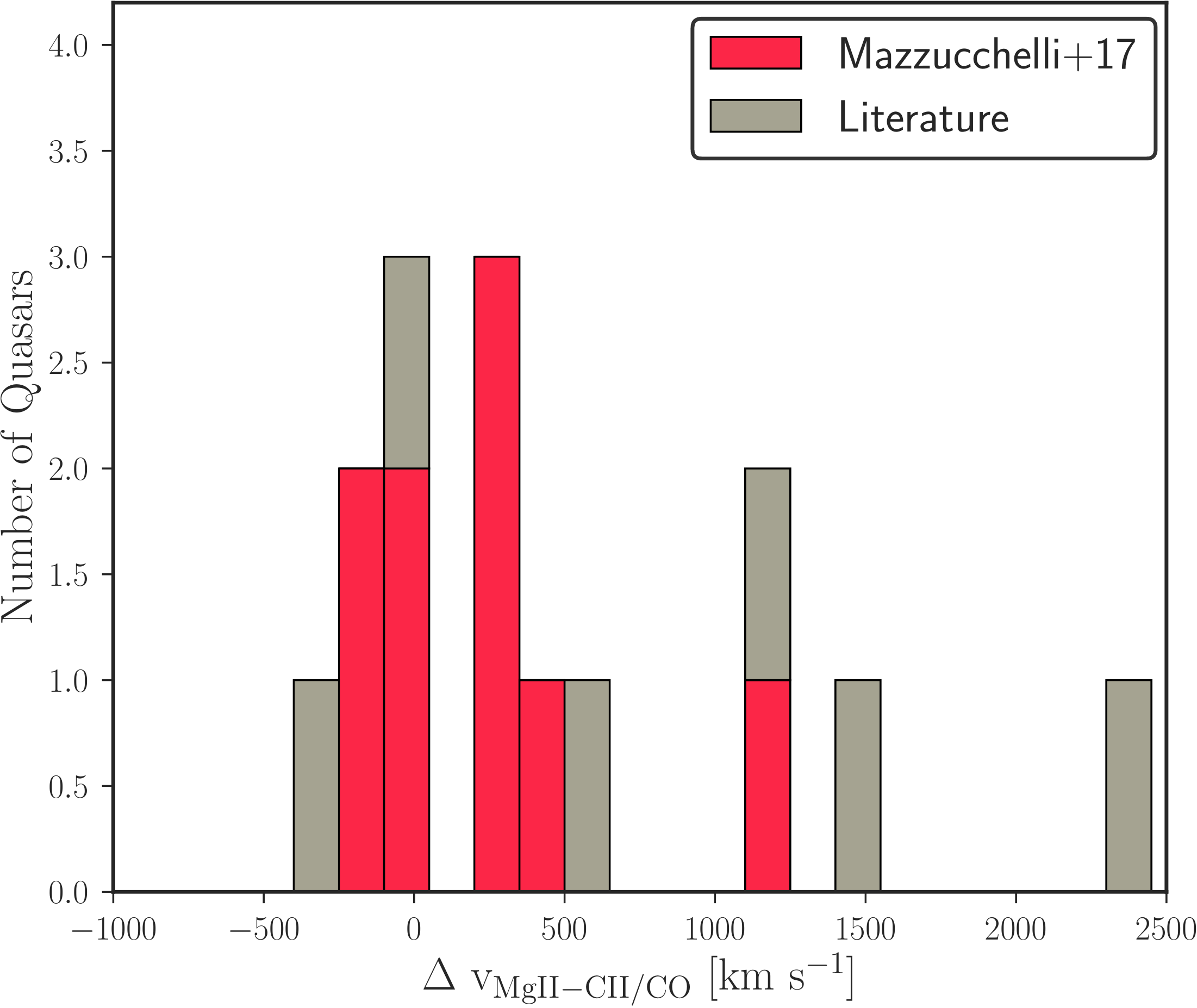}
\caption{Difference between the velocity measurements obtained from $\mathrm{Mg\,}${\scriptsize II} and [CII] or CO emission lines for a sample of $z\gtrsim$6 quasars. We consider 9 objects in this work for which we have both measurements (red histogram; see Tables \ref{tabSampleQSOs} and \ref{tabFitMgIIQSOs}) and six quasars from the literature (grey histogram; see text for references).
The positive sign indicates the blueshift of the Mg$\,${\scriptsize II} emission line. The offsets span a large range of values, with a mean and standard deviation of 485 $\pm$ 717 km s$^{-1}$, consistent with the results obtained by \cite{Venemans16}.}
\label{figShiftRed}
\end{figure}

The second best way to estimate redshifts is through the low-ionization $\mathrm{Mg\,}${\scriptsize II} $\lambda$2798.75 \AA $\,$ broad emission line, which is observable in the $K-$band at $z>$6.
This radiation is emitted from the broad line region (BLR), and therefore it provides a less precise measurement than the narrow emission from the cool gas traced by the [CII] emission.
Several studies, based on $z<$1 quasar samples, demonstrated that the $\mathrm{Mg\,}${\scriptsize II} emission is a far more reliable redshift estimator than other high-ionization emission lines (e.g. C$\,${\scriptsize IV} $\lambda$1549.06 \AA $\,$ and Si$\,${\scriptsize IV} $\lambda$1396.76 \AA), and it has a median shift of only 97 $\pm$ 269 km s$^{-1}$ with respect to the narrow [O$\,${\scriptsize III}] $\lambda$5008.24 \AA $\,$ emission line \citep{Richards02}.
We provide $z_{\mathrm{Mg II}}$ for HSC1205 and PSO247+24, for which we have no [CII] observations, as their best redshift estimates.
We also calculate $z_{\mathrm{Mg II}}$ for the remaining 9 quasars in our sample with NIR spectra (see Section \ref{secFitMgII} and Table \ref{tabFitMgIIQSOs}). Our new values are consistent, within 1$\sigma$ uncertainties, with the measurements from the literature for ULAS1120, VIK2348, VIK0109, VIK0305 \citep{DeRosa14}, and PSO036+03, PSO167-13, PSO338+29 (\citealt{Venemans15}).
However, it has been recently shown that, at $z\gtrsim$6, the mean and standard deviation of the shifts between $z_{\mathrm{Mg II}}$ and the quasar systemic redshift (as derived from the [CII] emission line), are significantly larger (480 $\pm$ 630 km s$^{-1}$) than what is found at low$-$redshift (see \citealt{Venemans16}).
We can study the distribution of the shifts between the redshifts measured from $\mathrm{Mg\,}${\scriptsize II} and [CII] (or CO) emission lines, considering both the newly discovered and/or newly analyzed sources in this sample, and quasars at $z\gtrsim 6$ with such information from the literature (six objects; the values of $z_{\mathrm{Mg II}}$ are taken from \citealt{Willott10b} and \citealt{DeRosa11}, while the $z_{\mathrm{[CII]}}$ measurements are from \citealt{Carilli10}, \citealt{Wang11}, \citealt{Willott13b}, \citeyear{Willott15}). The distribution of the shifts is shown in Figure \ref{figShiftRed}.
They span a large range of values, from +2300 km s$^{-1}$ to -265 km s$^{-1}$. We obtain a mean and median of 485 and 270 km s$^{-1}$, respectively, and a large standard deviation of 717 km s$^{-1}$. These results are in line with what was found by \cite{Venemans16}, although we measure a less extreme median value (270 km s$^{-1}$ against 467 km s$^{-1}$), and confirm that the $\mathrm{Mg\,}${\scriptsize II} emission line can be significantly blueshifted with respect to the [CII] emission in high$-$redshift quasars.
This effect is unlikely to be due to the infalling of [CII] in the quasars' host galaxies: resolved observations of the [CII] emission line in high$-$redshift quasars show that the gas is often displaced in a rotating disk (e.g.$\,$\citealt{Wang13}, \citealt{Shao17}), and no evidence is found in our sample to point at a different scenario. Also, the gas free-fall time would be too short ($\sim$few Myr, considering a typical galactic size of $\sim$2kpc and gas mass of $\sim$10$^{8}$ M$_{\odot}$, e.g.$\,$\citealt{Venemans16}) to allow the ubiquitous observation of [CII] in quasars at these redshifts. An alternative scenario explaining the detected blueshifts would be that the BLRs in these quasars are characterized by strong outflows/wind components.

Finally, for PSO261+19 and PSO011+09, only the optical spectra are available. We derive their redshifts from a $\chi^{2}$ minimization technique, comparing their spectrum with the low redshift quasar template from \cite{Selsing16}, and the composite of $z\sim$6 quasars presented by \cite{Fan06}; for further details on this procedure see \cite{Banados16}. The redshift measurements obtained in this case are the most uncertain, with $\mathrm{\Delta} z$=0.05.\\
We report all the redshifts, their uncertainties, the different adopted techniques and references in Table \ref{tabSampleQSOs}.
\subsection{Absolute magnitude at 1450\AA} \label{subsecM1450}
\begin{figure}
\centering
\includegraphics[width=\columnwidth]{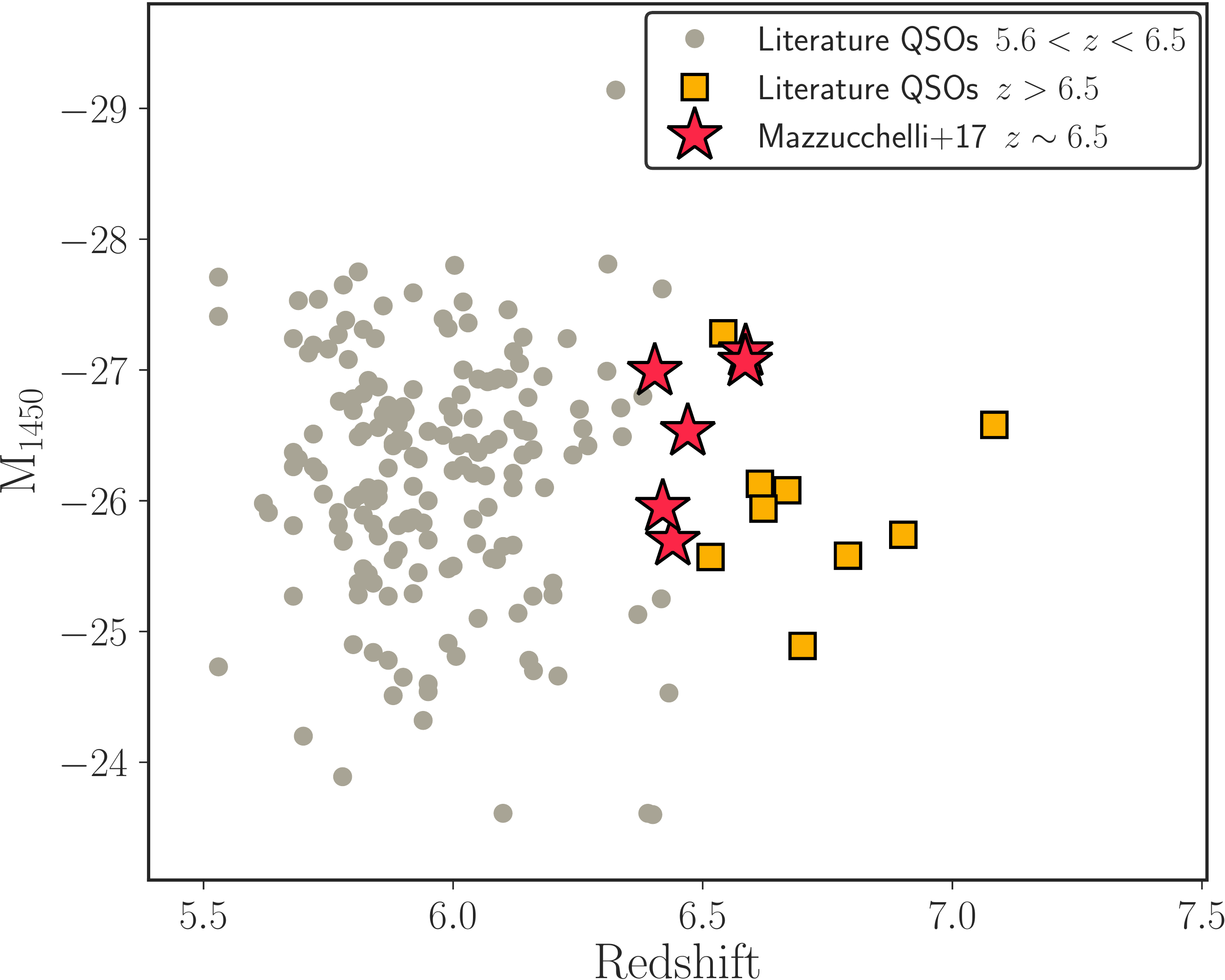}
\caption{Absolute magnitude at rest frame wavelength 1450 \AA, $M_{1450}$, against redshift, for quasars at $5.5 \lesssim z \lesssim 6.4$ from the literature (grey circles; see \citealt{Banados16}, Table 7, for references), and in the sample considered here, at $z \gtrsim 6.42$, both taken from the literature (for references see Table \ref{tabSampleQSOs}; yellow squares) and newly discovered in this work (red stars). All the $M_{1450}$ values were derived with a consistent methodology (see text). The magnitudes of the quasars presented here span a similar range as the ones at lower redshits.}
\label{figM1450vsRed}
\end{figure}

The apparent magnitude at rest-frame 1450 \AA$\,$($m_{1450}$) is a quantity commonly used in characterizing quasars. Following \cite{Banados16}, we extrapolate $m_{1450}$ from the $J$-band magnitude, assuming a power law fit of the continuum
($f\sim \lambda^{-\alpha}$), with $\alpha=-1.7$ \citep{Selsing16}\footnote{For PSO006+39 we use the $y_{\mathrm{P1}}-$band magnitude, since we do not have $J-$band information.}.
We derive the corresponding absolute magnitude ($M_{1450}$) using the redshifts reported in Table \ref{tabSampleQSOs}. In Figure \ref{figM1450vsRed} we show the distribution of $M_{1450}$, a proxy of the UV-rest frame luminosity of the quasars, as a function of redshift, for the sources in our sample and a compilation of quasars at $ 5.5 \lesssim z \lesssim 6.4 $ (see references in \citealt{Banados16}, Table 7). The highest-redshift objects considered here show similar luminosities to the ones at $z\sim$6. In Table \ref{tabM1450SlopQSOs} we report the values of $m_{1450}$ and $M_{1450}$ for the quasars analysed here.\\
\subsection{Quasar continuum} \label{secQSOCont}
The UV/optical rest-frame quasar continuum emission results from the superposition of multiple components: the non-thermal, power law emission from the accretion disk; the stellar continuum from the host galaxy; the Balmer pseudo-continuum; and the pseudo-continuum due to the blending of several broad Fe$\,${\scriptsize II} and Fe$\,${\scriptsize III} emission lines. In the literature, the continua of very luminous quasars such as the ones studied here, have been generally reproduced with a simple power-law, since the host galaxy emission is outshined by the radiation from the central engine.
Here, we first model the continuum with a single power law: 
\begin{equation}
\label{eqPLcont}
F_{\lambda} = F_{0} \left( \frac{\lambda}{2500 \mathrm{\AA}}\right)^{\alpha} 
\end{equation}
We consider regions of the rest-frame spectra which are free from strong emission lines: [1285$-$1295; 1315$-$1325; 1340$-$1375; 1425$-$1470; 1680$-$1710; 1975$-$2050; 2150$-$2250; and 2950$-$2990] \AA \quad (\citealt{Decarli10}). We slightly adjust these windows to take into account sky absorption, residual sky emission and regions with low S/N.
We use a $\chi^{2}$ minimization technique to derive the best values and corresponding uncertainties for $\alpha$ and $F_{0}$ (see Table \ref{tabM1450SlopQSOs}).

\cite{VandenBerk01} and \cite{Selsing16} report typical slopes of $\alpha=-1.5$ and -1.7, respectively, for composite templates of lower redshift ($z\sim2$) SDSS quasars. In our case, we find that $\alpha$ may significantly vary from object to object, with a mean of $\alpha=-1.6$ and a 1$\sigma$ dispersion of 1.0. This large range of values is in agreement with previous studies of lower$-$redshift quasars ($z<3$, \citealt{Decarli10}; $4 \lesssim z \lesssim 6.4$, \citealt{DeRosa11}, \citeyear{DeRosa14}).
However, we notice that the quasars for which we only have optical spectral information are poorly reproduced by a power-law model, and the slopes obtained are characterized by large uncertainties (see Table \ref{tabM1450SlopQSOs}). If we consider only the objects with NIR spectroscopy, we obtain a mean slope of $\alpha=-$1.2, with a 1$\sigma$ scatter of 0.4.

We use these power law continuum fits in the modeling of the C$\,${\scriptsize IV} broad emission line in our quasars with NIR coverage (see Section \ref{subsecCIV}). 
Afterwards, we implement a more accurate modeling of the spectral region around the Mg$\,${\scriptsize II} emission line, which, together with the Fe$\,${\scriptsize II} emission and the rest-frame UV luminosity, is a key tool commonly used to derive crucial quasar properties, e.g.$\,$black hole masses (see Section \ref{subsecBHM}).
\begin{deluxetable*}{lcccccc}[h]
\tabletypesize{\small}
\tablecaption{Parameters (slope and normalization) obtained from the power law fit of the spectra in our quasar sample (see Section \ref{secQSOCont}, eq. \ref{eqPLcont}). We report also: the apparent and absolute magnitude at rest frame wavelenght 1450 \AA$\,$(Section \ref{subsecM1450}, plotted as a function of redshift in Figure \ref{figM1450vsRed}); the C$\,${\scriptsize IV} blueshifts with respect to the Mg$\,${\scriptsize II} emission lines; the rest-frame C$\,${\scriptsize IV} EW (Section \ref{subsecCIV}). \label{tabM1450SlopQSOs}}
\tablewidth{0pt}
\tablehead{ \colhead{name}& \colhead{$\alpha$} & \colhead{$F_{0}$}  
 & \colhead{$m_{1450}$} &  \colhead{M$_{1450}$} & \colhead{$\Delta v_{\mathrm{CIV-MgII}}$} & \colhead{C$\,${\scriptsize IV} EW}
 \\
  & & [$10^{-17} \mathrm{erg\,s^{-1}\,cm^{-2}}$] & & & [km s$^{-1}$] & [\AA]
}
\startdata
PSO J006.1240+39.2219 & -3.92$\pm$ 0.03 & 0.060$^{+1.86}_{-4.006}$ & 20.00 & 25.94\footnote{Value taken from \cite{Tang17}} & $-$ & $-$
\\ 
PSO J011.3899+09.0325  & -3.75$^{+3.91}_{-0.01}$ & 0.051$^{+0.06}_{-0.001}$ & 20.85 & -25.95  &  $-$ &  $-$\\
VIK J0109--3047 & -0.96$^{+2.71}_{-0.04}$ & 0.141$^{+0.09}_{-0.075}$ & 21.30 & -25.58  & 4412$\pm$175 & 14.9$\pm$0.1 \\
PSO J036.5078+03.0498 & -1.61$^{+0.03}_{-0.07}$ & 0.610$\pm$0.05 & 19.55 & -27.28  &  5386$\pm$689 & 41.5$\pm$1.1\\
VIK J0305--3150 & -0.84$^{+0.02}_{-0.04}$ & 0.203$\pm$0.005 & 20.72 & -26.13   & 2438$\pm$137& 40.5$\pm$0.3\\
PSO J167.6415--13.4960  & -0.99$^{+1.12}_{-0.68}$ & 0.176$^{+0.055}_{-0.175}$ & 21.25 & -25.57  & - & -\\
ULAS J1120+0641 & -1.35$^{+0.24}_{-0.22}$ & 0.248$^{+0.086}_{-0.011}$ & 20.38 &  -26.58  & 2602$\pm$285 & 48.1$\pm$0.7\\
HSC J1205--0000  & -0.61$^{+0.01}_{-0.48}$ & 0.131$^{+0.075}_{-0.275}$ & 21.98 & -24.89  & $-$ & $-$\\
PSO J183.1124+05.0926  & -1.19$^{+0.13}_{-0.15}$ & 0.523$^{+0.02}_{-0.05}$ & 19.82 & -26.99  & $-$  &  $-$ \\
PSO J231.6576--20.8335 & -1.59$\pm 0.06$ & 0.504$^{+0.003}_{-0.075}$ & 19.70 & -27.14 & 5861$\pm$318 & 23.0$\pm$1.2\\
PSO J247.2970+24.1277  & -0.926$^{+0.15}_{-0.21}$ & 0.350$^{+0.102}_{-0.005}$ & 20.28 & -26.53  & 2391$\pm$110 & 29.1$\pm$0.7\\
PSO J261.0364+19.0286  & -2.01$^{+1.11}_{-0.01}$  & 0.166$^{+0.182}_{-0.024}$ & 21.12 & -25.69  & $-$ & $-$ \\
PSO J323.1382+12.2986 & -1.38$^{+0.20}_{-0.18}$ & 0.227$^{+0.005}_{-0.115}$ & 19.78 & -27.06  & 736$\pm$42 & 19.9$\pm$0.2\\
PSO J338.2298+29.5089 & -1.98$^{+0.87}_{-0.60}$ & 0.147$^{+0.035}_{-0.055}$ & 20.78 & -26.08  & 842$\pm$170 & 40.6$\pm$0.8 \\
VIK J2348--3054 & -0.65$^{+1.4}_{-0.6}$ & 0.155$^{+0.115}_{-0.134}$ & 21.17 & -25.74  & 1793$\pm$110 & 45.8$\pm$0.3
\enddata
\end{deluxetable*}
\subsection{C$\,${\scriptsize IV} blueshifts} \label{subsecCIV}
The peaks of high-ionization, broad emission lines, such as C$\,${\scriptsize IV}, show significant shifts bluewards with respect to the systemic redshifts in quasars at low$-$redshift (e.g. \citealt{Richards02}): this has been considered a signature of outflows and/or of an important wind component in quasars BLRs (e.g. \citealt{Leighly04}). Hints have been found of even more extreme blueshifts at high redshifts (e.g. \citealt{DeRosa14}).

\begin{figure}[h]
\epsscale{0.8}
\centering
\includegraphics[width=\columnwidth]{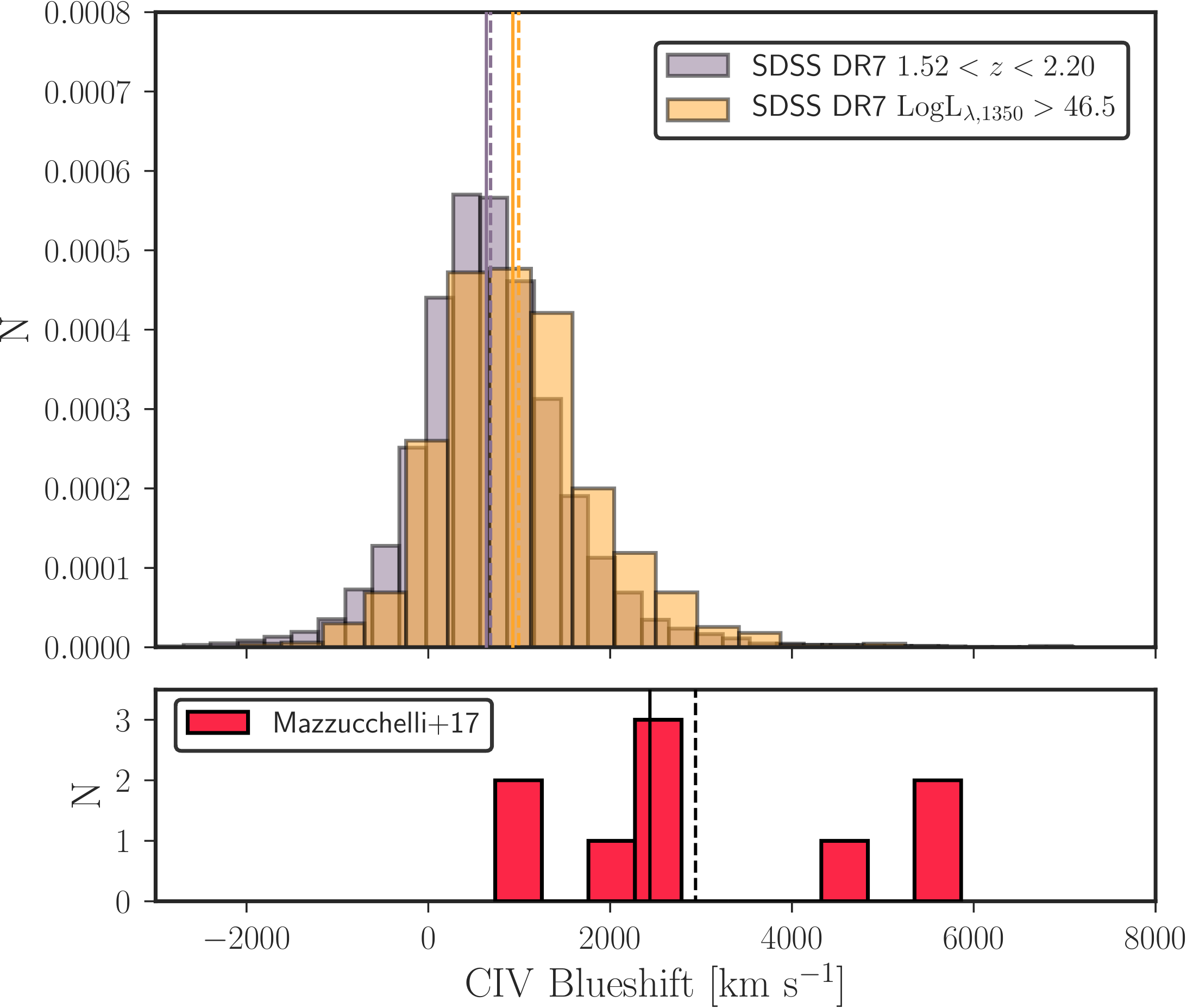}
\caption{Histogram of C$\,${\scriptsize IV} blueshifts with respect to the Mg$\,${\scriptsize II} emission line, for the objects in our sample (bottom panel, red histogram) and a collection of 1.52 $< z <$ 2.2 quasars from the SDSS DR7 catalog (upper panel, grey histogram; see text for details). A sub-sample of low redshift quasars with higher luminosities (L$_{\lambda, 1350} > 3 \times 10^{46}$ erg s$^{-1}$) is also reported (orange histogram). We adopt positive signs for blueshifts. The mean and median of the distributions are reported with continuous and dashed lines, respectively. Quasars at high redshift show much higher C$\,${\scriptsize IV} blueshifts (with values up to $\sim$5900 km s$^{-1}$) with respect to the sample at lower redshift. The histograms reported in the upper panel are normalized such that the underlying area is equal to one.}
\label{figCIVblueDist}
\end{figure}

Here, we investigate the presence of C$\,${\scriptsize IV} shifts in our high$-$redshift quasars by modeling the emission line with a single Gaussian function, after subtracting the continuum power law model obtained in Section \ref{secQSOCont} from the observed spectra. We report the computed C$\,${\scriptsize IV} shifts with respect to the Mg$\,${\scriptsize II} emission line (see Section \ref{subsecRedTempl} and Table \ref{tabFitMgIIQSOs}) in Table \ref{tabM1450SlopQSOs}. We consider here the Mg$\,${\scriptsize II} and not the [CII] line since we want to consistently compare our high$-$redshift sources to $z\sim$1 quasars (see below), for which the [CII] measurements are not always available. We adopt a positive sign for blueshifts. All quasars in our sample show significant blueshifts, from $\sim$730 to $\sim$5900 km s$^{-1}$. For the previously studied case of ULAS1120, the value found here is consistent with the ones reported in the literature (\citealt{DeRosa14}, \citealt{Greig17}). We neglect here: PSO167-13 and HSC1205, due to the low S/N; PSO183+05, for which we do not have a measurement of the Mg$\,${\scriptsize II} redshift (see Section \ref{secFitMgII}); and PSO011+09, PSO006+39 and PSO261+19, since we do not have NIR spectral coverage (see Section \ref{secFolUpObs} and Figure \ref{figSpecAll}); also, we still consider VIK2348, but with the caveat that this object was flagged as a possible broad absorption line (BAL) quasar \citep{DeRosa14}.
In Figure \ref{figCIVblueDist} we show the distribution of the blueshifts for high$-$redshift quasars in this work (\textit{bottom panel}) and for a sample of objects at lower redshift taken from the SDSS$-$DR7 catalog (\citealt{Shen11}; \textit{upper panel}). For comparison, we select a subsample of objects at low redshift, partially following \cite{Richards11}:
we consider quasars in the redshift range $1.52 < z < 2.2$ (where both the C$\,${\scriptsize IV} and Mg$\,${\scriptsize II} emission lines are covered), with significant detection of the
broad C$\,${\scriptsize IV} emission line (FWHM$\rm _{C\, IV} >$ 1000 km s$^{-1}$; FWHM$\rm _{C\, IV} >$ 2$\sigma_{\mathrm{FWHM_{C\, IV}}}$ ; EW$\rm _{C\, IV} >$ 5 \AA, EW$\rm _{C\, IV} >$ 2$\sigma_{\mathrm{EW_{C\, IV}}}$; where $\sigma_{\mathrm{FWHM}}$ and $\sigma_{\mathrm{EW}}$ are the uncertainties on the FWHM and EW, respectively), and of the Mg$\,${\scriptsize II} emission line (FWHM$\rm _{Mg\, II} >$ 1000 km s$^{-1}$; FWHM$\rm _{Mg\, II} >$ 2$\sigma_{\mathrm{FWHM_{Mg\, II}}}$ ; EW$\rm _{Mg\, II} >$ 2$\sigma_{\mathrm{EW_{Mg\, II}}}$), and that are not flagged as BAL quasars (BAL FLAG=0).
The total number of objects is $\sim$22700; the mean, median and standard deviation of the C$\,${\scriptsize IV} blueshift with respect to the Mg$\,${\scriptsize II} emission line in this lower$-$redshift sample are 685, 640 and 871 km s$^{-1}$, respectively. If we consider a sub sample of the brightest quasars (with luminosity at rest-frame wavelength 1350 \AA$\, \L_{\mathrm{\lambda, 1350}} > 3 \times 10^{46}$ erg s$^{-1}$; 1453 objects), we recover a higher mean and median values (994 and 930 km s$^{-1}$, respectively), but with large scatter (see Figure \ref{figCIVblueDist}). We also draw a sub-sample of SDSS quasars matched to the $L_{\mathrm{\lambda,1350}}$ distribution of the high$-$redshift sample (for details on the method, see Section \ref{subsecBHM}). In this case, the mean and median values of the C$\,${\scriptsize IV} blueshift are 790 and 732 km s$^{-1}$, respectively, with a standard deviation of 926 km s$^{-1}$.
The high$-$redshift quasar population is characterized by a mean, median and standard deviation of
2940, 2438 and 1761 km s$^{-1}$; C$\,${\scriptsize IV}
blueshifts tend to be much higher at high$-$redshift, as already observed for the Mg$\,${\scriptsize II} shifts with respect to the systemic quasars redshifts traced by CO/[CII] emission (see Section \ref{subsecRedTempl} and Figure \ref{figShiftRed}).

\begin{figure}[h]
\epsscale{0.8}
\centering
\includegraphics[width=\columnwidth]{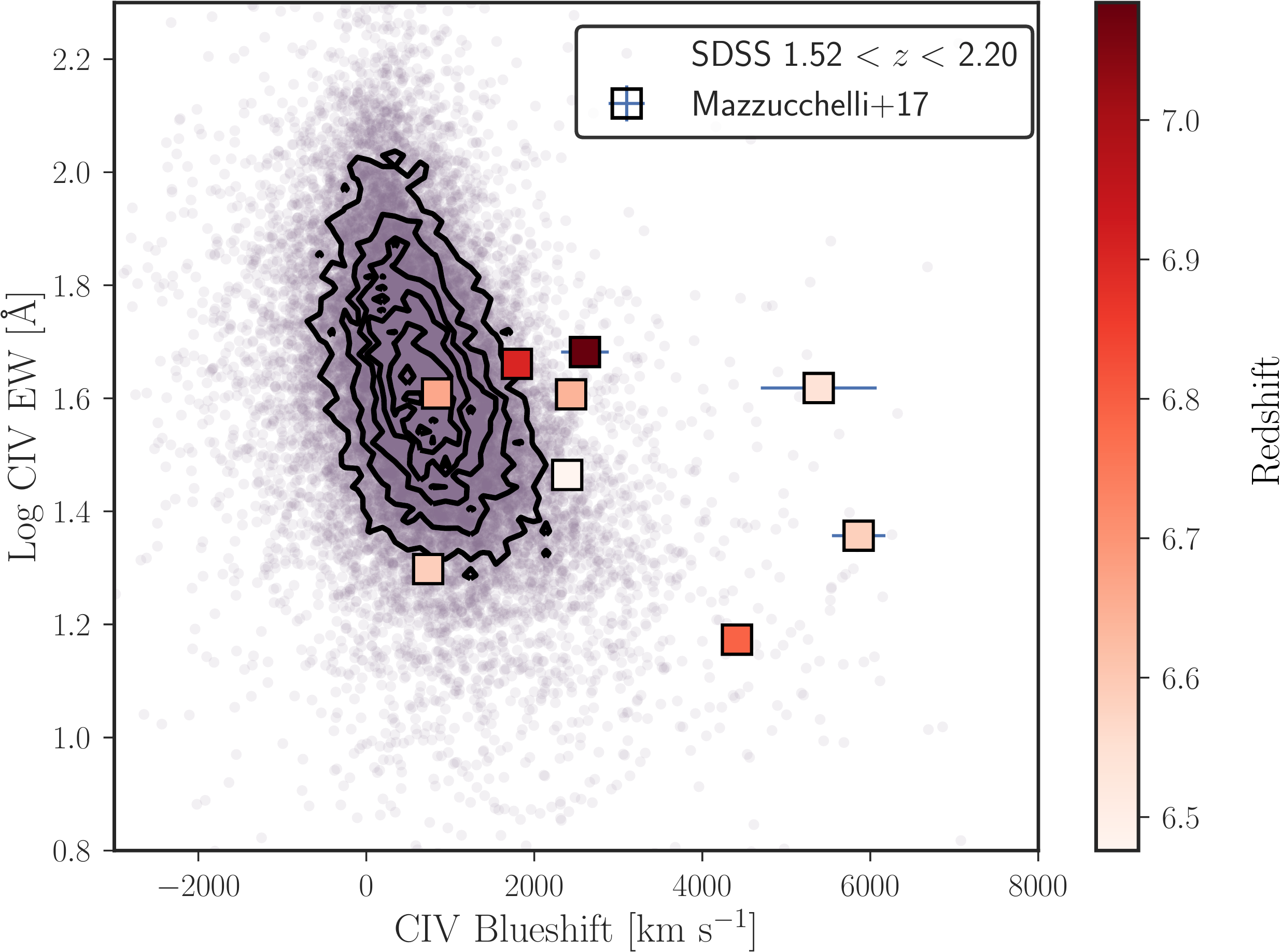}
\caption{Rest frame C$\,${\scriptsize IV} EW as a function of C$\,${\scriptsize IV} blueshift, for the quasars in our sample (big squares color-coded with respect to redshifts) and a sample of quasars at lower redshift from SDSS DR7 (\citealt{Shen11}; grey points and black contours; see text for details on the definition of this sub-sample). Quasars at low$-$redshift with very high blueshifts have small EW. The high redshift quasars are characterized by extreme blueshifts and small C$\,${\scriptsize IV} EW, following the trend at $z\sim 1$ but with larger scatter.}
\label{figCIVblueEW}
\end{figure}

In Table \ref{tabM1450SlopQSOs} we report the values of C$\,${\scriptsize IV} rest-frame EW of the quasars in the sample of this work, which are plotted as a function of C$\,${\scriptsize IV} blueshifts, together with objects at low$-$redshift, in Figure \ref{figCIVblueEW}. \cite{Richards11} show that C$\,${\scriptsize IV} blueshifts correlate with C$\,${\scriptsize IV} EW at $z \sim 1-2$: quasars with large EW are characterized by small blueshifts, while objects with small EW present both large and small blueshifts; no objects where found with strong C$\,${\scriptsize IV} line and high blueshift. The high redshift quasars studied here follow the trend of the low$-$redshift objects, with extreme C$\,${\scriptsize IV} blueshifts and EW equal or lower than the bulk of the SDSS sample. This is also in line with the higher fraction of weak emission line (WEL) quasars found at high redshifts (e.g. \citealt{Banados14}, \citeyear{Banados16}).

However, we note that C$\,${\scriptsize IV} blueshifts scale with quasars UV luminosities: this is linked to the anti-correlation between luminosity and emission lines EW (i.e. Baldwin effect; e.g. \citealt{Baldwin77}, \citealt{Richards11}). Also, the $z \gtrsim 6.5$ quasars presented here are biased towards higher luminosities (e.g. due to our selection criteria): we may then be considering here only the extreme cases of the highest redshift quasar population, and therefore missing the objects at lower luminosity and lower C$\,${\scriptsize IV} blueshifts.

\subsection{Mg$\,${\scriptsize II} and Fe$\,${\scriptsize II} emission modeling} \label{secFitMgII}
\begin{figure*}[h]
\epsscale{0.8}
\centering
\includegraphics[width=\textwidth]{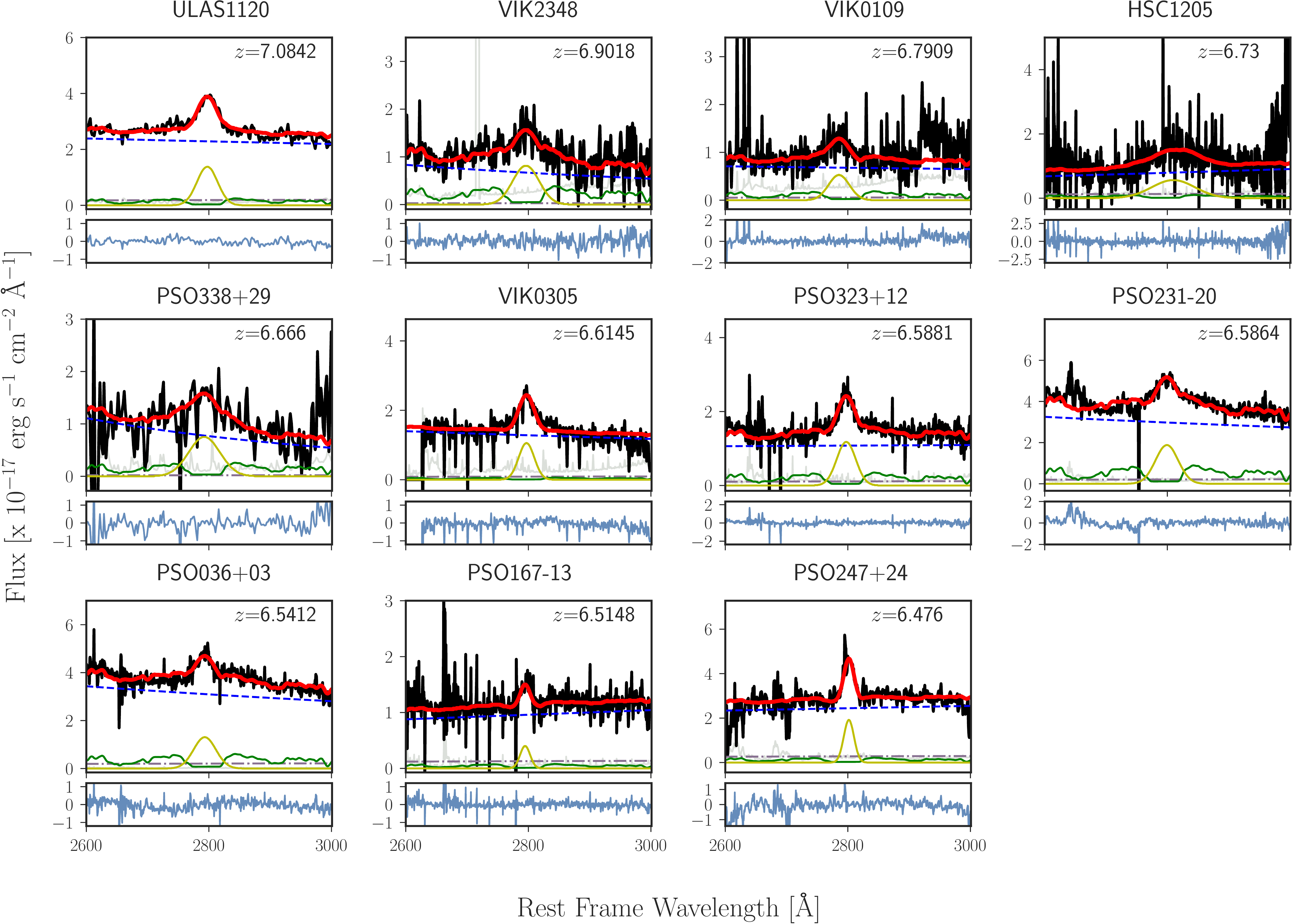}
\caption{Best fit of the spectral region around the Mg$\,${\scriptsize II} emission lines for the quasars in our sample for which we have K-band spectroscopy. We show the different components of the fit: the power law continuum (dashed blue), the Balmer (brown dot-dashed) and the Fe$\,${\scriptsize II} pseudo-continuum emission (solid green) and the gaussian Mg$\,${\scriptsize II} emission line (yellow solid line); the total fit is reported with a solid red line. In the bottom panels we show the residuals of the fit.
The derived quantities are listed in Table \ref{tabFitMgIIQSOs}.}
\label{figFitMgIIall}
\end{figure*}

We fit the quasar emission, in the rest-frame wavelength window 2100$< \lambda$/[\AA] $< $3200, as a superposition of multiple components:
\begin{itemize}
\item the quasar nuclear continuum emission, modeled as a power law (see Eq. \ref{eqPLcont});
\item the Balmer pseudo-continuum, modeled with the function provided by \citeauthor{Grandi82} (\citeyear{Grandi82}; see their Eq 7) and imposing that the value of the flux at $\lambda_{rf}=$3675 \AA$\,$ is equal to 10\% of the power law continuum contribution at the same wavelength;
\item the pseudo-continuum Fe$\,${\scriptsize II} emission, for which we use the empirical template by \cite{VestergaardWilkes01}; and
\item the Mg$\,${\scriptsize II} emission line, fitted with a single Gaussian function.
\end{itemize}
We use a $\chi^{2}$ minimization routine to find the best fitting parameters (slope and normalization) for the nuclear emission, together with the best scaling factor for the iron template; once we have subtracted the best continuum model from the observed spectra, we fit the Mg$\,${\scriptsize II} emission line (see for further details \citealt{Decarli10}). 
We apply this routine to all the quasars with NIR information in our sample.
We exclude PSO183+05 from this analysis, since this source is a weak emission line quasar (see Figure \ref{figSpecAll}) and the Mg$\,${\scriptsize II} fit is highly uncertain.
We show the obtained fit for the 11 remaining objects in Figure \ref{figFitMgIIall}.
In Table \ref{tabFitMgIIQSOs}, we list the derived monochromatic luminosities at rest-frame $\lambda_{rf}=$3000 $\rm \AA\,$($\rm \lambda L_{3000}$), calculated from the continuum flux  ($F_{\lambda,3000}$); the properties of the Mg$\,${\scriptsize II} line (full width at half maximum $-$FWHM$-$ and flux); the flux of the Fe$\,${\scriptsize II} emission and the redshift estimates $z_{\mathrm{Mg\,{\scriptsize II}}}$. We consider the 14th and 86th interquartiles of the $\chi^{2}$ distribution as our 1$\sigma$ confidence levels. 
\cite{DeRosa14} applied a similar analysis to the spectra of ULAS1120, VIK0305, VIK0109, VIK2348; their fitting procedures is however slightly different, since they fit all the spectral components at once, using the entire spectral range. Also, \cite{Venemans15} analyzed the NIR spectra of PSO036+03, PSO338+29 and PSO167-13, considering solely the nuclear continuum emission fitted with a power law and modeling the Mg$\,${\scriptsize II} emission line with a Gaussian function. The estimates that both studies obtain for $z_{\mathrm{Mg\,{\scriptsize II}}}$, black hole masses and bolometric luminosities are consistent, within the uncertainties, with the ones found here (see also Section \ref{subsecBHM}). 
\begin{deluxetable*}{lccccc}[h]
\tabletypesize{\small}
\tablecaption{Quantities derived from the fit of the spectral region around the Mg$\,${\scriptsize II} emission line: the monochromatic luminosity at rest-frame wavelength 3000 \AA$\,$($\rm \lambda L_{3000}$); FWHM, flux and redshift estimates of the Mg$\,${\scriptsize II} line, and the Fe$\,${\scriptsize II} flux. \label{tabFitMgIIQSOs}}
\tablewidth{0pt}
\tablehead{ \colhead{name}& \colhead{$\rm \lambda L_{3000}$} &  \colhead{$\mathrm{Mg_{II}}$ FWHM}  
 & \colhead{$\mathrm{Mg_{II}}$ Flux} &  \colhead{FeII Flux} &  \colhead{$z_{\mathrm{Mg\, II}}$}
 \\
  & [$10^{46} \mathrm{erg\,s^{-1}}$] &
  [km s$^{-1}$] & [$10^{-17} \mathrm{erg\,s^{-1}\,cm^{-2}}$] & [$10^{-17} \mathrm{erg\,s^{-1}\,cm^{-2}}$] 
}
\startdata
VIK J0109--3047 & 1.0$^{+0.1}_{-0.8}$ &  4313$^{+606}_{-560}$ & 22.5$^{+6.8}_{-6.2}$ & 45$^{+125}_{-0.15}$  & 6.763$\pm 0.01$\\
PSO J036.5078+03.0498 & 3.9$^{+0.4}_{-1.2}$  & 4585$^{+691}_{-461}$ & 59.4$^{+11.8}_{-9.2}$ & 147$^{+221}_{-81}$ & 6.533$^{+0.01}_{-0.008}$ \\
VIK J0305--3150 & 1.5$^{+0.2}_{-0.7}$ &  3210$^{+450}_{-293}$ & 41.0$^{+7.2}_{-5.0}$ & 42$^{+124}_{-15}$ & 6.610$^{+0.006}_{-0.005}$\\
PSO J167.6415--13.4960  & 0.9$^{+0.3}_{-0.4}$ & 2071$^{+211}_{-354}$ & 8.2$^{+1.4}_{-0.8}$ & $<$201 & 6.505$\pm 0.005$ \\
ULAS J1120+0641 & 3.6$^{+0.4}_{-1.4}$ &  4258$^{+524}_{-395}$ & 58.5$^{+9.3}_{-7.8}$ & 61$^{+225}_{-8}$ & 7.087$^{+0.007}_{-0.009}$\\
HSC J1205--0000  & 0.7$^{+0.3}_{-0.4}$ &  8841$^{+3410}_{-288}$ & 49.8$^{+5.9}_{-52.4}$ & $<$ 182 & 6.73$^{+0.01}_{-0.02}$ \\
PSO J231.6576--20.8335 & 3.7$^{+0.7}_{-0.9}$ & 4686$^{+261}_{-1800}$ & 87.6$^{+9.0}_{-28.2}$ & 216$^{+204}_{-128}$ & 6.587$^{+0.012}_{-0.008}$ \\
PSO J247.2970+24.1277  & 3.4$^{+0.1}_{-1.5}$ &  1975$^{+312}_{-288}$ & 40.2$^{+4.4}_{-5.8}$ & 54$^{+234}_{-0.2}$ & 6.476$\pm 0.004$ \\
PSO J323.1382+12.2986 & 1.6$^{+0.1}_{-1.0}$ & 3923$^{+446}_{-380}$ & 45.9$^{+7.4}_{-7.2}$ & 85$^{+109}_{-45}$ & 6.592$^{+0.007}_{-0.006}$\\
PSO J338.2298+29.5089 & 0.8$^{+0.4}_{-0.2}$ &  6491$^{+543}_{1105}$ & 47.7$^{+7.0}_{-9.0}$ & 76$^{+44}_{-54}$ & 6.66$^{+0.02}_{-0.01}$\\
VIK J2348--3054 & 0.9$^{+0.4}_{-0.3}$ & 5444$^{+470}_{-1079}$ & 44$^{+8.2}_{-8.5}$ & 95$^{+41}_{-72}$ & 6.902$\pm 0.01$
\enddata
\end{deluxetable*}
\subsection{Black Hole Masses} \label{subsecBHM}
We can estimate the quasar black hole masses ($M_{\mathrm{BH}}$) from our single epoch NIR spectra using the broad Mg$\,${\scriptsize II} emission line and $\lambda L_{\lambda,3000}$.
Under the assumption that the BLR dynamics is dominated by the central black hole gravitational potential, the virial theorem states:
\begin{equation}
M_{\mathrm{BH}}\sim \frac{R_{\mathrm{BLR}}v^{2}_{\mathrm{BLR}}}{G}
\end{equation}
where $G$ is the gravitational constant and $R_{\mathrm{BLR}}$ and $v_{\mathrm{BLR}}$ are the size and the orbital velocity of the emitting clouds, respectively. The velocity can be estimated from the width of the emission line:
\begin{equation}
v_{\mathrm{BLR}}=f \times \mathrm{FWHM}
\end{equation}
with $f$ a geometrical factor accounting for projection effects (e.g. \citealt{Decarli08}, \citealt{Grier13}, \citealt{Matthews17}).

Reverberation mapping techniques have been used to estimate the sizes of the BLRs from the H$\beta$ emission lines of nearby AGN \citep{Peterson04}. Several studies of this kind have shown that the continuum luminosity and $R_{\mathrm{BLR}}$ of AGN in the local universe correlate strongly (e.g. \citealt{Kaspi05}, \citealt{Bentz13}). Under the assumption that this relation holds also at high$-$redshift, we can use $\lambda L_{\lambda,3000}$ as a proxy of the BLR size.
We derive the mass of the black hole following \cite{VestergaardOsmer09}:
\begin{equation}
\label{eqMBH}
\frac{M_{\mathrm{BH}}}{\mathrm{M_{\odot}}} = 10^{6.86}\\ \left( \frac{\mathrm{FWHM}}{10^{3}\, \mathrm{km\, s^{-1}}} \right)^{2} \left( \frac{\lambda L_{\lambda,3000}}{10^{44}\, \mathrm{erg\, s^{-1}}} \right)^{0.5} 
\end{equation}
This relation has been obtained using thousands of high quality quasar spectra from SDSS-DR3 \citep{Schneider05}, and has been calibrated on robust reverberation mapping mass estimates \citep{Onken04}. The scatter on its zero point of 0.55 dex, which takes into account the uncertainty in the luminosity-$R_{\mathrm{BLR}}$ correlation, dominates the measured uncertainties on the black hole masses.

Also, from the black hole mass we can derive the Eddington luminosity ($L_{\mathrm{Edd}}$), the luminosity reached when the radiation pressure is in equilibrium with the gravitational attraction of the black hole:
\begin{equation}
\frac{L_{\mathrm{Edd}}}{\mathrm{erg\, s^{-1}}} = 1.3 \times 10^{38}\,  \frac{M_{\mathrm{BH}}}{\mathrm{M_{\odot}}} 
\end{equation}
Another useful quantity to derive is the Eddington ratio, the total measured bolometric luminosity of the quasar ($L_{\mathrm{bol}}$) divided by $L_{\mathrm{Edd}}$. 
We estimate $L_{\mathrm{bol}}$ using the bolometric correction by \cite{Shen08}:
\begin{equation}
\frac{L_{\mathrm{bol}}}{\mathrm{erg\, s^{-1}}} = 5.15 \times \frac{\lambda L_{\lambda,3000}}{\mathrm{erg\, s^{-1}}}
\end{equation}
The estimated values of black hole masses, bolometric luminosities and Eddington ratios for the quasars in our sample are shown in Table \ref{tabMBHQSOs}.

We notice that HSC1205, the faintest object in the sample, present a very broad Mg$\,${\scriptsize II} emission line: this leads to a high black hole mass ($\sim$5$\times$10$^{9}$ M$_{\odot}$) and a low Eddington ratio of 0.06. 
However, HSC1205 is also characterized by a red $J-W1$ color of 1.97, suggesting that the quasar has a red continuum, due to internal galactic extinction. This could affect our measurement of the quasar intrinsic luminosity and therefore we could observe a value of the Eddington ratio lower than the intrinsic one. We test this hypothesis by comparing the observed photometric information of this source with a suite of quasar spectral models characterized by different values of internal reddening $E(B-V)$. We obtain these models by applying the reddening law by \cite{Calzetti00} to a low$-$redshift quasar spectral template (\citealt{Selsing16}), redshifted at $z=6.73$ and corrected for the effect of the IGM absorption following \cite{Meiksin06}. We consider the $J$ magnitude provided by \cite{Matsuoka16}, $W1$ and $W2$ from \textit{WISE} (see Table \ref{tabPhotCatQSOs}) and $H$ and $K$ from the VIKING survey ($H$=21.38$\pm$0.21, $K$=20.77$\pm$0.14). A $\chi^{2}$ minimization routine suggests that this quasar has a large $E(B-V)$=0.3. The corrected monochromatic luminosity at $\lambda_{rf}=3000\,$ \AA $\,$ is 1.62$\times$10$^{46}$ erg s$^{-1}$, and the resulting black hole mass and Eddington ratio are 7.22$\times$10$^{9}$ M$_{\odot}$ and 0.09, respectively.
Therefore, even taking into account the high internal extinction, HSC1205 is found to host a very massive black hole and to accrete at the lowest rate in our sample.
\\[2mm]

We now place our estimates in a wider context, comparing them with the ones derived for low$-$redshift quasars.\\ 
We consider the SDSS-DR7 and DR12 quasar catalogs, presented by \cite{Shen11} and \cite{Paris17}, respectively; we select only objects in the redshift range $0.35 < z < 2.35$. In the DR7 release, we take into account the objects with any measurements of $\lambda L_{\lambda,3000}$ and Mg$\,${\scriptsize II} FWHM (85,507 out of $\sim$105,000 sources). We calculate $\lambda L_{\lambda,3000}$ for the quasars in the DR12 release, modeling a continuum power law with the index provided in the catalog (entry \texttt{ALPHA\_NU}), and normalizing it to the observed SDSS $i$ magnitude. We consider only the sources in DR12 with measurements of the power law index and of the Mg$\,${\scriptsize II} FWHM, and not already presented in DR7. Thus, out of the 297,301 sources in DR12, we select 68,062 objects: the total number of sources is 153,569.

\begin{figure}[h]
\epsscale{0.8}
\centering
\includegraphics[width=\columnwidth]{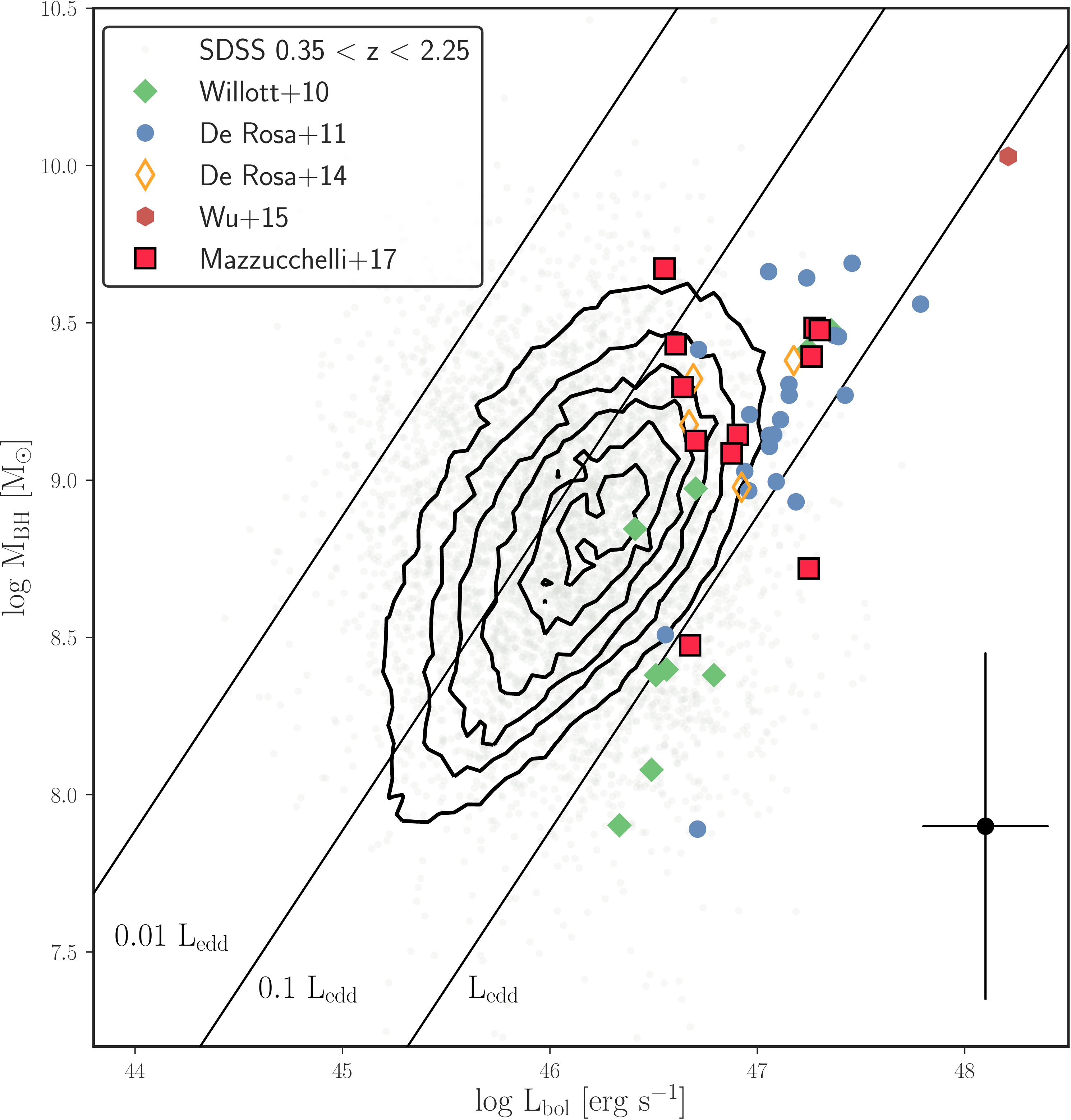}
\caption{Black hole mass as function of bolometric luminosity for several quasar samples. We report a sub-sample from the SDSS-DR7 and DR12 quasar catalogs (\citealt{Shen11} and \citealt{Paris17}, respectively) at $0.35 < z< 2.25 $ (grey points and countours). Also, we show measurements for quasars at higher redshifts, from \citeauthor{Willott10b} (\citeyear{Willott10b}, $z\sim6$; green filled diamonds), \citeauthor{DeRosa11} (\citeyear{DeRosa11}, $4 < z < 6.4$; blue points), \citeauthor{Wu15} (\citeyear{Wu15}, $z\sim6.3$; dark red hexagon). The objects presented in this study are reported with red, filled squares. We notice that four quasars (VIK0109, VIK0305, VIK2348 and ULAS1120) have also been analyzed by \citeauthor{DeRosa14} (\citeyear{DeRosa14}, orange empty diamonds): the two sets of measurements are consistent within the error bars. We show the method uncertainties on the black hole mass estimates and a representative mean error on the bolometric luminosity measurements (black point), and regions in the parameter space with constant Eddington luminosity (black lines). Quasars at high redshift are generally characterized by higher Eddington ratios than their lower$-$redshift counterparts, suggesting that they accreate at higher rates. However, the scatter in the $z \gtrsim 6.5$ sample is not negligible, with objects at $L_{\mathrm{bol}}/L_{\mathrm{Edd}}$ as low as $\sim$0.1.}
\label{figLbolMBH}
\end{figure}

\cite{DeRosa11} provide continuum luminosities and Mg$\,${\scriptsize II} measurements for 22 quasars at $4.0 \lesssim z \lesssim 6.4$ (observations collected from several studies: \citealt{Iwamuro02}, \citeyear{Iwamuro04}, \citealt{Barth03}, \citealt{Jiang07}, \citealt{Kurk07}, \citeyear{Kurk09}); \cite{Willott10b} present data for nine lower luminosity ($L_{\mathrm{bol}}<10^{47}$ erg s$^{-1}$) $z\sim6$ quasars; finally, \cite{Wu15} publish an ultra luminous quasar at $z\sim6.3$. In order to implement a consistent comparison among the various data sets, we re-calculate the black hole masses for all the objects in the literature using eq. \ref{eqMBH}. In Figure \ref{figLbolMBH} we show $M_{\mathrm{BH}}$ vs $L_{\mathrm{bol}}$, for the quasars presented here and the objects from the aforementioned studies. We highlight regions in the parameter space with constant Eddington ratio of 0.01, 0.1 and 1; we also show the typical errors on the black hole masses, due to the method uncertainties, and on the bolometric luminosities.

We note that the quasars at $z \gtrsim 4$ are generally found at higher bolometric luminosities ($L_{\mathrm{bol}} \gtrsim 10^{46}$ erg s$^{-1}$) than the objects at $z \sim 1$ (also due to selection effects, see below), but that the observed black hole masses span a similar range for both samples ($10^{8} \lesssim M_{\mathrm{BH}}/\mathrm{M_{\odot}} \lesssim 5 \times 10^{9}$). 
The bulk of the low redshift ($z\sim1$) quasar population shows lower Eddington ratios than the quasars at $z \gtrsim 4$.
As for the objects at $z>6.4$ presented in this sample, they occupy a parameter space similar to the sources from \cite{DeRosa11}, with a larger scatter in bolometric luminosities.  

\begin{figure}[h]
\epsscale{0.8}
\centering
\includegraphics[width=\columnwidth]{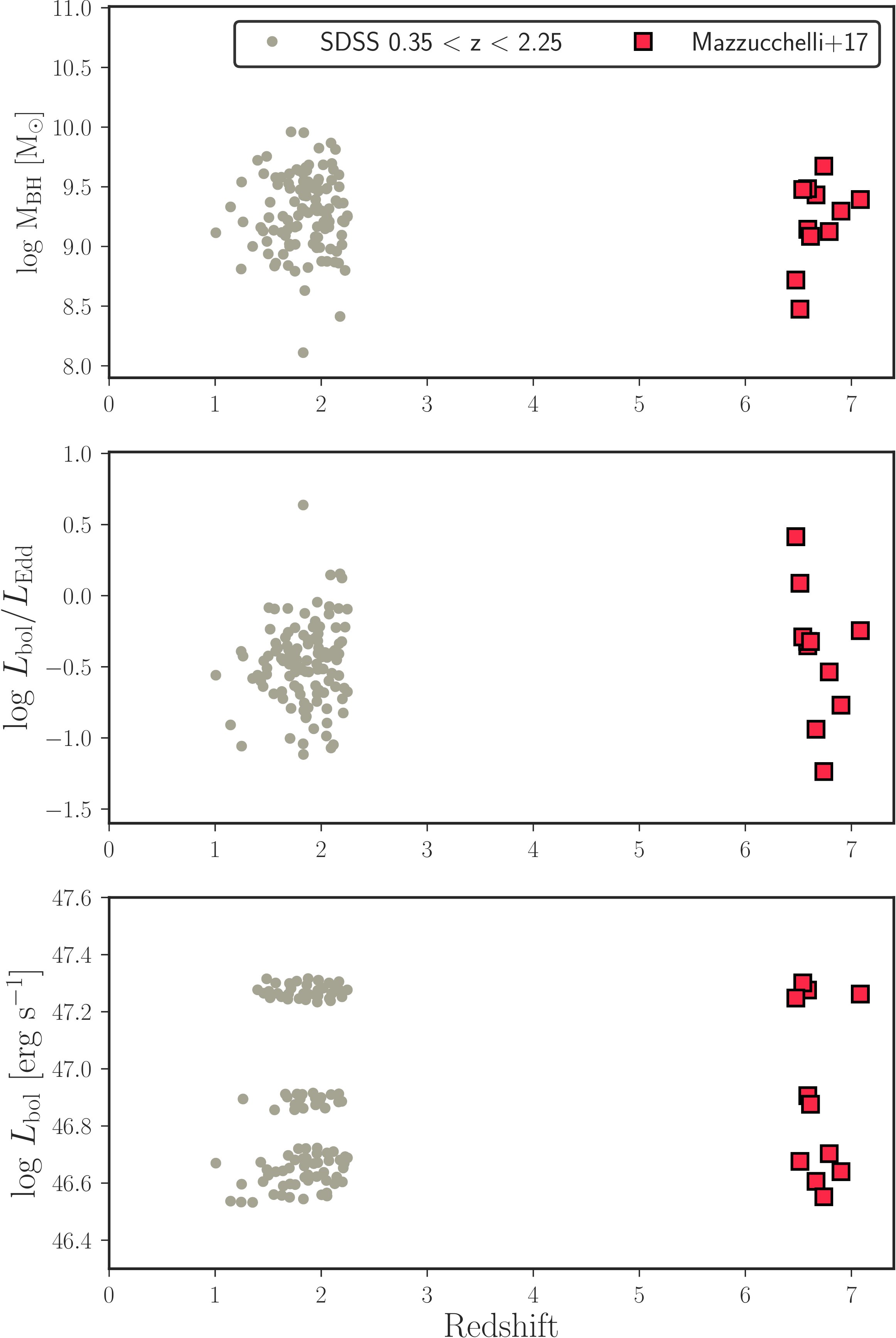}
\caption{Black hole mass (upper panel), Eddington ratio (central panel) and bolometric luminosity (lower panel) against redshift, for a bolometric luminosity matched quasar sample (see text for details on the selection of the subsample at $z\sim$1). We report the $z\gtrsim$6.5 quasars presented in this work with red squares, and the ones at lower redshift ($ 0.35 < z < 2.25$) from SDSS-DR7+DR12 (\citealt{Shen11}, \citealt{Paris17}) with grey points, respectively. The mean values of quasars black hole masses and Eddington ratios do not vary significantly with redshift (see also Figure \ref{figMBHR_hist}).}
\label{figMBHRedd}
\end{figure}

\begin{figure*}[h]
\epsscale{0.8}
\centering
\includegraphics[width=\textwidth]{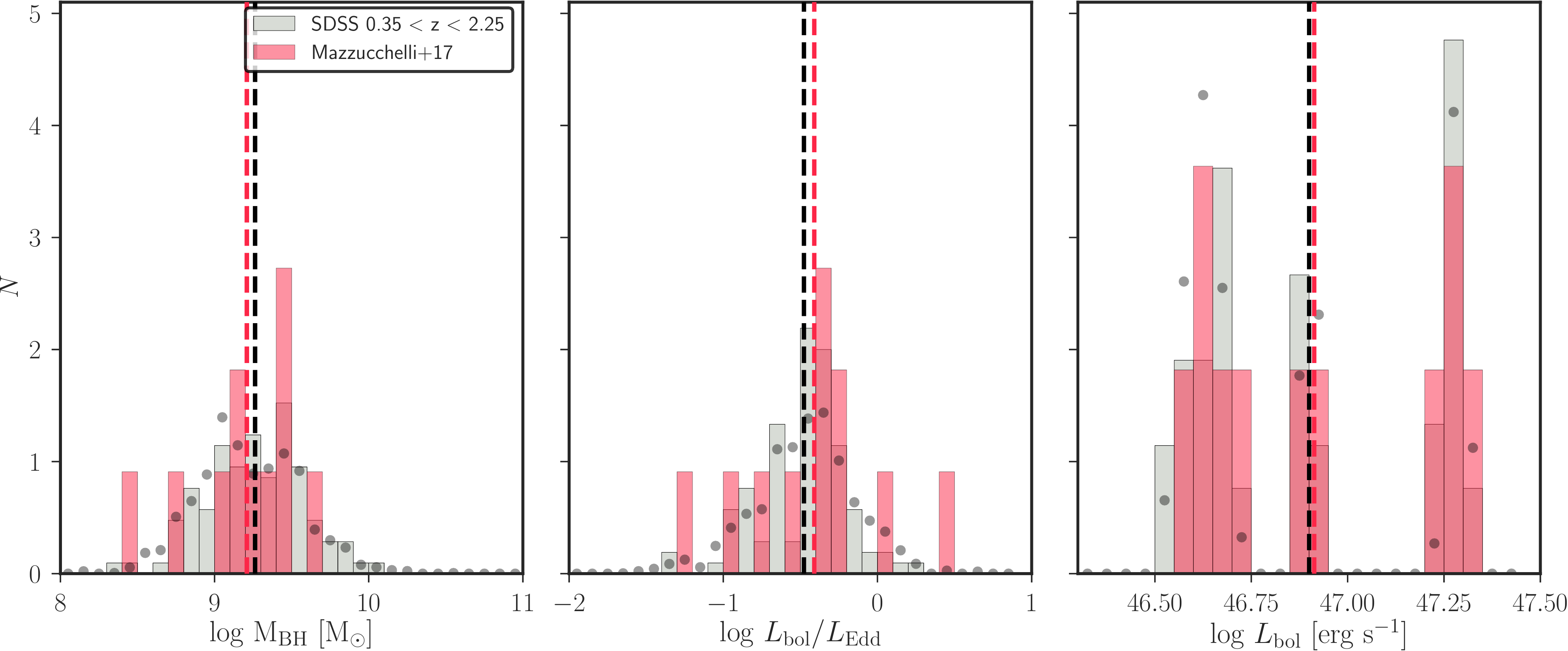}
\caption{Distribution of black hole masses (left panel), Eddington ratios (central panel) and bolometric luminosities (right panel) for one of the 1000 bolometric luminosity matched sub-samples drawn from low$-$redshift SDSS quasars (grey histograms; \citealt{Shen11} and \citealt{Paris17}; see text for details), and for the $z\gtrsim 6.5$ quasars presented here (red histograms). The grey points represent the mean of black hole masses, Eddington ratios and bolometric luminosities in each bin, resulting from all the 1000 trials at $z \sim$1. The mean of each quantity, for the low and high redshift populations, are shown in each panel with black and red dashed lines, respectively. We note that the mean black hole masses and Eddington ratios of the two samples are consistent, suggesting a non evolution of accretion rate with cosmic time. The histograms are normalized such that the underlying area is equal to one.}
\label{figMBHR_hist}
\end{figure*}

In order to provide a consistent comparison, we study the evolution of the black hole masses and Eddington ratios, as a function of redshift, for a quasar sample matched in bolometric luminosity.
Since the high$-$redshift quasars studied here are highly biased towards higher luminosities, mainly due to our selection criteria, a simple luminosity cut would not produce a truly luminosity matched sample.
In order to reproduce the same luminosity distribution as the one of the high$-$redshift sources, we sample the low$-$redshift SDSS quasars by randomly drawing sources with comparable $L_{\mathrm{bol}}$ to $z \gtrsim 6.5$ quasars (within 0.01dex); we repeat this trial for 1000 times. We show in Figure \ref{figMBHRedd} the black hole masses, bolometric luminosities and Eddington ratios, as a function of redshift, for the quasars presented in this work and for objects in one of the samples drawn at $z\sim 1$. The distributions of these quantities are also reported in Figure \ref{figMBHR_hist}.
We consider, as representative values for black hole mass and Eddington ratio of a bolometric luminosity matched sample at $z \sim$1, the mean of the means and the mean of the standard deviations calculated from the 1000 sub-samples.
We then obtain $\langle \mathrm{log}(M_{\mathrm{BH}}) \rangle$=9.21
and $\langle \mathrm{log}(L_{\mathrm{bol}}/L_{\mathrm{Edd}}) \rangle$=-0.47, with a scatter of 0.34 and 0.33, respectively. 
These values are consistent, also considering the large scatter, with the estimates obtained for $z \gtrsim$6.5 quasars:
$\langle \mathrm{log}(M_{\mathrm{BH}}) \rangle$=9.21 and $\langle \mathrm{log}(L_{\mathrm{bol}}/L_{\mathrm{Edd}}) \rangle$=-0.41, with a scatter of 0.34 and 0.44, respectively. Therefore, considering a bolometric luminosity matched sample, we do not find convincing evidence for an evolution of quasars accretion rate with redshift.

Finally, we caution that we have witnessed evidence suggesting the presence of a strong wind component in the BLR (see Sections \ref{subsecCIV} and \ref{subsecRedTempl}). In case of non-negligible radiation pressure by ionizing photons acting on the BLRs, the black hole masses derived by the simple application of the virial theorem might be underestimated (e.g.$\,$\citealt{Marconi08}). This effect depends strongly on the column density ($N_{\mathrm{H}}$) of the BLR, and on the Eddington ratio. \cite{Marconi08} show that, in case of 0.1$\lesssim L_{\mathrm{bol}}/L_{\mathrm{Edd}} \lesssim$1.0, as found in $z \gtrsim$6.5 quasars, and for typical values of 10$^{23}< N_{\mathrm{H}}/[\mathrm{cm^{-2}}] <$10$^{24}$, the true black hole masses would be $\sim$2-10$\times$ larger than the virial estimates. This would lead to an even stronger challenge for the current models of primordial black holes formation and growth.
An in depth discussion of this effect, given the uncertainties on the contribution of the possible wind and on the BLR structure itself, is beyond the scope of this paper.
\\[2mm]
\subsection{Black Hole Seeds}
Measurements of black hole masses and Eddington ratios of high$-$redshift quasars help us constrain formation scenarios of the first supermassive black holes in the very early universe.
While the black hole seeds from Pop III stars are expected to be relatively small ($\sim 100$ M$_{\odot}$; e.g. \citealt{Valiante16}), direct collapse of massive clouds can lead to the formation of more massive seeds ($\sim 10^{4}-10^{6}$ M$_{\odot}$; for a review see \citealt{Volonteri10}).
In general, the time in which a black hole of mass $M_{\mathrm{BH,f}}$ is grown from an initial seed $M_{\mathrm{BH,seed}}$, assuming it accretes with a constant Eddington ratio for all the time, can be written as (\citealt{Shapiro05}, \citealt{VolonteriRees05}):
\begin{equation}
\label{eqTBH}
\frac{t}{\mathrm{Gyr}} = t_{\mathrm{s}} \times \left[ \frac{\epsilon}{1-\epsilon} \right] \times \frac{L_{\mathrm{Edd}}}{L_{\mathrm{bol}}} \times \ln \left( \frac{M_{\mathrm{BH,f}}}{M_{\mathrm{BH,seed}}} \right) 
\end{equation}
where $t_{\mathrm{s}}=0.45$ Gyr is the Salpeter time and $\epsilon\sim 0.07$ is the radiative efficiency \citep{Pacucci15}. The average $M_{\mathrm{BH}}$ and $L_{\mathrm{bol}}/L_{\mathrm{Edd}}$ of all the $z\gtrsim 6.5$ quasars in the sample presented here (11 objects, not considering any luminosity cut, see Table \ref{tabMBHQSOs}) are 1.62$\times$10$^{9}$ M$_{\odot}$ and 0.39,
respectively. If we insert these values in Eq. \ref{eqTBH}, we can calculate the time needed by a black hole seed of $M_{\mathrm{BH,seed}}=[10^{2},10^{4},10^{5},10^{6}]$ M$_{\odot}$ to grow to the mean $M_{\mathrm{BH}}$ found here, assuming that it always accretes at an average Eddington rate of $\sim$0.39. We find that this time is
$t=$[1.44,1.04,0.84,0.64] Gyr. As the age of the universe at z$\sim$6.5 is only $\sim$0.83 Gyr, this implies that only very massive seeds ($\sim$10$^{6}$ M$_{\odot}$) would be able to form the observed supermassive black holes.

\begin{figure}[h]
\epsscale{0.8}
\centering
\includegraphics[width=\columnwidth]{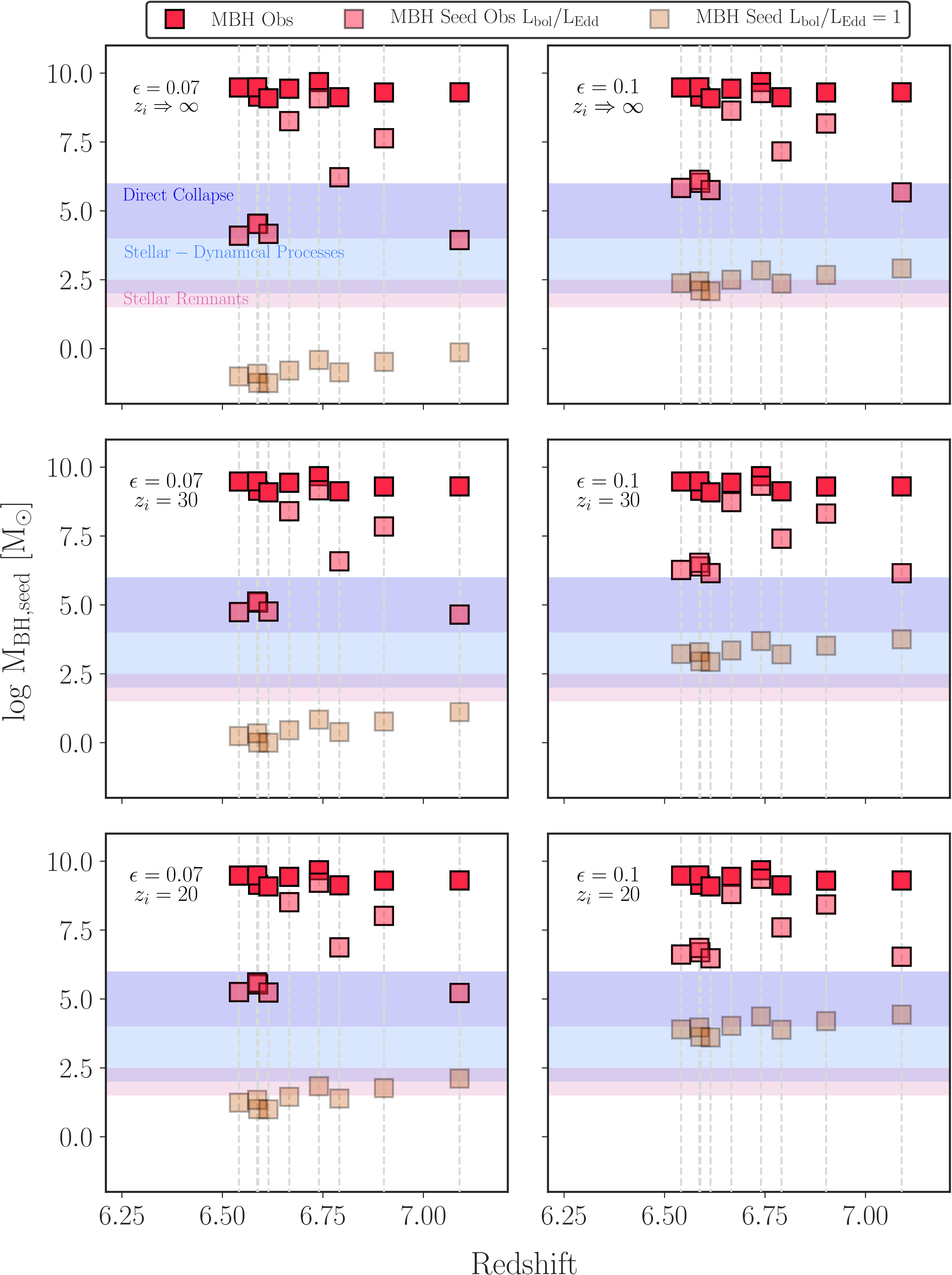}
\caption{Masses of the black hole seeds required to obtain the observed black hole masses in our quasar sample (dark red squares).
We here vary the efficiency ($\epsilon$=0.07/0.1, left and right columns) and the redshift of the seed formation ($z \Rightarrow \infty$/30/20, from top to bottom).
For each case, we assume that the sources accrete constantly with the observed Eddington ratio (light red squares; see also Table \ref{tabMBHQSOs}), and at Eddington rate (yellow squares).
The range of black hole seeds predicted by current theoretical models are shown in orange, light blue and deep blue shaded areas (see text for references). Black hole seeds with masses $\gtrsim$10$^{2}$ M$_{\odot}$ can produce the observed high$-$redshift quasars in all cases with $\epsilon$=0.07 and $L_{\mathrm{bol}}/L_{\mathrm{Edd}}$=1, and in case of [$\epsilon$=0.1, $L_{\mathrm{bol}}/L_{\mathrm{Edd}}$=1 and $z_{i} \Rightarrow \infty$]. If the efficiency is higher (10\%), and the seeds form at $z\sim$30$-$20, their predicted masses are correspondingly larger ($\sim$10$^{3-4}$ M$_{\odot}$, at Eddington accretion).}
\label{figBHSeedAll}
\end{figure}

Alternatively, we can invert Eq.$\,$\ref{eqTBH} and derive the initial masses of the black hole seeds required to obtain the observed black holes.
This result depends on the assumptions made, e.g., on the redshift of the seed formation ($z_{i}$), on the accretion rate ($L_{\mathrm{bol}}/L_{\mathrm{Edd}}$), and on the radiative efficiency ($\epsilon$; see Eq.$\,$\ref{eqTBH})\footnote{The efficiency depends in turn on the black hole spin and can be as high as $\sim$40\% in the case of maximally spinning black holes. The spin is still an elusive parameter; it has been observationally measured only in $\sim$20 sources in the local Universe (through the relativistic broadening of the Fe K$\alpha$ line; \citealt{Brenneman11} and \citealt{Reynolds14}). Thanks to stacked \textit{Chandra} deep observations of $\sim$30 lensed quasars \cite{Walton15} detected a broadened component of the K$\alpha$ line up to $z\sim$4.5; however the low S/N prevented a measurement of the single quasars' black hole spins. Current semi-analytical models place only weak constraints on the spin value at $z\gtrsim$5, which depends on the gas accretion mode, galactic morphology and black hole mass (e.g. \citealt{Sesana14}). However, since the spin decreases with black hole mass, we do not expect large values for our sample of quasars with $M_{\mathrm{BH}} \gtrsim $10$^{8}$ M$_{\odot}$.}.
We here consider different values for these parameters: We assume that the black holes accrete constantly with the observed Eddington ratios or with $L_{\mathrm{bol}}/L_{\mathrm{Edd}}$=1; also, we consider that they grow for a period of time equal to the age of the universe at their redshifts (i.e.$\,z_{i} \Rightarrow \infty$), and from $z_{i}$=30 or 20 (see different rows in Figure 13). Finally, we assume an efficiency of 7\% or 10\% (left and right columns in Figure 13). The derived values of black hole seeds for all the combinations of these parameters are shown in Figure \ref{figBHSeedAll}.
In all the cases considered here with $\epsilon$=0.07 and Eddington accretion (and in case of $\epsilon$=0.1, $z_{i} \Rightarrow \infty$ and $L_{\mathrm{bol}}/L_{\mathrm{Edd}}$=1), the calculated seed masses ($\gtrsim$10$^{2}$ M$_{\odot}$) are consistent with being formed by stellar remnants. Alternatively, a scenario of higher efficiency ($\epsilon$=0.1), later seeds birth (i.e.$\,z=$30 or 20), and accretion at $L_{\mathrm{bol}}/L_{\mathrm{Edd}}$=1, would require more massive seeds ($\sim$10$^{3-4}$ M$_{\odot}$) as progenitors of the observed $z\gtrsim$6.5 quasars.\\
\begin{deluxetable*}{lcccc}[h]
\tabletypesize{\small}
\tablecaption{Estimated quantities for the quasars in our sample: bolometric luminosities, black hole masses, Eddington ratios, Fe$\,${\scriptsize II}-to-Mg$\,${\scriptsize II} flux ratios. \label{tabMBHQSOs}}
\tablewidth{0pt}
\tablehead{ \colhead{name}& \colhead{$\rm L_{bol}$} & \colhead{$\rm M_{BH}$} & \colhead{$\rm L_{bol}/L_{edd}$}  
 & \colhead{Fe$\,${\scriptsize II}/Mg$\,${\scriptsize II}}  
 \\
  & [$10^{47} \mathrm{erg\,s^{-1}}$] & [$\times 10^{9} \mathrm{M_{\odot}}$ ] & & 
}
\startdata
VIK J0109--3047 & 0.51$^{+0.05}_{-0.06}$ & 1.33$^{+0.38}_{-0.62}$ & 0.29$^{+0.88}_{-2.59}$ & 2.02$^{+5.56}_{-0.65}$\\
PSO J036.5078+03.0498 & 2.0$^{+0.22}_{-0.64}$ & 3.00$^{+0.92}_{-0.77}$ & 0.51$^{+0.17}_{-0.21}$ &  2.47$^{+3.71}_{-1.36}$ \\
VIK J0305--3150 & 0.75$^{+0.10}_{-0.34}$ & 0.90$^{+0.29}_{-0.27}$ & 0.64$^{+2.20}_{-3.42}$ & 1.03$^{+3.04}_{-0.37}$ \\
PSO J167.6415--13.4960 &  0.47$^{+0.16}_{-0.22}$ & 0.30$^{+0.08}_{-0.12}$  & 1.22$^{+0.51}_{-0.75}$ & $<$3.1 \\
ULAS J1120+0641 & 1.83$^{+0.19}_{-0.072}$ & 2.47$^{+0.62}_{-0.67}$ & 0.57$^{0.16}_{0.27}$ & 1.04$^{+3.84}_{-0.14}$  \\
HSC J1205--0000  & 0.36$^{+0.18}_{-0.20}$ & 4.7$^{+1.2}_{-3.9}$ & 0.06$^{+0.32}_{-0.58}$ & $<$0.50 \\
PSO J231.6576--20.8335 & 1.89$^{+0.34}_{-0.45}$ & 3.05$^{+0.44}_{-2.24}$ & 0.48$^{+0.11}_{-0.39}$ & 2.64$\pm$1.7\\
PSO J247.2970+24.1277 & 1.77$^{+0.06}_{-0.76}$ & 0.52$^{+0.22}_{-0.25}$ & 2.60$^{+0.08}_{-0.15}$ & 1.33$^{+5.82}_{-0.01}$\\
PSO J323.1382+12.2986 & 0.81$^{+0.07}_{-0.50}$ & 1.39$^{+0.32}_{-0.51}$ & 0.44$^{+1.09}_{-3.19}$ & 1.85$^{+2.37}_{-0.97}$  \\
PSO J338.2298+29.5089 & 4.04$^{+2.14}_{-0.90}$ & 2.70$^{+0.85}_{-0.97}$ & 0.11$^{+0.71}_{-0.49}$ & 1.29$^{+2.1}_{-0.74}$ \\
VIK J2348--3054 & 0.43$^{+0.20}_{-0.13}$ & 1.98$^{+0.57}_{-0.84}$ & 0.17$^{+0.92}_{-0.88}$ & 2.13$^{+0.93}_{-1.54}$   
\enddata
\end{deluxetable*}
\subsection{Fe$\,${\scriptsize II}/ Mg$\,${\scriptsize II} } \label{subsecFeMgII}
The estimate of the relative abundances of metals in high redshift sources is a useful proxy in the investigation of the chemical composition and evolution of galaxies in the early universe. In this context, the Mg$\,${\scriptsize II}/Fe$\,${\scriptsize II} ratio is of particular interest: $\alpha$-elements, such as Mg, are mainly produced via type II supernovae (SNe) involving massive stars, while type Ia SNe from binary systems are primarily responsible for the provision of iron \citep{Nomoto97}. Given that SNe Ia are expected to be delayed by $\sim$1 Gyr (\citealt{Matteucci&Greggio86}) with respect to type II SNe, estimating the relative abundances of $\alpha$-elements to iron provides important insights on the stellar population in the galaxy, and on the duration and intensity of the star formation burst. Tracking the evolution of the Mg$\,${\scriptsize II}/Fe$\,${\scriptsize II} ratio as a function of redshift allows us to reconstruct the evolution of the galactic star formation history over cosmic time.

\begin{figure}[h]
\epsscale{0.8}
\centering
\includegraphics[width=\columnwidth]{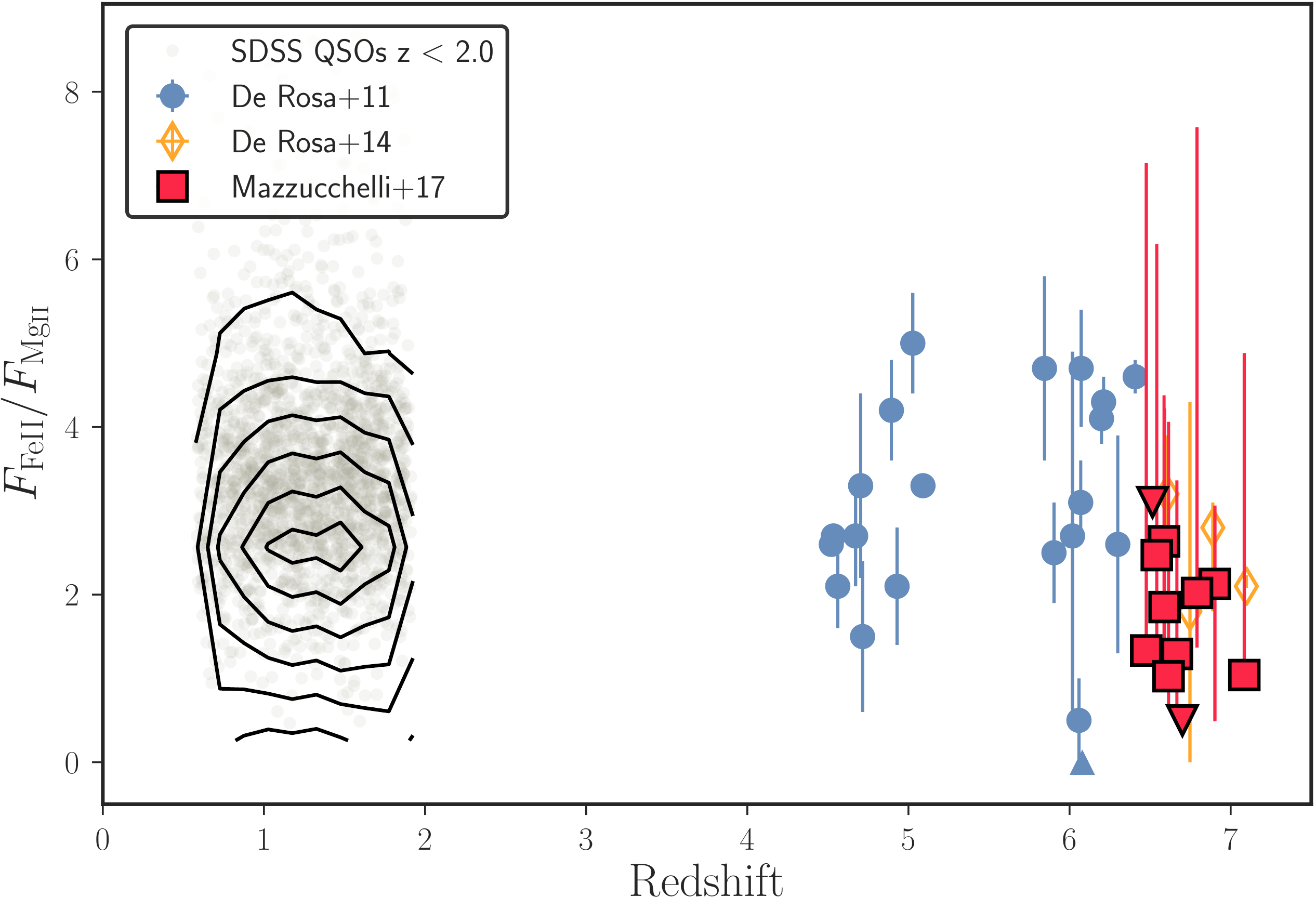}
\caption{Fe$\,${\scriptsize II}-to-Mg$\,${\scriptsize II} flux ratio, considered as a first-order proxy for the relative abundance ratio, versus redshift. We show the quasars in our sample (red squares and, in case of upper limits, down-pointing triangles) and taken from the literature: \citealt{DeRosa11} (blue points) and $z\lesssim 2$ SDSS quasars (\citealt{Calderone16}, grey points and black contours). We show with orange empty diamonds the measurements of \cite{DeRosa14} for four of our quasars (VIK0109, VIK0305, VIK2348 and ULAS1120); they have been derived with a slightly different fitting routine (see text for details) but are consistent, within the errors, with the estimates obtained here. Our measurements are systematically lower than that of samples at lower redshifts; however, taking into account the large uncertainties, we find no statistical evidence for an evolution of the flux ratio with redshift.}
\label{figFeMgred}
\end{figure}

Many studies in the literature investigate the Mg$\,${\scriptsize II}/Fe$\,${\scriptsize II} ratio in the BLR of quasars, by estimating the ratio of the Fe$\,${\scriptsize II} and Mg$\,${\scriptsize II} fluxes ($F_{\mathrm{Fe\, II}}/F_{\mathrm{Mg\, II}}$), considered a first-order proxy of the abundance ratio (e.g. \citealt{Barth03}, \citealt{Maiolino03}, \citealt{Iwamuro02},\citeyear{Iwamuro04}, \citealt{Jiang07}, \citealt{Kurk07}, \citealt{Sameshima09}, \citealt{DeRosa11},\citeyear{DeRosa14}). In particular, \cite{DeRosa11} and (\citeyear{DeRosa14}) present a consistent analysis of $\sim$30 quasar spectra in the redshift range $4 \lesssim z \lesssim 7.1$, and find no evolution of their $F_{\mathrm{Fe\, II}}/F_{\mathrm{Mg\, II}}$ with cosmic time.
We estimate the Fe$\,${\scriptsize II} and Mg$\,${\scriptsize II} fluxes for the quasars in our study following \cite{DeRosa14}: for the former we integrate the fitted iron template over the rest-frame wavelength range $2200 < \lambda\, [\mathrm{\AA}] < 3090$, and for the latter we compute the integral of the fitted Gaussian function (see Table \ref{tabFitMgIIQSOs} and \ref{tabMBHQSOs} for the estimated flux values). In Figure \ref{figFeMgred}, we plot $F_{\mathrm{Fe\, II}}/F_{\mathrm{Mg\, II}}$ as a function of redshift, for both the quasars in our sample and sources from the literature. We consider the sample by \cite{DeRosa11} and (\citeyear{DeRosa14}), and  a sample of low$-$redshift quasars ($z \lesssim 2.05$) from \cite{Calderone16}. They consistently re-analyzed a sub-sample of quasars ($\sim$70,000) from the SDSS-DR10 catalog, and provide measurements of the flux for Mg, Fe and the continuum emission at rest-frame $\lambda_{rf}=$3000 \AA\footnote{http://qsfit.inaf.it/}. Here, we take only the sources with no flag on the quantities above ($\sim$44,000 objects), and we correct the Fe$\,${\scriptsize II} flux to account for the different wavelength ranges where the iron emission was computed\footnote{\cite{Calderone16} integrates the iron template in the rest-frame wavelength range $2140 < \lambda [\mathrm{\AA}] < 3090$, while we use the range $2200 < \lambda [\mathrm{\AA}] < 3090$.}.
From Figure \ref{figFeMgred}, we see that the flux ratios of the quasars in our sample are systematically lower than the ones of the sources at lower redshift, both from \cite{DeRosa11} and from \cite{Calderone16}: this suggests a possible depletion of iron at $z\gtrsim6.5$, and therefore the presence of a younger stellar population in these quasar host galaxies. However, our estimates are also characterized by large uncertainties, mainly due to the large uncertainties on the iron flux estimates (see Table \ref{tabFitMgIIQSOs}). Within the errors, our measurements are consistent with a scenario of non$-$evolving $F_{\mathrm{Fe\, II}}/F_{\mathrm{Mg\, II}}$ over cosmic time, in agreement with \cite{DeRosa14}.
We test whether the systematic lower values of $F_{\mathrm{Fe\, II}}/F_{\mathrm{Mg\, II}}$ for the highest redshift quasar population is statistically significant. We associate to each of our measurements a probability distribution, built by connecting two half-Gaussian functions with mean and sigma equal to the calculated ratio and to the lower (or upper) uncertainty, respectively. We sum these functions to obtain the total probability distribution for the objects at high$-$redshift.
We compare this function with the distribution of the $F_{\mathrm{Fe\, II}}/F_{\mathrm{Mg\, II}}$ values for the quasars at $z\sim 1$. We randomly draw nine sources from the two distributions (the objects in our sample excluding the limits) and we apply a Kolmogorov-Smirnoff test to check if these two samples could have been taken from the same probability distribution; we repeat this draw 10000 times.
We obtain that the $p$-value is greater than 0.2 (0.5) in the 51\% (27\%) of the cases: this highlights that, considering the large uncertainties, we do not significantly measure a difference in the total probability distribution of $F_{\mathrm{Fe\, II}}/F_{\mathrm{Mg\, II}}$ at low and high$-$redshift.
Data with higher S/N in the Fe$\,${\scriptsize II} emission line region are needed to place more stringent constraints on the evolution of the abundance ratio. 
\subsection{Infrared and [CII] luminosities} \label{subsecCII}
We observed four quasars in our sample with NOEMA (see Section \ref{secFolUpObs}).
We extract their spectra, and fit the continuum+[CII] line emission with a flat+Gaussian function (see Figure \ref{figCIIline}). We estimate the line properties, e.g. the peak frequency, the width, amplitude and flux, and we calculate the continuum flux at rest frame wavelength 158 $\mu$m from the continuum map. We report these values in Table \ref{tabNOEMAdata}.

We can derive the far infrared properties of the observed quasars, following a number of assumptions commonly presented in the literature (e.g. \citealt{Venemans12}, \citeyear{Venemans16}). We approximate the shape of the quasar infrared emission with a modified black body: $f_{\nu} \propto B_{\nu}(T_{d})(1-e^{\tau_{d}})$, where $B_{\nu}(T_{d})$ is the Planck function and $T_{d}$ and $\tau_{d}$ are the dust temperature and optical depth, respectively \citep{Beelen06}. Under the assumption that the dust is optically thin at wavelength $\lambda_{rf}>40 \mu$m ($\tau_{d} << 1$), we can further simplify the function above as $f_{\nu} \propto B_{\nu}(T_{d}) \nu^{\beta}$, with $\beta$ the dust emissivity power law spectral index. We take $T_{d}=47$ K and $\beta=1.6$, which are typical values assumed in the literature \citep{Beelen06}. 
We scale the modified black body function to the observed continuum flux at the rest frame frequency $\nu_{rf}=1900$ GHz; we then calculate the FIR luminosity ($L_{\mathrm{FIR}}$) integrating the template in the rest frame wavelength range 42.5 $\mu$m$-$122.5$\mu$m \citep{Helou88}. The total infrared (TIR) luminosity ($L_{\mathrm{TIR}}$) is defined instead as the integral of the same function from 8 $\mu$m to 1000 $\mu$m.
We note that these luminosity values are crucially dependent on the assumed shape of the quasar infrared emission, which, given the poor photometric constraints available, is highly uncertain.
We can also calculate the luminosity of the [CII] emission line ($L_{\mathrm{[CII]}}$) from the observed line flux ($S_{\mathrm{[CII]}}\Delta v$; \citealt{Carilli13}):
\begin{equation}
\frac{L_{\mathrm{[CII]}}}{L_{\odot}}=1.04 \times 10^{-3} \frac{S_{\mathrm{[CII]}}\Delta v}{\mathrm{Jy\, km\, s^{-1}}}  \left( \frac{D_{\mathrm{L}}}{\mathrm{Mpc}} \right)^{2} \frac{\nu_{obs}}{\mathrm{GHz}}
\end{equation}
where $D_{\mathrm{L}}$ is the luminosity distance and $\nu_{obs}$ is the observed frequency.
In Table \ref{tabNOEMAdata}, we list our estimates for the [CII], FIR and TIR luminosities.

\begin{figure}[h]
\epsscale{0.8}
\centering
\includegraphics[width=\columnwidth]{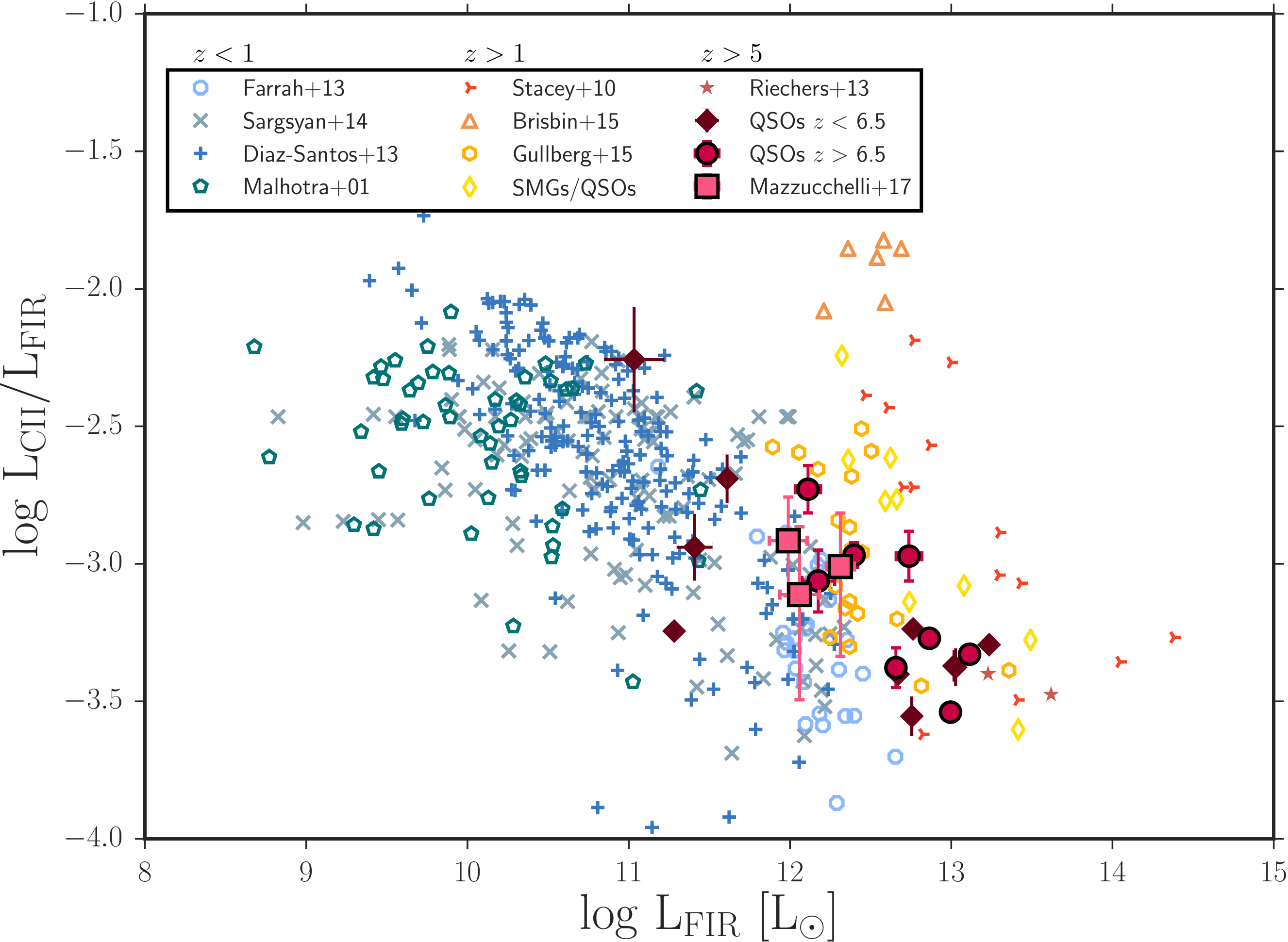}
\caption{[CII]-to-FIR luminosity ratio as function of FIR luminosity. With open blue/green symbols we report objects at $z<1$: star-forming galaxies (\citealt{Malhotra01}, \citealt{Sargsyan14}), LIRGs \citep{DiazSantos13} and ULIRGS \citep{Farrah13}. Values for sources at $1 < z < 5$ are shown with open yellow/orange symbols: star-forming galaxies (\citealt{Stacey10}, \citealt{Brisbin15}, \citealt{Gullberg15}) and a collection of $3\lesssim z\lesssim 5$ sub-millimeter galaxies and quasars (\citealt{Cox11},\citealt{Wagg10}, \citeyear{Wagg12}, \citealt{Ivison10}, \citealt{DeBreuck11}, \citealt{Valtchanov11}, \citealt{Walter12}, \citealt{Maiolino09}). The $z=6.3$ SMG presented in \cite{Riechers13} is shown as a light-red star. Quasars at $5\lesssim z \lesssim 6.5$ (\citealt{Maiolino05}, \citealt{Wang13}, \citealt{Willott15}) are shown with filled dark-pink diamonds. Quasars in the sample presented here are reported with filled light-pink squares (new observations for PSO338+29, PSO323+12, PSO006+39), and with filled pink circles (data taken from the literature; ULASJ1120, \citealt{Venemans12}; PSO036+03, \citealt{Banados15b}; VIKJ0109, VIKJ0305, VIKJ2348, \citealt{Venemans16}; PSO231-20, PSO183+05,PSO167-13, \citealt{DecarliSub}). Local sources show a decrease in the [CII]-to-FIR ratio at high FIR luminosities, whereas the values of this ratio for the high redshift sample have a large scatter. The $z>6$ quasars whose mm observations are presented in this work are characterized by values of $L_{\mathrm{[CII]}}/L_{\mathrm{FIR}}$ comparable to local ULIRGs. The range of [CII]-to-FIR luminosity ratio of the general population of $z>6$ quasars however hints to an intrinsic diversity among their host galaxies.}
\label{figLfirRatio}
\end{figure}

In Figure \ref{figLfirRatio}, we plot $L_{\mathrm{[CII]}}/L_{\mathrm{FIR}}$ vs $L_{\mathrm{FIR}}$ for the quasars studied here and for a variety of sources from the literature.
At low redshift ($z<1$) both star-forming galaxies (\citealt{Malhotra01}, \citealt{Sargsyan14}) and more extreme objects, e.g. LIRGS and ULIRGS (\citealt{DiazSantos13}, \citealt{Farrah13}), show lower luminosity ratios at higher FIR luminosities: this phenomenon is known as the ``CII-deficit''.
At $z>1$, the scenario is less clear, where the scatter in the measurements of $L_{\mathrm{[CII]}}/L_{\mathrm{FIR}}$ for star-forming galaxies (\citealt{Stacey10}, \citealt{Brisbin15}, \citealt{Gullberg15}), SMGs and quasars increases. Quasars at $z>5$ present a variety of $L_{\mathrm{[CII]}}/L_{\mathrm{FIR}}$ values, mostly depending on their far-infrared brightness. \cite{Walter09} and \cite{Wang13} observe quasars with high $L_{\mathrm{FIR}}$, and show that they are characterized by low luminosity ratios, comparable to local ULIRGS ($\langle \mathrm{log} (L_{\mathrm{[CII]}}/L_{\mathrm{FIR}}) \rangle \sim -3.5$). On the other hand, quasars with lower far infrared luminosities and black hole masses ($M_{\mathrm{BH}}<10^{9}$ M$_{\odot}$ ; \citealt{Willott15}) are located in a region of the parameter space similar to the one of regular star forming galaxies ($\langle \mathrm{log} (L_{\mathrm{[CII]}}/L_{\mathrm{FIR}}) \rangle \sim -2.5$). In the literature, the decrease of $L_{\mathrm{[CII]}}$ in high-redshift quasars has been tentatively explained invoking a role of the central AGN emission, which is heating the dust. The problem is however still under debate, and several other alternative scenarios have been advocated, e.g.$\,$C$^{+}$ suppression due to X-ray radiation from the AGN \citep{Langer15}, or the relative importance of different modes of star formation on-going in the galaxies \citep{GraciaCarpio11}.
The quasars whose new infrared observations are presented here, with $L_{\mathrm{FIR}} \sim 10^{12}$ L$_{\sun}$, are characterized by values of the luminosity ratio in between the ones of FIR bright quasars and of the sample by \citeauthor{Willott15} (\citeyear{Willott15}; $\langle \mathrm{log} (L_{\mathrm{[CII]}}/L_{\mathrm{FIR}}) \rangle \sim -3.0$). This is similar to what was found by \citeauthor{Venemans12} (\citeyear{Venemans12}, \citeyear{Venemans17}) for ULAS1120, and suggests that the host galaxies of these quasars are more similar to ULIRGS.
\begin{deluxetable*}{lcccc}[h]
\tabletypesize{\small}
\tablecaption{Results from our NOEMA observations: we report the [CII] line and continuum emission quantities obtained from our fit (i.e. flux and line width), and the [CII] line, FIR and TIR luminosities. \label{tabNOEMAdata}}
\tablewidth{0pt}
\tablehead{ & \colhead{HSC J1205-0000} & \colhead{PSO J338.2298+29.5089} &  \colhead{PSO J006.1240+39.2219}  & \colhead{PSO J323.1283+12.2986}}
\\
\startdata
$z_{\mathrm{[CII]}}-z_{\mathrm{MgII}}$ [km s$^{-1}$] & $-$ & 818$^{+168}_{-138}$ & $-$ & 230 $\pm$ 13\\
$\mathrm{\rm [CII]}$ line width [km s$\mathrm{^{-1}}$] & $-$ & 740$\mathrm{^{+541}_{-313}}$ & 277$\mathrm{^{+161}_{-141}}$ & 254$\mathrm{^{+48}_{-28}}$ \\
$\rm [CII]$ flux [Jy km s$^{-1}$] & $-$ & 1.72$^{+0.91}_{-0.84}$ & 0.78$^{+0.54}_{-0.38}$ & 1.05$^{+0.33}_{-0.21}$\\
Continuum flux density [mJy] & 0.833 $\pm$ 0.176 & 0.972 $\pm$ 0.215 & 0.548 $\pm$ 0.178 & 0.470 $\pm$ 0.146\\ 
$L_{\mathrm{[CII]}}$ [10$^{9}\, L_{\odot}$] & $-$ & 2.0 $\pm$ 0.1 & 0.9$^{+0.6}_{-0.4}$ & 1.2$^{+0.4}_{-0.2}$ \\
$L_{\mathrm{FIR}}$ [10$^{12}\, L_{\odot}$] & 1.9 $\pm$ 0.4 & 2.1 $\pm$ 0.5 & 1.1 $\pm$ 0.4 & 1.0 $\pm$ 0.3 \\
$L_{\mathrm{TIR}}$ [10$^{12}\, L_{\odot}$] & 2.6 $\pm$ 0.5 & 2.8 $\pm$ 0.6 & 1.5 $\pm$ 0.5 & 1.3 $\pm$ 0.4 \\
$L_{\mathrm{[CII]}}/L_{\mathrm{FIR}}$ [10$^{-3}$] & $-$ & 0.98$^{+0.55}_{-0.52}$ & 0.77$^{+0.59}_{-0.45}$ & 1.2$^{+0.54}_{-0.45}$ 
\enddata
\end{deluxetable*}
\subsection{Near Zones} \label{subsecNZ}
Near zones are regions surrounding quasars where the IGM is ionized by the UV radiation emitted from the central source. 
Taking into account several approximations, e.g. that the IGM is partially ionized and solely composed of hydrogen, and that photoionization recombination equilibrium is found outside the ionized region \citep{Fan06}, the radius  of the ionized bubble can be expressed as:
\begin{equation} \label{eqNZRs}
R_{s} \propto \left( \frac{\dot N_{Q} t_{Q}}{f_{\mathrm{HI}}} \right) ^{1/3}  
\end{equation}
where $\dot N_{Q}$ is the rate of ionizing photons produced by the quasar, $t_{Q}$ is the quasar lifetime, and $f_{\mathrm{HI}}$ is the IGM neutral fraction.
Several studies provide estimates of near zone radii for samples of $z>5$ quasars, and investigate its evolution as a function of redshift, in order to investigate the IGM evolution (\citealt{Fan06}, \citealt{Carilli10}, \citealt{Venemans15}, \citealt{Eilers17}).

\begin{figure*}
\epsscale{0.8}
\centering
\includegraphics[width=\textwidth]{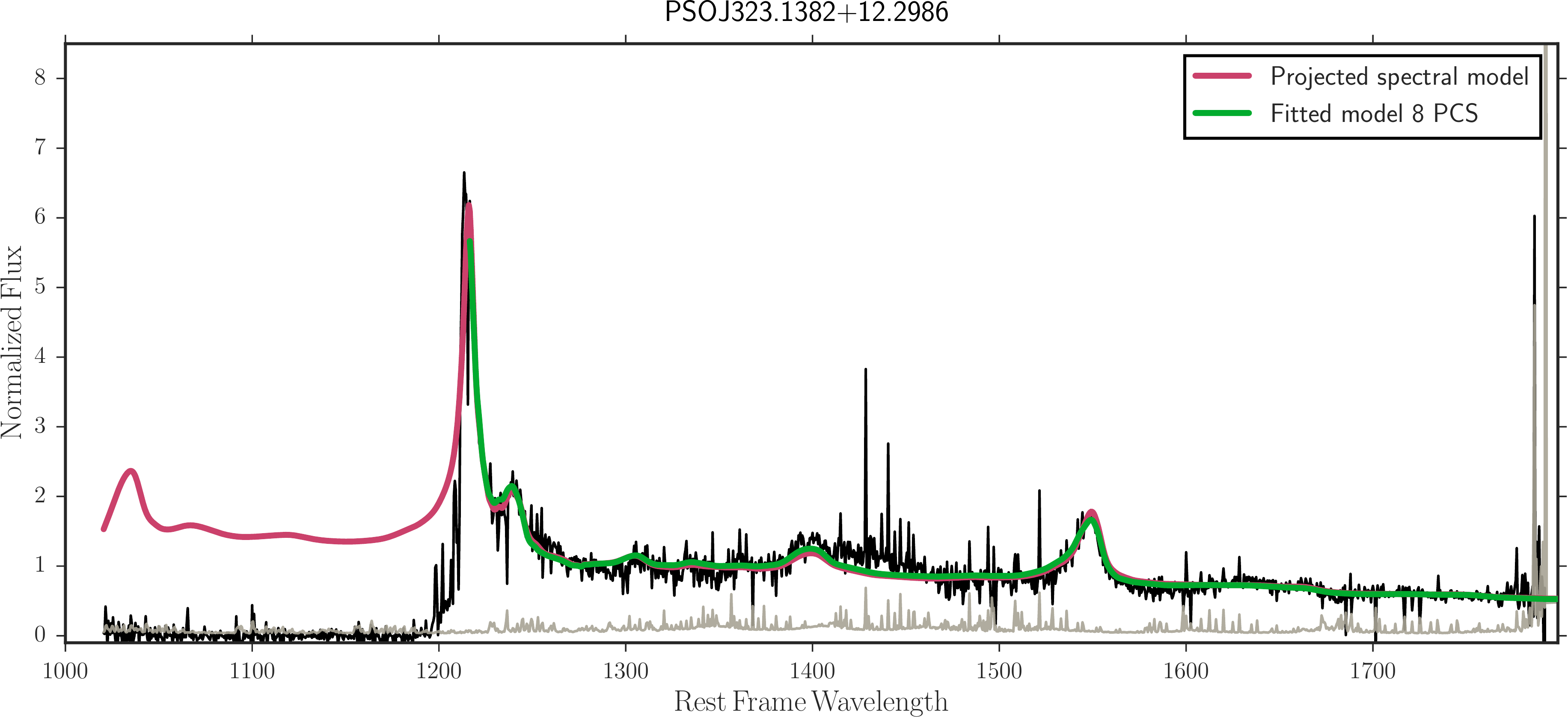}
\caption{Example of quasar continuum emission fit with the PCA method for one of the sources in our sample (PSO323+12). We show the fitted model at wavelength greater than the Ly$\alpha$ emission line (with 8 PCAs; green line), and the projected model on the entire spectrum (purple line).}
\label{figPCA}
\end{figure*}

However, it is not straightforward to derive the exact values of $R_{s}$ from the observed spectra; instead, we calculate here the near zone radii ($R_{\mathrm{NZ}}$) for the sources in our sample. We follow the definition of \cite{Fan06}, i.e. $R_{\mathrm{NZ}}$ is the distance from the central source where the transmitted flux drops below 0.1, once the spectrum has been smoothed to a resolution of 20\, \AA. The transmitted flux is obtained by dividing the observed spectrum by a model of the intrinsic emission.
We here model the quasar emission at $\lambda_{rf}<1215.16$ \AA$\,$ using a principal component analysis (PCA) approach:
the total spectrum, $q(\lambda)$, is represented as the sum of a mean spectrum, $\mu(\lambda)$, and $n=1,..,N$ principal component spectra (PCS), $\xi_{n}(\lambda)$, each weighted by a coefficient $w_{n}$:
\begin{equation} 
q(\lambda) = \mu(\lambda) + \sum _{n=1} ^{N} w_{n} \xi_{n}(\lambda)
\end{equation}
\cite{Paris11} and \cite{Suzuki06} apply the PCA to a collection of 78 $z\sim3$ and 50 $z\lesssim1$ quasars from SDSS, respectively. In our study, we follow the approach by \cite{Eilers17}, mainly referring to \cite{Paris11} who provide PCS functions within the rest frame wavelength window $1020 < \lambda /[\mathrm{\AA}] < 2000$. After normalizing our spectra to the flux at $\lambda_{rf}=1280$ \AA, we fit the region redwards than the Ly$\alpha$ emission line ($\lambda_{rf}>1215.16$ \AA) to the PCS by \cite{Paris11}, and we derive the best coefficients by finding the maximum likelihood. We then obtain the best coefficients which reproduce the entire spectrum by using the projection matrix presented by \cite{Paris11}. For further details on this modeling procedure, see \cite{Eilers17}. We show in Figure \ref{figPCA} an example of PCA for one of the quasars in our sample.
Also, in this way we provide an analysis of the near zone sizes consistent with \cite{Eilers17}, making it possible to coherently compare the results obtained from the two data sets.

\begin{figure*}
\epsscale{0.8}
\centering
\includegraphics[width=\textwidth]{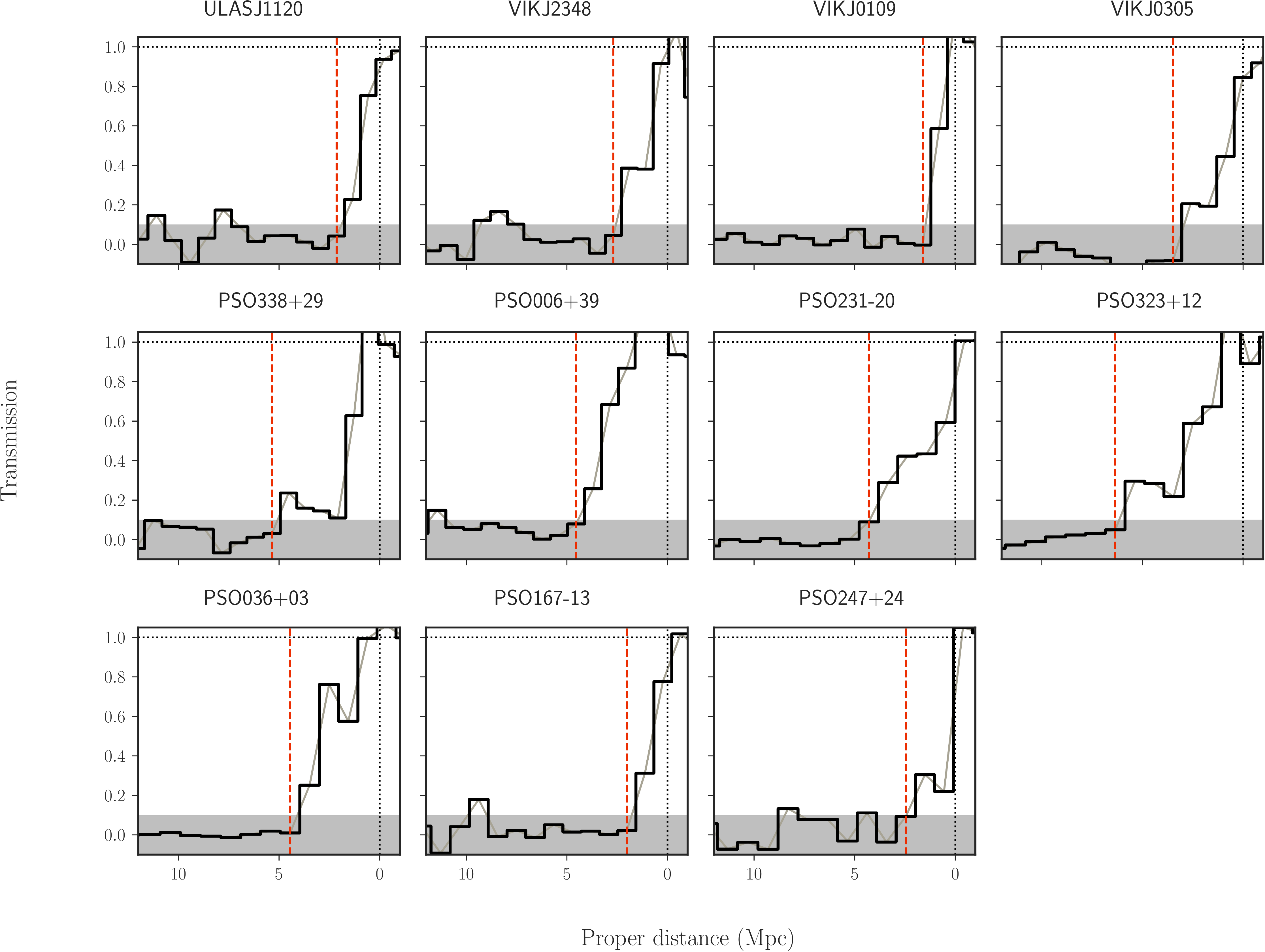}
\caption{Transmission fluxes of the quasars in our sample, obtained normalizing the observed spectra by the emission model from the PCA method (see Section 4), as a function of proper distance from the source. We identify the near zone radius (dashed red line) as the distance at which the flux drops below 10\%, after smoothing each spectrum to a common resolution of 20 \AA. We do not consider in our analysis HSC1205, PSO183+05, PSO011+09 or PSO261+19 (see text).}
\label{figAllNZ}
\end{figure*}

The near zone sizes depend also on quasar luminosity (through the $\dot N_{Q}$ term in equation \ref{eqNZRs}): if we want to study their evolution with redshift, we need to break this degeneracy. We re-scale the quasar luminosities to the common value of M$_{1450}$=-27 (following previous studies, e.g \citealt{Carilli10}, \citealt{Venemans15}), and we use the scaling relation obtained from the most recent numerical simulations presented in \cite{Eilers17} and Davies et al.$\,$(in prep). They simulate radiative transfer outputs for a suite of $z=$6 quasars within the luminosity range -24.78$<$M$_{1450}<$-29.14, and constantly shining over 10$^{7.5}$ yr, considering two scenarios in which the surrounding IGM is mostly ionized (as supported at $z\sim$6 by recent studies, e.g. \citealt{McGreer15}) or mostly neutral. They obtain comparable results for the two cases, which are both in agreement with the outcome obtained by fitting the observational data (see \citealt{Eilers17}, Figure 5). Following the approach of \cite{Eilers17}, we consider the case of a mostly ionized IGM: they fit the simulated quasar near zone sizes against luminosity with the power law:
\begin{equation} \label{eqNZM}
R_{\mathrm{NZ}}=5.57\,\mathrm{pMpc} \times 10^{0.4(\mathrm{M_{1450}})/2.35}
\end{equation}
with pMpc being proper Mpc, from which they derive the following scaling relation, that we also use here:
\begin{equation} \label{eqNZM}
R_{\mathrm{NZ,corr}}=R_{\mathrm{NZ}}\, 10^{0.4(27+\mathrm{M_{1450}})/2.35}
\end{equation}
We report in Table \ref{tabNZQSOs} the derived quantities, and the transmission fluxes are shown in Figure \ref{figAllNZ}. We do not consider in our analysis the following quasars: HSC1205, due to the poor quality of the spectrum in the Ly$\alpha$ emission region (see Figure \ref{figSpecAll}); PSO183+05, since this quasar is believed to present a proximate ($z\approx6.404$) DLA (see \citealt{Chen16}, \citealt{BanadosInPrep}); PSO011+09 and PSO261+19. These last two objects were discovered very recently and the only redshift measurements are provided by the Ly$\alpha$ emission line: the lack of any other strong emission line, and the broad shape of the Ly$\alpha$ line, do not permit us to rule out that these quasars are BAL objects. The redshifts of the remaining objects are mainly derived from [CII] observations
(see Table \ref{tabNZQSOs}).

\begin{figure}
\epsscale{0.8}
\centering
\includegraphics[width=\columnwidth]{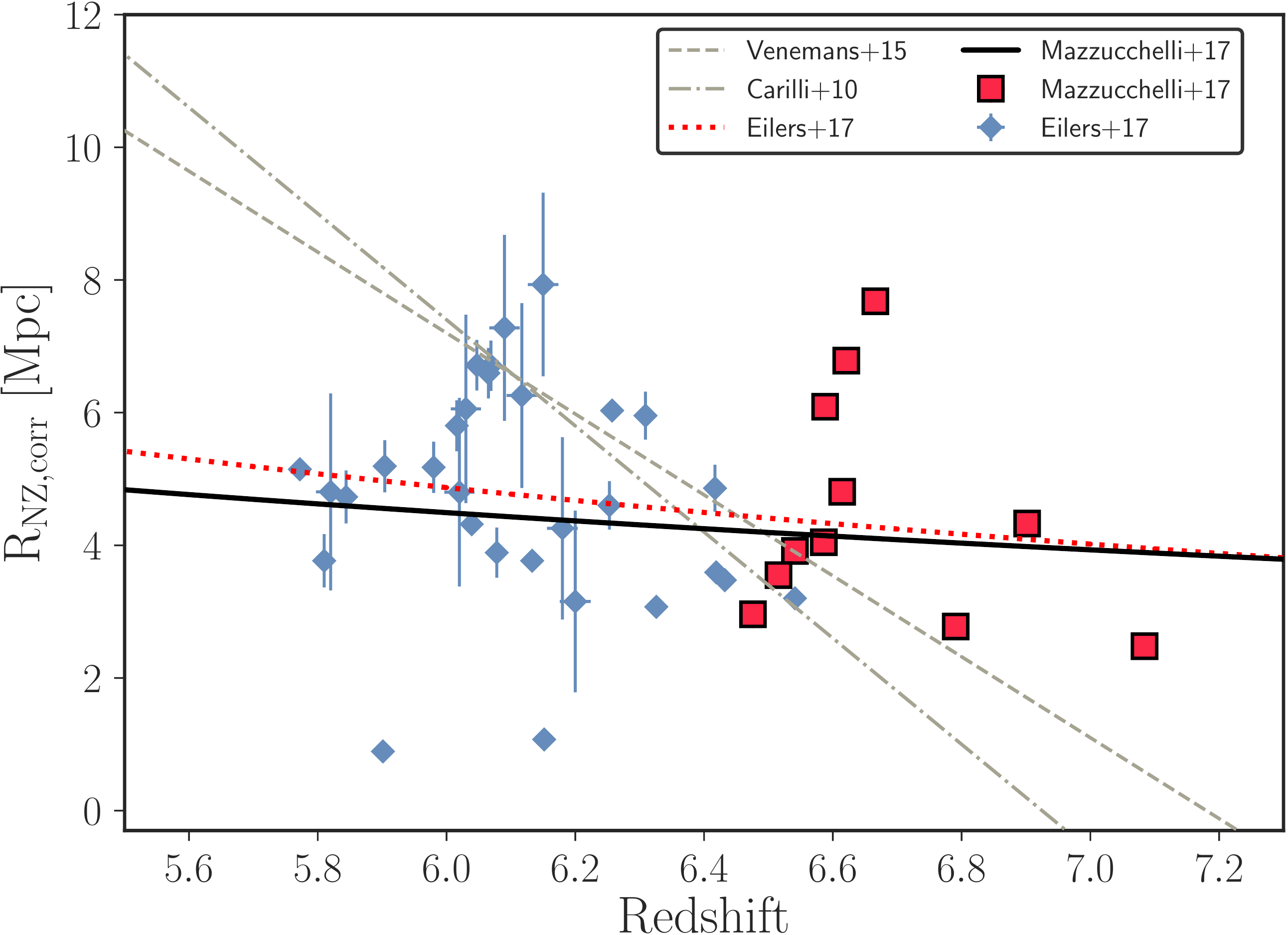}
\caption{Near zone size as a function of redshift for the objects in our sample (red squares), and the ones taken from \citeauthor{Eilers17} (\citeyear{Eilers17}; blue diamonds). The errors plotted are only due to the uncertainties on the redshifts, and for the quasars in this work are particularly small due to our precise $z_{\mathrm{{[CII]}}}$ measurements (see Table \ref{tabNZQSOs}). We fit the data with a power law function (solid black line): the amplitude and slope values obtained are in line with the results presented by \citeauthor{Eilers17} (\citeyear{Eilers17}; red dotted line; see text and Eq. \ref{eqNZred}). We find a redshift evolution of the near zone radii much shallower than that obtained in previous literature, e.g.$\,$by \citeauthor{Carilli10} (\citeyear{Carilli10}; dot-dashed line) and by \citeauthor{Venemans15} (\citeyear{Venemans15}; grey dashed line). This could be explained by the fact that $R_{\mathrm{NZ,corr}}$ depends more strongly on the individual quasar properties which vary from object-to-object, rather than on the overall characteristics of the IGM.}
\label{figNZredshift}
\end{figure}

We show the evolution of $R_{\mathrm{NZ,corr}}$ as a function of redshift in Figure \ref{figNZredshift}. We compare our data with estimates at lower redshift ($5.6 \lesssim z \lesssim 6.6 $) presented by \cite{Eilers17}. The best fit of the evolution of $R_{\mathrm{NZ,corr}}$ with $z$, modeled as a power law function, gives the following:
\begin{equation} \label{eqNZred}
R_{\mathrm{NZ,corr}} = (4.49 \pm 0.92) \times \left( \frac{1+z}{7} \right) ^{-1.00 \pm 0.20}
\end{equation}
The values obtained are consistent, within the errors, with the results of \cite{Eilers17}\footnote{$R_{\mathrm{NZ,corr}} \approx 4.87 \times [(1+z)/7] ^{-1.44}$; see also their Figure 6.}. In agreement with both measurements from observations and radiative transfer simulations presented by \cite{Eilers17}, we find a weak evolution of the quasar near zone sizes with cosmic time: this evolution is indeed much shallower than what was obtained by previous works (\citealt{Fan06}, \citealt{Carilli10}, \citealt{Venemans15}), which argued that the significant decrease of $R_{\mathrm{NZ}}$ with redshift could be explained by a steeply increasing IGM neutral fraction between $z\sim$5.7 and 6.4\footnote{We note that these studies considered a smaller and lower$-z$ quasar sample, whose redshift measurements (mainly from the Mg$\,${\smallskip II} or Ly$\alpha$ emission lines, with only a minority of objects observed in CO or [CII]) have larger errors, and that they fit the redshift evolution of the near zone sizes with a linear relation.}.
The different trend of near zone sizes with redshift with respect to what was found in the literature may be due to several reasons, i.e. we consider higher quality spectra and a larger sample of quasars, we take into consideration a consistent definition of $R_{\mathrm{NZ}}$ and we do not exclude the WEL quasars at $z\sim$6 (see \citealt{Eilers17} for an in depth discussion of the discrepancies with previous works).
We argue that the shallow evolution is due to the fact that $R_{\mathrm{NZ,corr}}$ does not depend entirely or only on the external IGM properties, but it correlates more strongly with the quasar characteristics (e.g. lifetime, regions of neutral hydrogen within the ionized zone), which are highly variable from object to object. 
\begin{deluxetable*}{lcccc}
\centering
\tabletypesize{\small}
\tablecaption{Near zone sizes of 11 quasars in the sample presented here. The corrected vaues have been calculated with Eq. \ref{eqNZM}, and take into account the dependency on their luminosity. We also report the number of PCS adopted in the continuum fit. \label{tabNZQSOs}}
\tablewidth{0pt}
\tablehead{ \colhead{name} & \colhead{$R_{\mathrm{NZ}}$} & \colhead{$R_{\mathrm{NZ,corr}}$} & \colhead{$R_{\mathrm{NZ,corr,err}}$} & \colhead{PCS}
  \\
 & \colhead{[Mpc]} & \colhead{[Mpc]} & \colhead{[Mpc]} } 
\startdata
PSO J006.1240+39.2219  & 4.47 & 6.78 & 0.09 & 5\\
VIK J0109--3047 & 1.59 & 2.78 &  0.03 & 8\\
PSO J036.5078+03.0498  & 4.37 & 3.91 & 0.08 & 8\\
VIK J0305--3150 & 3.417 & 4.81 & 0.006 & 10 \\
PSO J167.6415--13.4960  & 2.02 & 3.55 & 0.03 & 8\\
ULAS J1120+0641 & 2.10 & 2.48 & 0.02 & 9\\
PSO J231.6576--20.8335 & 4.28 & 4.05 & 0.03 & 8 \\
PSO J247.2970+24.1277  & 2.46 & 2.96  & 0.24 &  5\\
PSO J323.1382+12.2986 & 6.23  & 6.09 & 0.01 & 8\\
PSO J338.2298+29.5089  & 5.35 & 7.68 & 0.25 & 5\\
VIK J2348--3054 & 2.64 & 4.33 & 0.05 & 8
\enddata
\end{deluxetable*}
  \\
\section{Discussion and Conclusions} \label{secDiscConc}
In this work we present our search for $z-$dropouts in the third internal release of the Pan-STARRS1 stacked catalog (PS1 PV3), which led to the discovery of six new $z\sim$6.5 quasars.

We complement these newly found quasars with all the other $z\gtrsim$ 6.5 quasars known to date, and perform a comprehensive analysis of the highest redshift quasar population.
In particular, we provide new optical/NIR spectroscopic observations for the six newly discovered quasars and for three sources taken from the literature (PSO006+39, PSO338+29 and HSC1205); we also present new millimeter observations of the [CII] 158 $\mu$m emission line and the underlying continuum emission from NOEMA, for four quasars (PSO006+39, PSO323+12,PSO338+12 and HSC1205).

Our main results are:
\begin{itemize}
\item We calculate  C$\,${\scriptsize IV} rest$-$frame EWs, and blueshifts with respect to the Mg$\,${\scriptsize II} emission line, for 9 sources in our sample. We derive that all the $z\gtrsim$6.5 quasars considered here show large blueshifts (740$-$5900 km s$^{-1}$), and that they are outliers with respect to a SDSS quasar sample at $z\sim1$; they also have EW values equal or lower to the ones of the low$-$redshift quasars. This evidence hints to a strong wind/outflow component in the BLRs of the highest redshift quasars known.  

\item We derive bolometric luminosities, black hole masses and accretion rates estimates by modeling the Mg$\,${\scriptsize II} emission line region (2100$< \lambda\,[\mathrm{\AA}] <$3200) for 11 objects with available NIR spectroscopic observations. Comparing those measurements with the ones of a bolometric luminosity matched quasar sample at lower redshift (0.35$< z <$2.35), we find that high$-$redshift quasars accrete their material at a similar rate, with a mean of $\langle \log (L_{\mathrm{bol}}/L_{\mathrm{Edd}}) \rangle \sim-$0.41 and a 1$\sigma$ scatter of $\sim$0.4 dex, than their low$-$redshift counterparts, which present a mean of $\langle \log (L_{\mathrm{bol}}/L_{\mathrm{Edd}})\rangle \sim -0.47$ and a scatter of $\sim$0.3 dex.
We also note that the high$-$redshift sample is biased towards higher luminosities:
a more homogeneous coverage of the quasar parameter space at high redshift will help us investigating this evolution in the future.
\item We estimate the black hole seed masses ($M_{\mathrm{BH,seed}}$) required to grow the observed $z\gtrsim$6.5 quasars studied here, assuming that they accrete at the constant observed Eddington ratio or with an Eddington ratio of unity, for a time equal to the age of the universe at the observed redshift, and with a constant radiative efficiency of 7\%. In the first case, we derive $M_{\mathrm{BH,seed}} \gtrsim 10^{4}$ M$_{\odot}$ (higher than what is expected in the collapse of Pop III stars), while in the second case we obtain a lower value,
consistent with all current theoretical models; this is valid even in the scenario where the seeds are formed at $z$=20.
Also, in the case the black hole seeds accrete at the Eddington rate with an efficiency of 10\% from the beginning of the universe, their predicted masses are consistent with being formed by Pop III stars. Alternatively, we calculate that, if they seeds are created at $z\sim20-30$ and accrete with $\epsilon$=0.1, they would need to be as massive as $\gtrsim$10$^{3-4}$ M$_{\odot}$ (see Figure \ref{figBHSeedAll}).


\item We calculate the Fe$\,${\scriptsize II}/Mg$\,${\scriptsize II} flux ratio, as a first-order proxy of the abundance ratio.
We derive values systematically lower than the ones obtained for lower redshift quasars, implying a decrease of the iron abundance at $z\gtrsim 6.4$. However our measurements are hampered by large uncertainties, and, within these errors, we are consistent with a scenario of no evolution of the abundance ratio with redshift, as previously found by \citeauthor{DeRosa11} (\citeyear{DeRosa11}, \citeyear{DeRosa14}) from a smaller sample of high$-$redshift quasars.

\item From new millimeter observations reported here for four objects, we derive precise redshift estimates ($\Delta z\lesssim$0.004), and [CII] emission line and continuum luminosities, from which we obtain near infrared and total infrared luminosities. We study the $L_{\mathrm{[CII]}}/L_{\mathrm{FIR}}$ ratio as a function of $L_{\mathrm{FIR}}$ for these sources, and we place them in the context of present measurements from the literature, for both high and low redshift objects, normal star forming galaxies, LIRGS, ULIRGS and quasars. We find that the values obtained cover a parameter space similar to the one of ULIRGS.
\item
We calculate the near zone sizes of 11 objects. We study these measurements, together with the ones for a 5.6$\lesssim z \lesssim$6.5 quasar sample from \cite{Eilers17}, as a function of redshift. The two data sets are analyzed with a consistent methodology; in agreement with \cite{Eilers17}, we find a much shallower evolution of the near zone sizes with cosmic times than what was found by previous work (e.g. \citealt{Carilli10}, \citealt{Venemans15}). This result is also in line with recent radiative transfer simulations (Davies et al. in prep), and, as argued by \cite{Eilers17}, may be due to the much stronger dependency of the near zone sizes on the particular quasar characteristics (e.g. age and/or islands of neutral gas located inside the ionized spheres) than on the general IGM properties.
\end{itemize}

The analysis presented here highlights the large variety of physical properties of the quasars at the highest redshifts accessible today, and shows how these observations can  address a number of crucial open issues.
In the future, further statistical studies, supported by a larger sample of quasars (e.g.$\,$fainter sources and objects at higher$-$redshift) and by observations with new transformational facilities (e.g.$\,$ALMA and \textit{JWST}), will play a key role in our understanding of the universe at the very dawn of cosmic time.

\acknowledgments
We thank the anonymous referee for providing positive feedback and useful and constructive comments.

We acknowledge the assistance of Mayte Alfaro and Nicolas Goody in some of the observations presented here. 

EPF, BPV and FW acknowledge funding through the ERC grant ``Cosmic Dawn''. Support for RD was provided by the DFG priority program 1573 ``The physics of the interstellar medium''. CM thanks the IMPRS for Astronomy and Cosmic Physics at the University of Heidelberg. 

The Pan-STARRS1 Surveys (PS1) have been made possible through contributions of the Institute for Astronomy, the University of Hawaii, the Pan-STARRS Project Office, the Max-Planck Society and its participating institutes, the Max Planck Institute for Astronomy, Heidelberg and the Max Planck Institute for Extraterrestrial Physics, Garching, The Johns Hopkins University, Durham University, the University of Edinburgh, Queen's University Belfast, the Harvard-Smithsonian Center for Astrophysics, the Las Cumbres Observatory Global Telescope Network Incorporated, the National Central University of Taiwan, the Space Telescope Science Institute, the National Aeronautics and Space Administration under Grant No. NNX08AR22G issued through the Planetary Science Division of the NASA Science Mission Directorate, the National Science Foundation under Grant No. AST-1238877, the University of Maryland, and Eotvos Lorand University (ELTE).

The present work is based on observations taken with ESO Telescopes at the La Silla Paranal Observatory, under the programs: 092.A-­‐0339(A), 092.A-­‐0150(A), 092.A-­‐0150(B), 093.A-0863(A), 095.A-9001(A), 095.A-0375(A), 095.A-0535(A), 095.A-0535(B), 096.A-0420(A), 096.A-9001(A), 097.A-9001(A), 097.A-0094(A), 097.A-0094(B),

Part of the funding for GROND (both hardware and personnel) was generously granted from the Leibniz Prize to Prof G. Hasinger (DFG grant HA 1850/28-1).

Part of the data presented herein were obtained at the W.M. Keck Observatory, which is operated as a scientific partnership among the California Institute of Technology, the University of California and the National Aeronautics and Space Administration. The Observatory was made possible by the generous financial support of the W.M. Keck Foundation. The authors wish to recognize and acknowledge the very significant cultural role and reverence that the summit of Mauna Kea has always had within the indigenous Hawaiian community.  We are most fortunate to have the opportunity to conduct observations from this mountain. 

Some of the data here reported is based on observations collected at the Centro Astron{\'o}mico Hispano Alem{\'a}n at Calar Alto, jointly operated by the Max Planck Institute for Astronomy and the Instituto de Astrofis{\'i}ca de Andaluc{\'i}a. 

This paper includes data gathered with the 6.5 meter Magellan Telescope located at Las Campanas Observatory, Chile. The FIRE observations were supported by the NFS under grant AST$-$1109915

Part of the observations reported here were obtained at the MMT Observatory, a joint facility of the University of Arizona and the Smithsonian Institution.  

The LBT is an international collaboration among institutions in the United States, Italy and Germany. LBT Corporation partners are: The University of Arizona on behalf of the Arizona university system; Istituto Nazionale di Astrofisica, Italy; LBT Beteiligungsgesellschaft, Germany, representing the Max-Planck Society, the Astrophysical Institute Potsdam, and Heidelberg University; The Ohio State University, and The Research Corporation, on behalf of The University of Notre Dame, University of Minnesota and University of Virginia.
This paper used data obtained with the MODS spectrographs built with
funding from NSF grant AST-9987045 and the NSF Telescope System
Instrumentation Program (TSIP), with additional funds from the Ohio
Board of Regents and the Ohio State University Office of Research.

This paper makes use of the following ALMA data: ADS/JAO.ALMA\#2012.1.00882.S; ADS/JAO.ALMA\#2015.1.01115.S. ALMA is a partnership of ESO (representing its member states), NSF (USA) and NINS (Japan), together with NRC (Canada), NSC and ASIAA (Taiwan), and KASI (Republic of Korea), in cooperation with the Republic of Chile. The Joint ALMA Observatory is operated by ESO, AUI/NRAO and NAOJ.

This work includes observations carried out with the IRAM NOEMA Interferometer. IRAM is supported by INSU/CNRS (France), MPG (Germany) and IGN (Spain).

This publication makes use of data products from the Wide-field Infrared Survey Explorer, which is a joint project of the University of California, Los Angeles, and the Jet Propulsion Laboratory/California Institute of Technology, funded by the National Aeronautics and Space Administration.

This work is based in part on data obtained as part of the UKIRT Infrared Deep Sky Survey.
  
This research has benefited from the SpeX Prism Library, maintained by Adam Burgasser at http://www.browndwarfs.org/spexprism.

This research made use of Astropy, a community-developed core Python package for Astronomy (Astropy Collaboration, 2013;  http://www.astropy.org).

\textit{Facilities:} PS1 (GPC1), VLT:Antu (FORS2), NTT (EFOSC2, SofI), Max Planck.2.2m (GROND), Magellan:Baade (FIRE), Keck:I (LRIS), Hale (DBSP), CAO:3.5m (Omega2000), CAO:2.2m (CAFOS), DuPont (Retrocam), MMT (Red Channel Spectrograph), LBT (MODS).





\appendix

\section{Filters} \label{secAppFilt}
We list here in Table \ref{tabApFilt} the broad band filters used throughout this work, both from public surveys and follow up campaigns. 
\begin{deluxetable*}{lccc}[h]
\tabletypesize{\small}
\tablecaption{List of broad band filters used in this work and their characteristics (Telescope/Survey, central wavelength and width).\label{tabApFilt}}
\tablewidth{0pt}
\tablehead{ \colhead{Filter name} & \colhead{Instrument/Survey} & \colhead{$\rm \lambda_{c}$} & \colhead{$\rm \Delta \lambda$}
 \\
  & & [$\mu$m] & [$\mu$m]
}
\startdata
$g_{\mathrm{P1}}$ & PS1 & 0.487 & 0.117\\
$r_{\mathrm{P1}}$ & PS1 & 0.622 & 0.132\\
$i_{\mathrm{P1}}$ & PS1 & 0.755 & 0.124\\
$z_{\mathrm{P1}}$ & PS1 & 0.868 & 0.097\\
$y_{\mathrm{P1}}$ & PS1 & 0.963 & 0.062\\
$g_{\mathrm{decam}}$ & DECaLS & 0.475 & 0.152 \\
$r_{\mathrm{decam}}$ & DECaLS & 0.640 & 0.143\\
$z_{\mathrm{decam}}$ & DECaLS & 0.928 & 0.147\\
$Y$  & UKIDSS/VHS & 1.000 & 0.120\\
$J$  & UKIDSS/VHS & 1.250 & 0.213\\
$H$  & UKIDSS/VHS & 1.650 & 0.307\\
$K$  & UKIDSS/VHS & 2.150 & 0.390\\
$z_{\mathrm{O2K}}$ & CAHA 3.5m/Omega2000 & 0.908 & 0.158\\
$Y_{\mathrm{O2K}}$ & CAHA 3.5m/Omega2000 & 1.039 & 0.205\\
$J_{\mathrm{O2K}}$ & CAHA 3.5m/Omega2000 & 1.234 & 0.164\\
$I_{\mathrm{E}}$ & NTT/EFOSC2 & 0.793 & 0.126\\
$Z_{\mathrm{E}}$ & NTT/EFOSC2 & $>$0.840 & $-$\\
$J_{\mathrm{S}}$ & NTT/SofI & 1.247 & 0.290\\
$i_{\mathrm{w}}$ & CAHA 3.5m/CAFOS & 0.762  & 0.139\\
$i_{\mathrm{MMT}}$ & MMT/MMTCam & 0.769  & 0.130\\
$Y_{\mathrm{retro}}$ & su Pont/Retrocam & 1.000 & 0.120\\
$g_{\mathrm{G}}$ & MPG 2.2m/GROND & 0.459 & 0.137\\
$r_{\mathrm{G}}$ & MPG 2.2m/GROND & 0.622 & 0.156\\
$i_{\mathrm{G}}$ & MPG 2.2m/GROND & 0.764 & 0.094\\
$z_{\mathrm{G}}$ & MPG 2.2m/GROND & 0.899 & 0.128\\
$J_{\mathrm{G}}$ & MPG 2.2m/GROND & 1.240 & 0.229\\
$H_{\mathrm{G}}$ & MPG 2.2m/GROND & 1.647 & 0.264\\
$K_{\mathrm{G}}$ & MPG 2.2m/GROND & 2.171 & 0.303\\
$W1$ & ALLWISE    & 3.353 & 0.663\\
$W2$ & ALLWISE    & 4.603 & 1.042
\enddata
\end{deluxetable*}
\section{Spectroscopically rejected objects} \label{secApSpecRej}
We report in Table \ref{tabSpecRej} the Galactic contaminants found in our spectroscopic follow-up observations, which satisfied our selection criteria considering the PS1 PV3 database information (three sources). We list names, coordinates, $z_{\mathrm{P1}}$, $y_{\mathrm{P1}}$, $Y$ and $J$ magnitudes. An accurate spectral classification of the sources is beyond the scope of this work.
\begin{deluxetable*}{lccccccc}[h]
\tabletypesize{\small}
\tablecaption{Objects spectroscopically confirmed to not be high redshift quasars.\label{tabSpecRej}}
\tablewidth{0pt}
\tablehead{ \colhead{Name} & \colhead{RA (J2000)} & \colhead{DEC (J2000)} &
 \colhead{$z_{\mathrm{P1}}$} & \colhead{$y_{\mathrm{P1}}$} & \colhead{$Y$} & \colhead{$J$}
}
\startdata
 PSO229.40365$-$22.37078 & 229.403651 & -22.3707877 & $>$22.36 & 20.36 $\pm$ 0.14 & $-$ & 20.95 $\pm$ 0.27\\ 
 PSO267.27554+15.6457  & 267.2755422 & 15.64579622 & 22.48 $\pm$ 0.31 & 20.69 $\pm$ 0.13 & $-$ & 20.31 $\pm$ 0.18\\ 
 PSO357.24231+25.77427 & 357.2423123 & 25.77427024 & $>$22.81 & 20.72 $\pm$ 0.13 & 21.52 $\pm $0.2 & 21.16 $\pm$ 0.14 
\enddata
\end{deluxetable*}

\clearpage

\clearpage

\end{document}